\documentclass{JHEP3}

\usepackage{graphicx}
\usepackage{amssymb}
\usepackage{slashed}
\usepackage{amsmath}
\usepackage{tabularx}
\usepackage{url}
\usepackage{feynmf}
\usepackage{dcolumn}
\usepackage{bm}
\usepackage{verbatim}
\usepackage{float}
\usepackage{slashed}
\usepackage{appendix}

\newcommand{\be}{\begin{equation}}
\newcommand{\eq}{\end{equation}}
\newcommand{\bea}{\begin{eqnarray}}
\newcommand{\eea}{\end{eqnarray}}
\newcommand{\ba}{\begin{eqnarray*}}
\newcommand{\ea}{\end{eqnarray*}}
\newcommand{\barr}{\begin{array}}       
\newcommand{\earr}{\end{array}}
\def\nn{\nonumber}

\newcommand{\bi}{\begin{itemize}}
\newcommand{\ei}{\end{itemize}}

\newcommand{\draftnote}[1]{\textbf{#1}}

\newcommand{\gsim}{\gtrsim}

\renewcommand{\l}{\lambda}
\renewcommand{\k}{\kappa}

\newcommand{\amuSUOL}{a_\mu^{\rm SUSY,1L}}

\newcommand{\amuneu}{a_\mu^{\chi^0}}
\newcommand{\amucha}{a_\mu^{\chi^\pm}}

\newcommand{\Usmu}{U^{\tilde{\mu}}}
\newcommand{\MSUSY}{M_{\rm SUSY}}

\title{Low-energy Observables and General Gauge Mediation in the MSSM and NMSSM.\\}

\author{Arun M. Thalapillil\\ 
Enrico Fermi Institute and Department of Physics, \\
University of Chicago, 5640 South Ellis Avenue, Chicago, IL 60637 \\
\ E-mail:  \email{madhav@uchicago.edu} }
\abstract{
We study constraints on the general gauge mediation (GGM) parameter space arising from low-energy observables in the MSSM and NMSSM. Specifically, we look at the dependence of the spectra and observables on the correlation function ratios in the hidden sector where supersymmetry is presumably broken. Since these ratios are not a priori constrained by theory, current results from the muon anomalous magnetic moment and flavor physics can potentially provide valuable intuition about allowed possibilities. It is found that the muon anomalous magnetic moment and flavor-physics observables place significant constraints on the GGM parameter space with distinct dependences on the hidden sector correlation function ratios. The particle spectra arising in GGM, with the possibility of different correlation function ratios, is contrasted with common intuition from regular gauge mediation (RGM) schemes (where the ratios are always fixed). Comments are made on precision gauge coupling unification, topography of the NLSP space, correlations of the muon anomalous magnetic moment with other observables, and approximate scaling relations in sparticle masses with respect to the high-scale correlation function ratios. 
}


\preprint{EFI 10-32}

\begin{document}
% ----------------------------------------------------------------------------------------------------------

\section{Introduction}
Supersymmetry (SUSY) is considered to be one of the most promising extensions to the standard model (SM), since it potentially explains many of the problems in it. In all supersymmetric extensions of the SM it is required, for reasons of viability, that SUSY be broken in a hidden sector which then gets communicated to the visible sector (see for example \cite{Martin:1997ns} and references therein). An appealing implementation of this SUSY breaking paradigm has been regular gauge mediation (RGM) \cite {Giudice:1998bp} since it solves the flavor problem and at the same time is parsimonious compared to, for instance, gravity mediation. The phenomenology of the RGM implementations have been studied extensively in the MSSM and its extensions (see for example \cite {Giudice:1998bp}  and related citations).
\par
Recently it was pointed out in \cite{Meade:2008wd} that the parameter space of gauge mediation is potentially larger and a convenient parametrization to account for and study this was introduced there with the terminology of general gauge mediation (GGM). Since then there have been many studies exploring the phenomenology of this enlarged parameter space \cite{{Carpenter:2008he},{Rajaraman:2009ga},{Meade:2009qv},{Ruderman:2010kj},{Abel:2009ve},{Abel:2010vb},{Kobayashi:2009rn}}. Also, on the theoretical side, the ideas have since been extended \cite{{Buican:2008ws},{Intriligator:2010be},{Dumitrescu:2010ha},{McGarrie:2010kh},{McGarrie:2010qr},{McGarrie:2010yk}}. The mass spectra in GGM was first investigated in \cite{Carpenter:2008he} where it was pointed out that the spectra could be very distinct from minimal gauge mediation. For the MSSM, an investigation of the NLSP topography for various GGM correlation function ratios was instigated in \cite{Rajaraman:2009ga} and more recently the prompt decays of NLSPs and co-NLSPs were investigated in \cite {Meade:2009qv} and \cite{Ruderman:2010kj}. Low-energy observables with fixed ratios of the correlation functions were investigated in \cite{Abel:2009ve} and in \cite{Abel:2010vb} benchmarks points in GGM were discussed (again for fixed ratios), in the context of LHC searches. 
\par
We wish to complement the studies of ~\cite{{Carpenter:2008he},{Rajaraman:2009ga},{Abel:2009ve},{Abel:2010vb}}, by exploring the effects of different hidden sector correlation function ratios on low-energy observables and the NLSP topography, in the context of the Minimal Supersymmetric Standard Model (MSSM) and the Next-to-Minimal Supersymmetric Standard Model (NMSSM). The hope in the present study is that since these ratios are not a priori constrained by theory, current results from the muon anomalous magnetic moment and flavor-physics observables can be useful guides. If low-energy SUSY is discovered at the LHC and if the soft masses can be measured to very good accuracy ($\sim1\%$ uncertainty) sometime in the future, then it may even be possible to determine the correlation functions at the high scale to good precision using renormalization group invariants~\cite{{Carena:2010gr}, {Carena:2010wv}}. More recently, the role of the GGM messenger scale in the context of mass sum rules and RG invariants was considered in \cite{{Jaeckel:2011ma},{Jaeckel:2011qj}}
\par
In the next two subsections we introduce our notations and definitions for the MSSM and NMSSM, and then proceed to review relevant features of RGM and GGM that are the main focus of our investigation. Then in section 2 we briefly review some of the low-energy observables and collider bounds that we use to constrain GGM and to explore correlation function dependences. In section 3 we present our numerical study along with results and observations. Section 4 is the summary.

\subsection{The MSSM and NMSSM}
MSSM is a supersymmetric extension of the SM where the effects of SUSY breaking are parametrized by super-renormalizable soft terms. For reasons of anomaly cancellation and holomorphicity of the superpotential there are two Higgs doublets in the MSSM apart from the sparticles. In this subsection we fix our notation.
\par
The superfields (denoted by a ` $\hat{ }$ ' ) in the MSSM are defined as
\bea\label{superfields}
\widehat{Q}(3,2,1/3) &=& \left(\barr{c} \widehat{U}_L \\ \widehat{D}_L
\earr \right) ,~\widehat{U}_{R}^{c}(\bar{3},1,-4/3),~ \widehat{D}_{R}^{c}(\bar{3},1,2/3) \; ,  \\ \nn
\widehat{L}_\alpha(1,2,-1) &=& \left(\barr{c} \widehat{\nu}^\alpha_{L} \\ \widehat{E}^\alpha_L
\earr \right) ,~\widehat{E}_{R}^{c}(1,1,2)  \; ,  \\ \nn
\widehat{H}_u(1,2,1)  &=& \left(\barr{c} \widehat{H}_u^+ \\ \widehat{H}_u^0
\earr \right) , \
\widehat{H}_d (1,2,-1) = \left(\barr{c} \widehat{H}_d^0 \\ \widehat{H}_d^-
\earr \right) ,
\eea
where all the generation indices have been suppressed and the respective $SU(3)_{C}\times SU(2)_{L} \times U(1)_{Y}$ assignments are shown in brackets.
The MSSM superpotential (with all the generation indices again suppressed) is given by

\bea\label{WMSSM}
\mathcal{W}_\text{\tiny{MSSM}} = Y_u\, \widehat{H}_u \cdot \widehat{Q}\;
\widehat{U}^c_R + Y_d\, \widehat{H}_d \cdot \widehat{Q}\;
\widehat{D}^c_R + Y_e\, \widehat{H}_d \cdot \widehat{L}\;
\widehat{E}_R^c +  \mu \,\widehat{H}_u \cdot 
\widehat{H}_d   \; .
\eea
\par
where we define $A\cdot B= \epsilon_{ij}A^iB^j$. The corresponding soft SUSY breaking terms in the MSSM may be parametrized as 
\bea
-{\cal L}_\mathrm{soft}^\text{\tiny{MSSM}} &=&
-\frac{1}{2}(M_1 \,\tilde{B}\tilde{B}+M_2\, \tilde{W}\tilde{W}+M_3\, \tilde{g}\tilde{g}+ \mathrm{h.c.}) 
+ m_{H_u}^2 | H_u |^2 + m_{H_d}^2 | H_d |^2  \nn \\
&&+m_{\tilde{Q}}^2|Q^2| + m_{\tilde{U}}^2|U_R^2|+m_{\tilde{D}}^2|D_R^2| +m_{\tilde{L}}^2|L^2| +m_{\tilde{E}}^2|E_R^2|
\nn \\
&&+ (Y_u A_u\; Q \cdot H_u\; U_R^c - Y_d A_d\; Q \cdot H_d\; D_R^c 
- Y_{e} A_{e}\; L \cdot H_d\; E_R^c+ \mathrm{h.c.})\nn \\ &&
+(B\mu\, H_u \cdot H_d\;  + \mathrm{h.c.}) \; ,
\label{SOFTMSSM}
\eea
where $\tilde{B},\tilde{W}$ and $\tilde{g}$ are the gauginos corresponding to $ U(1)_{Y},SU(2)_{L}$ and $SU(3)_{C}$ respectively. 
We have the usual definition for $\tan \beta$
\be
\tan\beta=\frac{v_u}{v_d}  \; ,  
\eq
and the soft mass terms, $A$-terms and $B\mu$ are defined in the standard way~\cite{Martin:1997ns}. The MSSM Higgs potential has four free parameters
\be
\mu,B\mu, m_{H_u}^2,m_{H_d}^2  \; .
\eq
Using the minimization conditions we may trade some of these parameters for others, for instance $\tan\beta$ and $M_{Z^0}$.
\par
Augmenting the MSSM with a singlet chiral superfield $\widehat{N}$ we may define the NMSSM superpotential as (see for example~\cite{Ellwanger:2009dp})
\bea\label{WNMSSM}
\mathcal{W}_\text{\tiny{NMSSM}} =  \mathcal{W}_\text{\tiny{MSSM}}+\lambda \widehat{N}\,\widehat{H}_u \cdot 
\widehat{H}_d + \xi_F \widehat{N} + \frac{1}{2} \mu' \widehat{N}^2 +
\frac{\kappa}{3} \widehat{N}^3  \; .
\eea
\par
The corresponding soft SUSY breaking masses and couplings in the NMSSM are
\bea
-{\cal L}_\mathrm{soft}^\text{\tiny{NMSSM}} &=&
-\frac{1}{2}(M_1 \,\tilde{B}\tilde{B}+M_2\, \tilde{W}\tilde{W}+M_3\, \tilde{g}\tilde{g}+ \mathrm{h.c.}) +
m_{H_u}^2 | H_u |^2 + m_{H_d}^2 | H_d |^2 
+ m_{N}^2 | N |^2 \nn \\
&&+m_{\tilde{Q}}^2|Q^2| + m_{\tilde{U}}^2|U_R^2|+m_{\tilde{D}}^2|D_R^2| +m_{\tilde{L}}^2|L^2| +m_{\tilde{E}}^2|E_R^2|
\nn \\
&&+ (Y_u A_u\; Q \cdot H_u\; U_R^c - Y_d A_d\; Q \cdot H_d\; D_R^c 
- Y_{e} A_{e}\; L \cdot H_d\; E_R^c+ \mathrm{h.c.})\nn \\ &&
+(\lambda A_\lambda\, H_u \cdot H_d\; N + \frac{1}{3} \kappa A_\kappa\,
N^3 + m_3^2\, H_u \cdot H_d + \frac{1}{2}m_{N}'^2\, N^2 + \xi_N\, N 
+ \mathrm{h.c.}) \;  .
\label{SOFTNMSSM}
\eea
Usually a simpler NMSSM potential is considered by imposing scale invariance whereby the parameters $\mu,\mu',\xi_{F},m^2_3, m'^2_N$ and $\xi_N$ are vanishing. This potential has a discrete $\mathbb{Z}_3$ symmetry and the Higgs sector is described by seven parameters
\be
\lambda,\kappa,A_\lambda,A_\kappa,m_{H_u}^2,m_{H_d}^2,m_{N}^2  \; .
\eq
One may again use the minimization conditions for the NMSSM Higgs potential to trade some of these for other quantities. 
% ----------------------------------------------------------------------------------------------------------

\draftnote{}

\subsection{Regular Gauge Mediation and General Gauge Mediation}
\par
The need for a separate SUSY breaking sector to accommodate viable phenomenology is well known. In this paradigm the SUSY breaking is expected to occur in a separate hidden sector and the SUSY breaking is then communicated to the visible sector directly/indirectly. If the mediators between the hidden and visible sector are physical particles then they constitute the messengers of the SUSY breaking. The mass scale associated with these messenger superfields is termed the \textit{messenger scale}. A particular realization of this scheme is gauge mediation where the messenger fields are assumed to be flavor blind. 
\par
In \textit{regular gauge mediation} (RGM), the gaugino and sfermion soft masses arise from loops involving messenger fields and are given by~\cite{Giudice:1998bp} 
\begin{equation}
\begin{array}{l}
M_r = \frac{\alpha_r}{4 \pi} N_m \frac{F}{M_{m}} f(x)  \; ,  
\\
m_{\tilde{f}}^2 = 2 N_m \left| \frac{F}{M_{m}} \right|^2 \sum\limits_r 
\left( \frac{\alpha_r}{4 \pi} \right)^2 \mathcal{C}_2 (f\vert r) g(x),
\label{eqn:GMSBgenmasses}
\end{array} 
\end{equation}
where
\begin{equation}\begin{array}{l}
f(x) = \left. \frac{1}{x^2} \right.
\left[ (1+x) \log (1+x) + (1-x) \log (1-x) \right]  \; ,  	\\
g(x) = \left. \frac{1+x}{x^2} \right.
\left[ \log (1+x) - 2 \text{Li}_2 (x/[1+x]) +
\frac{1}{2} \text{Li}_2 (2x / [1+x]) \right] + (x \rightarrow -x) \; ;
\\
\label{eqn:GMSBgenmassesfg}
\end{array}
\end{equation}
$x = F/M_{m}^2$, $\mathcal{C}_2(f\vert r)$ are the quadratic Casimirs, $M_{m}$ is the messenger scale and $N_m$ is the number of copies of the messenger particles in the loop (or in other words the sum of the Dynkin indices). The RGM includes both minimal gauge mediation, where $N_m=1$, and non-minimal gauge mediation where $N_m > 1$. 
\par
For small $x$, $f(x)\rightarrow\,1$ and $g(x) \rightarrow\,1$ in Eq. (\ref{eqn:GMSBgenmasses}). This implies that for $N_m=1$
\be
M_r \approx m_{\tilde{f}}  \; ,  
\eq
since the sfermion mass squared is at 2-loop while the gaugino mass is at 1-loop. In RGM we can certainly tune the gaugino to sfermion mass ratio to some extent by increasing $N_m$, but beyond a point it is very difficult without further inputs to get a viable model along with gauge coupling unification. We will see that in GGM it is easier to get large or small gaugino to sfermion mass ratios naturally.
\par
In RGM, the gravitino is always the LSP since
\be
m_{LSP}=m_{3/2} \sim \frac{F}{M_{pl}}   \; ,  
\eq
and  $M_{m}\ll M_{pl}$. At low (electroweak) scale $\alpha_{1},\alpha_ {2}<\alpha_ {3}$ and hence the NLSP is usually the $\tilde{\chi}$ or $\tilde{l}$ in large regions of the parameter space.

\par
Moreover due to gauge coupling unification and the fact that $M_{a}/g_{a}^2$ is an RG invariant (to one-loop order) it is always true, at any scale, that
\be
\frac{M_{1}}{g_1^2}=\frac{M_{2}}{g_2^2}=\frac{M_{3}}{g_3^2}   \; ,  
\eq
at this order.
\par
The corresponding expressions in \textit{general gauge mediation} (GGM) are~\cite{Meade:2008wd}
\begin{equation}
\begin{array}{l}
M_r = g_r^2 M_{s} \tilde{B}^{1/2}_{r} (0)  \; ,  \\
m_{\tilde{f}}^2 = g_1^2 Y_f \zeta + \sum\limits_{r = 1}^3 g_r^4 \mathcal{C}_2
(f\vert r) M_{s}^2 \tilde{A}_r  \; ,  
\end{array}
\label{GGMmass}
\end{equation}
where
\begin{equation}
\begin{array}{lll}
\tilde{A}_r &=& - \int \frac{d^4 p}{(2 \pi)^4} \frac{1}{M_{s}^2 p^2} \left(
3 \tilde{C}_1^{(r)} (p^2 / M_{s}^2) - 4 \tilde{C}_{1/2}^{(r)} (p^2 / M_{s}^2)
+ \tilde{C}_0^{(r)} (p^2 / M_{s}^2) \right) \\
 &=& - \frac{1}{16 \pi^2} \int dy \left( 3 \tilde{C}_1^{(r)} (y)
 - 4 \tilde{C}_{1/2}^{(r)} (y) + \tilde{C}_0^{(r)} (y) \right)  \; .  \\
\end{array}
\end{equation}
$\tilde{B}^{1/2}_{r} (0),\ \tilde{C}_{\rho}^{(r)}$ (where $\rho$ is a vector, fermion or scalar index) are associated with the current-current correlators in the hidden sector, $\mathcal{C}_2(f\vert r)$ are again the quadratic Casimirs, $\zeta$ is a possible Fayet-Iliopoulos term (D-term) and $M_{s}$ is some characteristic soft SUSY-breaking scale associated with the hidden sector. With this definition note that the correlation functions $(\tilde{B}_r,\tilde{A}_r)$ are dimensionless. Also note that the subscript `$r$' labels the associated SM gauge group. Henceforth we will abbreviate  $\tilde{B}^{1/2}_{r} (0)$ as just $ \tilde{B}_{r}$, not to be confused with the notation for a bino ($\tilde{B}$) that has no subscript index. 
\par
 In the GGM parametrization the interpretation of the messenger sector is enlarged to accommodate more general scenarios, for instance, cases where the sector is strongly coupled and there are no explicit messenger particles. Note that the RGM limit can be obtained from the GGM scenario by considering points in the vicinity of
\be
\tilde{B}_{r}=\sqrt{\frac{\tilde{A}_{r}}{2}}  \; ,  
\eq
for the case when there is no inter or intra hierarchies among the $\tilde{B}_r$ and $\tilde{A}_r$. By considering points near the above region we can accommodate both minimal and non-minimal RGM cases.
\par
Thus now we have the varied possibilities
\bea
M_r ~\approx ~m_{\tilde{f}} , ~M_r ~\ll~m_{\tilde{f}} ,~ M_r ~\gg~ m_{\tilde{f}} \; ,
\eea
depending on the correlation function hierarchies (i.e. hierarchies among $\tilde{B}_{r}$ and $\tilde{A}_{r}$). Also note that it is the case that in general now
\be
\frac{M_{1}}{g_1^2}\neq\frac{M_{2}}{g_2^2}\neq\frac{M_{3}}{g_3^2}   \; ,  
\eq
at any scale, even when there is gauge coupling unification, due to possibly different $\tilde{B}_{r}=M_{r}/g_{r}^2$.
\par
On a final note, the Fayet-Iliopoulos term ($\zeta$) that makes an appearence in the expression for the sfermion masses  is potentially dangerous since it could lead to tachyonic masses. It is usually set to zero due to this by imposing some discrete symmetry like messenger parity~\cite{Meade:2008wd}.

% ----------------------------------------------------------------------------------------------------------

\draftnote{}
% ----------------------------------------------------------------------------------------------------------

\section{Low-energy observables and constraints}

\subsection{Anomalous magnetic moment of the muon}
% ----------------------------------------------------------------------------------------------------------
The anomalous magnetic moment of the muon has been measured to very good precision. Current measurements yield a value~\cite{Bennett:2006fi}
\begin{equation}
\begin{array}{l}
\left[a_{\mu}\right]_{ \text{\tiny Exp}}=(11\,659\,2080.0 \pm 54_{\rm stat} \pm 33_{\rm syst})\times 10^{-11} 
\end{array}
\end{equation}
\par
The theoretical contributions may be divided as~\cite{{Prades:2009qp},{Bennett:2006fi},{Davier:2009zi},{Prades:2009tw}}
\begin{equation}
\begin{array}{l}
\left[a_{\mu}\right]_{ \text{\tiny QED}}=(11\,658\,4718.1 \pm 1.6)\times 10^{-11}  \; ,  \\ [2mm]
\left[a_{\mu}\right]^{e^{+}e^{-}}_{ \text{\tiny Had. LO}}=( 6955 \pm 40_{\rm exp}\pm  7_{\rm QCD}) \times 10^{-11} \; ,  \\ [2mm]
\left[a_{\mu}\right]^{e^{+}e^{-}}_{ \text{\tiny Had. NLO}}=(-97.9 \pm 0.8_{\rm exp} \pm 0.3_{\rm rad})\times 10^{-11}  \; ,  \\ [2mm]
\left[a_{\mu}\right]_{ \text{\tiny LBL}}=(105\pm 26)\times 10^{-11} \; ,   \\ [2mm]
\left[a_{\mu}\right]_{ \text{\tiny EW}}=(154 \pm 1_{\rm had} \pm 2_{\rm Higgs})\times 10^{-11} \; ,  \\ [2mm]
\left[a_{\mu}\right]_{ \text{\tiny Tot. SM}}=(11\,659\,1834 \pm 41_{\rm \, LO\,had}  \pm 26_{\rm \, NLO\,had}  \pm 2_{\rm other} )\times 10^{-11}  \; .
\end{array}
\end{equation}
In the above `had.' means hadronic and `LBL' stands for light-by-light scattering. For the hadronic vacuum polarization corrections we have quoted the value from $\sigma(e^+ e^-\rightarrow Hadrons)$ low-energy data. 
Comparing the theoretical prediction and the most current experimental result gives a discrepancy
\begin{equation}
\begin{array}{l}
\left[\Delta a_{\mu}\right]^{\text{\tiny Exp.}}_{\text{\tiny Th}.}=~(246\pm 80)\times 10^{-11}  \; .
\end{array}
\end{equation}
%%%%%%%%%%%%%%%%%%%%%%%%%%%%%%%%%
\FIGURE
{
\includegraphics[scale=0.420]{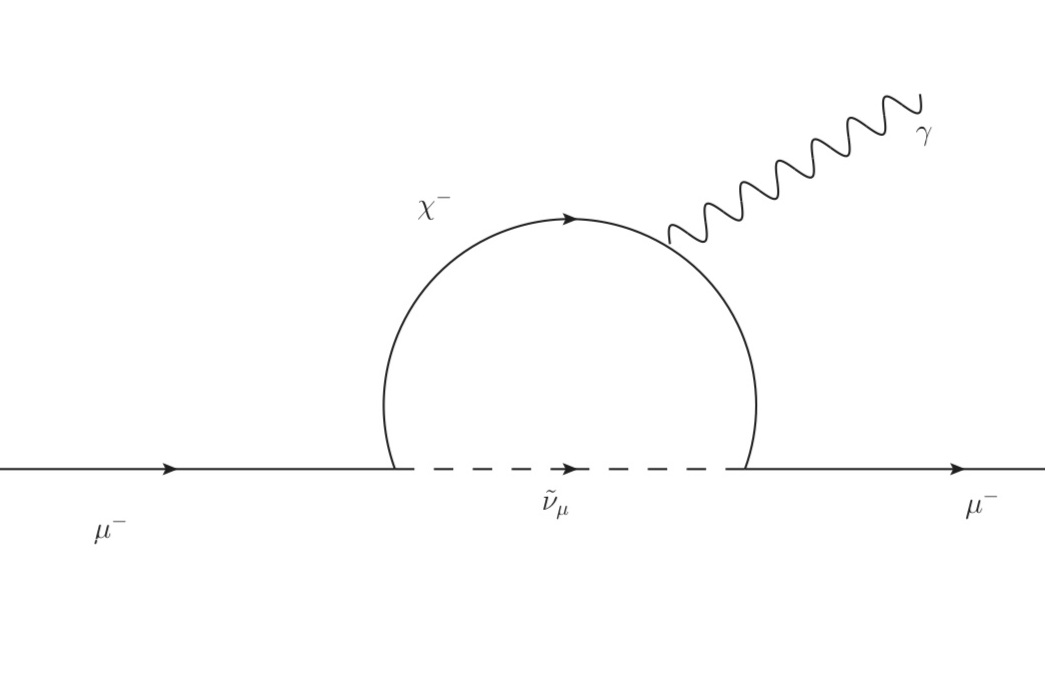}
\includegraphics[scale=0.420]{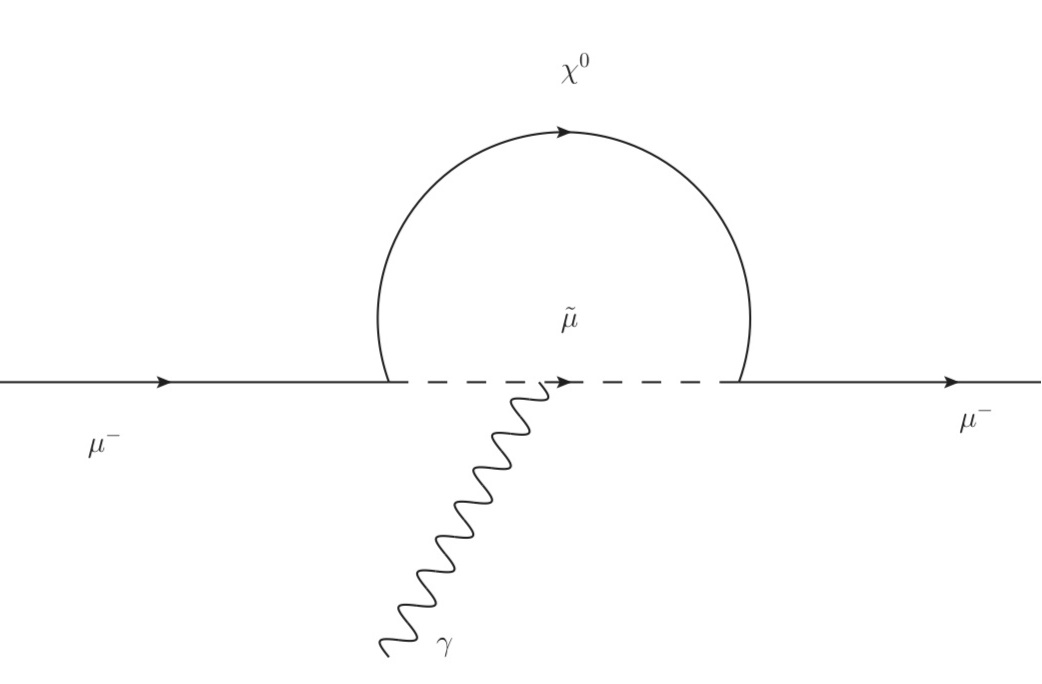}
\caption{The well known 1-loop contributions to $(g-2)_\mu$ in the MSSM. The dominant contribution comes from the chargino diagram. For brevity the diagrams are shown in terms of the mass eigenstates.}
}
%%%%%%%%%%%%%%%%%%%%%%%%%%%%%%%%%%
\par
In the MSSM the 1-loop contributions to the muon anomalous magnetic moment come from loop diagrams with a chargino-sneutrino and neutralino-smuon in the loop and are given by ~\cite{MartinWells}
\begin{eqnarray}
\label{amuSUOL}
\amuSUOL &=\amucha+\amuneu  \; ,  
\end{eqnarray}
with
\begin{eqnarray}
\amucha&=&\frac{m_\mu}{16\pi^2}\sum_{k}
\left[
\frac{m_\mu}{12m_{\tilde{\nu}_\mu}^2}(|q_{k}^L|^2+|q_{k}^R|^2)
X_1(x_{k})
+\frac{2m_{\chi_k^\pm}}{3m_{\tilde{\nu}_\mu}^2}
{\rm Re}[q_{k}^Lq_{k}^R] X_2(x_{k})
\right]
\label{amuchipm}
 \; ,  \\
 \amuneu&=&\frac{m_\mu}{16\pi^2}\sum_{i,m}
\left[
-\frac{m_\mu}{12m_{\tilde{\mu}_m}^2}(|p_{im}^L|^2+|p_{im}^R|^2)
Y_1(x_{im})
+\frac{m_{\chi_i^0}}{3m_{\tilde{\mu}_m}^2}
{\rm Re}[p_{im}^Lp_{im}^R] Y_2(x_{im})
\right]
 \; ,  
\label{amuchi0}
\end{eqnarray}
where $i=1\ldots4$ and $k=1,2$ denote the $\chi_0$ and $\chi_{\pm}$, $m=1,2$ denotes the $\tilde{\mu}$ index. Also we have the definitions
\begin{eqnarray}
q_{k}^L &=& - g_2 V_{k1} \; ,  \\
q_{k}^R &=& y_\mu U_{k2} \; ,  \\
p_{im}^L &=& \frac{1}{\sqrt2}(g_1 N_{i1}+g_2 N_{i2})\Usmu_{m1}{}^*
-y_\mu N_{i3}\Usmu_{m2}{}^* \; ,  \\
p_{im}^R &=& \sqrt2 g_1 N_{i1} \Usmu_{m2} + y_\mu N_{i3}\Usmu_{m1}  \; .
\end{eqnarray}
The matrices $U,V,N,U^{\tilde{f}}$ relate the mass eigenstates to the interaction eigenstates for the gauginos, Higgsinos and sfermions and are defined as
\bea
\chi_a^{+}&=&V_{ab}\psi^{+}_b \; ,  \\ \nn
\chi_a^{-}&=&U_{ab}\psi^{-}_b \; ,  \\ \nn
\chi_a^{0}&=&N_{ab}\psi^{0}_b \; ,  \\ \nn
\tilde{f}_a &=& U^{\tilde{f}}_{ah} \tilde{f}_{h}  \; .
\eea
where the LHS states are mass eigenstates (charginos, neutralinos and sfermions) and $h$ labels the chirality state.
With the definitions $x_{im}=m_{\chi_i^0}^2/m_{\tilde{\mu}_m}^2$ and
$x_k=m_{\chi_k^\pm}^2/m_{\tilde{\nu}_\mu}^2$ the loop functions
are ~\cite{MartinWells}
\begin{eqnarray}
X_1(x) &=& \frac{ 2 }{(1-x)^4 }\big[
2+3x-6x^2+x^3+6x\log x
\big] \; ,  \\
X_2(x) &=& \frac{ 3 }{(1-x)^3 }\big[
-3+4x-x^2-2\log x
\big] \; ,  \\
Y_1(x) &=& \frac{ 2 }{(1-x)^4 }\big[
1-6x+3x^2+2x^3-6x^2\log x
\big] \; ,  \\
Y_2(x) &=& \frac{ 3 }{(1-x)^3 }\big[
1-x^2+2x\log x
\big] \; .
\end{eqnarray}
\par
To get some intuition about the above 1-loop contributions let us make some simplifying assumptions. Let us take $\tan\beta$ to be large with all soft masses roughly equal to a common scale $M_{SUSY}$; then the above expressions may be approximated as
\begin{eqnarray}
\amuneu&\simeq& \frac{g_1^2-g_2^2}{192\pi^2}\frac{m_\mu^2}{\MSUSY^2}
\ {\rm sign}(\mu M_2)\ \tan\beta \; ,  \\
\amucha&\simeq& \ \frac{g_2^2}{32\pi^2}\ \ \frac{m_\mu^2}{\MSUSY^2}
\ {\rm sign}(\mu M_2)\ \tan\beta \; ,  
\label{amucha}
\end{eqnarray}
where real parameters and equal signs of $M_1$ and $M_2$ have been
assumed. From the above approximate expressions it is clear that the $\chi^\pm$ contribution to $a_\mu$ is numerically larger than the $\tilde{\chi}$ contribution. It is also noted that the sign of the SUSY contribution is determined by the sign of $\mu$.
\par
\par
We can estimate a large class of 2-loop contributions by noting that the large logarithms arising from QED corrections to one-loop SUSY diagrams may be quantified as 
\begin{equation}
\delta a_{\mu}^{SUSY+QED}\simeq\delta a_{\mu}^{1-Loop
(SUSY)}\left(1-\frac{4\alpha}{\pi}\ln \frac{m_{SUSY}}{m_{\mu}}\right)  \; .
\end{equation} 
This leads to a \textit{reduction} by a few percent of the LO contributions.

\par
The contributions to $a_{\mu}$ at 1-loop in NMSSM are identical to those in the MSSM provided that we take into account the additional singlino state in the $\tilde{\chi}$ calculation. The CP-even and CP-odd parts of the singlet field $N$ in the NMSSM will mix with the neutral components of the Higgs doublets. In certain regions of the parameter space it is therefore possible to have a relatively light CP-odd Higgs scalar which can provide a significant contribution to $a_\mu$.
\par
In the SM or MSSM, Higgs effects are usually negligibly small because of the current lower bounds on the Higgs masses. The SM 1-loop Higgs diagram, for instance, is about 2-3 orders of magnitude
below experimental sensitivity for $m_h\gsim114$~GeV.
\par
 In the NMSSM on the other hand the mass bounds are relaxed relative to the MSSM. In the NMSSM the lightest CP-odd Higgs boson $a_1$ can be as light as a few GeV and still satisfy constraints from flavor physics, especially from $Br(B_s\rightarrow \mu^+\mu^-)$, for low values of $\tan\beta$ or when the
loop-induced coupling is small.
\par
The SUSY Higgs contributions (CP-even, CP-odd
and charged) to $a_\mu$ may be calculated as~\cite{Leveille:1977rc} 
\begin{eqnarray}
\delta a_{\mu}^{1L\ CP-even}&=&\frac{G_\mu
m_{\mu}^2}{4\sqrt{2}\pi^2\cos^2\beta} \sum_{i} X_{i2}^2\int_0^1{\frac{x^2(2-x)\,dx}{x^2+
\left(\frac{m_{h_i}}{m_{\mu}}\right)^2(1-x)}} \; ,  \\
\delta a_{\mu}^{1L\ CP-odd}&=&-\frac{G_\mu
m_{\mu}^2\tan^2\beta}{4\sqrt{2}\pi^2}
\sum_{i} 
Y^2_{i1}\int_0^1{\frac{x^3\,dx}{x^2+
\left(\frac{m_{a_i}}{m_{\mu}}\right)^2(1-x)}}\label{amuHiggs1L} \; ,  \\
\delta a_{\mu}^{1L\ charged}&=&
\frac{G_\mu
m_{\mu}^2\tan^2\beta }{4\sqrt{2}\pi^2}
\int_0^1{\frac{x(x-1)\,dx}{x-1+
\left(\frac{m_{H^{\pm}}}{m_{\mu}}\right)^2}} \; ,  
\end{eqnarray}
where $X_{ij},Y_{ij}$  are the Higgs mixing matrices for the CP-even and CP-odd cases.
\par
It has been noticed previously that the 1-loop and 2-loop Higgs contributions come with opposite signs. Moreover it is seen from the above formula that the 1-loop CP-odd scalar contribution is negative. This will be potentially relevant to the NMSSM case where there is an extra CP-odd Higgs that could be relatively light.
% ----------------------------------------------------------------------------------------------------------

\subsection{Flavor-physics constraints}

\begin{table}[!h]
\begin{center}
\hspace*{-1.95cm}\begin{tabularx}{18cm}{|c|c|c|X}
\cline{1-3}
\textbf{Flavor-physics Observables} & \textbf{Measurements} & \textbf{2$\sigma$ bounds}& \\[2mm]
\cline{1-3}&&\\
${\rm Br}(B \to X_s \gamma) $& $ (3.52\pm0.23\pm0.09)\times10^{-4}$ &  $2.15\times 10^{-4}\leq \mathrm{Br(}b\to s\gamma\mathrm{)}\leq 4.89\times 10^{-4}  $\\[2mm]
\cline{1-3}&&\\
${\rm Br}(B_s \to \mu^+ \mu^-)$ &$ < 5.8 \times 10^{-8}$ &$ {\rm Br}(B_s \to \mu^+ \mu^-) < 6.6 \times 10^{-8}$\\[2mm]
\cline{1-3}&&\\
$\Delta M_{d}$ &$ (5.07\pm0.04) \times 10^{-1}$ &$ 4.99\times 10^{-1} < \Delta M_d <  5.15\times 10^{-1}$\\[2mm]
\cline{1-3}&&\\
$\Delta M_{s}$ &$ (1.777\pm0.012) \times 10^{+1}$ &$ 1.753\times 10^{+1} < \Delta M_s <  1.801\times 10^{+1}$\\[2mm]
\cline{1-3}&&\\
$\Delta_0(B \to K^* \gamma)$ & $(3.1 \pm 2.3)\times 10^{-2}$  &$-1.7\times 10^{-2} < \Delta_0 < 8.9\times 10^{-2}$\\[2mm]
\cline{1-3}&&\\
$\dfrac{{\rm Br}(K \to \mu \nu)}{{\rm Br}(\pi \to \mu \nu)}$ & $0.6358\pm0.0011$  & $0.6257 < \dfrac{{\rm Br}(K \to \mu \nu)}{{\rm Br}(\pi \to \mu \nu)} < 0.6459$\\[2mm]
$R_{\ell 23}$ & $1.004 \pm 0.007$  &$ 0.990 < R_{\ell 23} < 1.018$ \\[2mm]
\cline{1-3}&&\\
${\rm Br}(B_u \to \tau \nu_\tau)$ &$ (1.41 \pm 0.43) \times 10^{-4}$  & $ 0.39 \times 10^{-4} < {\rm Br}(B_u \to \tau \nu_\tau) < 2.42 \times 10^{-4}  $\\[2mm]
$R_{\tau\nu_\tau}$ & $ 1.28\pm 0.38 $  &$ 0.52 < R_{\tau\nu_\tau} < 2.04$ \\[3mm]
\cline{1-3}&&\\
${\rm Br}(B \to D^0 \tau \nu_\tau)$ &$ (8.6 \pm 2.4 \pm 1.1 \pm 0.6) \times 10^{-3} $ &$ 2.9\times 10^{-3} < {\rm Br}(B \to D^0 \tau \nu_\tau) < 14.2\times 10^{-3} $\\[2mm]
$\xi_{D \ell \nu}$ &$ 0.416 \pm 0.117 \pm 0.052$ &$ 0.151 < \xi_{D \ell \nu} < 0.681 $\\[2mm]
\cline{1-3}&&\\
${\rm Br}(D_s \to \tau \nu_\tau)$ &$ (5.7 \pm 0.4) \times 10^{-2}$  & $ 4.8 \times 10^{-2} < {\rm Br}(D_s \to \tau \nu_\tau) < 6.6 \times 10^{-2}$\\[2mm]
${\rm Br}(D_s \to \mu \nu_\mu)$ & $ 5.8\pm 0.4  \times 10^{-3}$  &$ 4.9 \times 10^{-3} < {\rm Br}(D_s \to \mu \nu_\mu) < 6.7 \times 10^{-3}$ \\[2mm]
\cline{1-3}
\end{tabularx}
\end{center}
\caption{Flavor-physics observables~\cite{{Nakao:2004th},{Barberio:2008fa},{Aubert:2007dsa},{Aaltonen:2007kv},{Akeroyd:2009tn},{Aubert:2008cy},{Antonelli:2008jg}}. Most of the intervals are adopted from the relevant table in the $\mathtt{SuperIso}$ manual \cite{Mahmoudi:2008tp}. For details on the derivation of these intervals the reader is pointed to references therein. The flavor observables listed in the table are defined in the text below.}
\label{flavor_physics}
\end{table}
Flavor physics can place strong constraints on supersymmetry since flavor-changing (FC) contributions that are in general loop-suppressed can become important due to $\tan\beta$ enhancement.
We use the flavor observables shown in Table\,\ref{flavor_physics} in our study of the GGM parameter space. The idea is that the hidden sector correlation functions ($\tilde{B}_r$, $\tilde{A}_r$) determine the sparticle spectra at some large scale which then RG evolve down to the low scale and potentially contribute to the Wilson coefficients of the various meson transitions.
\par
The rare branching ratios $Br(B\to X_s \gamma)$, $Br(B_s \to \mu^+ \mu^-)$, $Br(B\to\tau\nu)$ and  $Br(B \to D \tau \nu)$ are used with their $2\sigma$ intervals to check for viability of points. Note that the best bounds on a charged Higgs in the MSSM are from $Br(B\to X_s \gamma)$ as opposed to collider studies. This is a telling feature of the complementary nature of collider and low-energy constraints. 
\par
\par
The $b\to s\gamma$ transitions may be parametrized by
\begin{equation}
{\cal H}_{eff}=\frac{G_F}{\sqrt{2}}\sum_{p=u,c} V^*_{ps}V_{pb} \sum_{i=1}^8 C_i(\mu) \, O_i\;,
\end{equation}
where $O_i(\mu)$ are the relevant operators and $C_i(\mu)$ are the Wilson coefficients evaluated at the scale $\mu$ corresponding to these operators. They may be expressed in the standard operator basis \cite{chetyrkin} by:
\begin{equation}
\label{standard_basis}
\begin{array}{rl}
O_1 ~= & (\bar{s} \gamma_{\mu} T^a \hat{P}_L c)(\bar{c} \gamma^{\mu} T^a \hat{P}_L b)\;,\\[2mm]
O_2 ~= & (\bar{s} \gamma_{\mu} \hat{P}_L c)(\bar{c} \gamma^{\mu} \hat{P}_L b)\;,\\[2mm]
O_3 ~= & (\bar{s} \gamma_{\mu} \hat{P}_L b) {\displaystyle\sum_q} (\bar{q} \gamma^{\mu} q)\;,\\[2mm]     
O_4 ~= & (\bar{s} \gamma_{\mu} T^a \hat{P}_L b) {\displaystyle\sum_q} (\bar{q} \gamma^{\mu} T^a q)\;,\\[2mm]    
O_5 ~= & (\bar{s} \gamma_{\mu_1}\gamma_{\mu_2}\gamma_{\mu_3} \hat{P}_L b) {\displaystyle\sum_q} (\bar{q} \gamma^{\mu_1}\gamma^{\mu_2}\gamma^{\mu_3} q)\;,\\[2mm]
O_6 ~= & (\bar{s} \gamma_{\mu_1}\gamma_{\mu_2}\gamma_{\mu_3} T^a \hat{P}_L b) {\displaystyle\sum_q} (\bar{q} \gamma^{\mu_1}\gamma^{\mu_2}\gamma^{\mu_3} T^a q)\;,\\[2mm]
O_7 ~= & \dfrac{e}{16\pi^2} \left[ \bar{s} \sigma^{\mu \nu} (m_s \hat{P}_L + m_b \hat{P}_R) b \right] F_{\mu \nu}\;,\\[3mm]
O_8 ~= & \dfrac{g}{16\pi^2} \left[ \bar{s} \sigma^{\mu \nu} (m_s \hat{P}_L + m_b \hat{P}_R) T^a b \right] G_{\mu \nu}^a\;,
\end{array}
\end{equation}
where $\hat{P}_{R,L}=(1\pm \gamma_5)/2$. The SUSY contributions to the $b\to s\gamma$ transitions are encoded in the Wilson coefficients as additional contributions with respect to SM values.
\par
The values for the rare branching ratio $Br(B_s \to l^+ l^-)$ in the SM to NLO are given by (see for example ~\cite{Isidori:2010gz} and references therein)
\bea
Br( B_s \to \mu^+ \mu^-)_{\rm SM} &=& 3.1 \times 10^{-9} 
\left( \frac{|V_{ts}|}{0.04}  \right)^2 \times \left( \frac{f_{B_s}}{0.21~\mbox{GeV}} \right)^2 \!\!  \; ,  \\ 
 \qquad \frac{ Br( B_s \to e^+ e^-)_{\rm SM} }{Br( B_s \to \mu^+ \mu^-)_{\rm SM} } &=& 2.4 \times 10^{-5}~~,~~\frac{Br( B_s \to \tau^+ \tau^-)_{\rm SM} }{  
 Br( B_s \to \mu^+ \mu^-)_{\rm SM} } = 215 ~.
\eea
The current bound on the leptonic branching ratio is
\be
Br(B_{s}\to \mu^+ \mu^-) < 6.6 \times 10^{-8} \quad (2\sigma) \; .
\eq
Note that the analogous $B_d$ channels will be suppressed relative to the $B_s$ case by 
\linebreak $ ( f^2_{B_d}/f^2_{B_s})\, |V_{td}/V_{ts}|^2 \approx 0.03$. As we have already mentioned, in the NMSSM the lightest CP-odd boson $a_1$ can be as light as a few GeV and still satisfy constraints from $Br(B_s \to \mu^+ \mu^-)$, for low to intermediate values of $\tan\beta$.
\par
 We also check the contributions to the mass splittings 
\bea
\Delta M_X &=& 2\, |M_{12}| \left[ 1 + 
  {\cal O} \left( \left| \frac{\Gamma_{12}}{M_{12}} \right|^2 \right) \right]~,
  \label{mgsol:a} \\
  \eea
from B-mixing ($\Delta M_s$ and $\Delta M_d$). 
\par
The isospin asymmetry $\Delta_{0} (B \to K^* \gamma)$ is defined as
\begin{equation}
\Delta_{0} (B \to K^* \gamma)= \frac{\Gamma(\overline{B}^0 \to \overline{K}^{*0}\gamma)-\Gamma(B^{\pm} \to K^{*\pm}\gamma)}
{\Gamma(\overline{B}^0 \to \overline{K}^{*0}\gamma)+\Gamma(B^{\pm} \to K^{*\pm}\gamma)}  \; .
\end{equation}
\par
The ratio of leptonic kaon decays to pion decays in SUSY is
\begin{eqnarray}
\dfrac{\rm{Br}(K \to \mu \nu_\mu)}{\rm{Br}(\pi \to \mu \nu_\mu)}&=& 
\frac{\tau_K}{\tau_\pi}\left|\frac{V_{us}}{V_{ud}} \right|^2 \frac{f^2_K}{f^2_\pi} \frac{m_K}{m_\pi}\left(\frac{1-m^2_\ell/m_K^2}{1-m^2_\ell/m_\pi^2}\right)^2 \nonumber \\
&& \times \left[1-\frac{m^2_{K^+}}{M^2_{H^+}}\left(1 - \frac{m_d}{m_s}\right)\frac{\tan^2\beta}{1+\epsilon'_0\tan\beta}\right]^2 \left(1+\delta_{\rm em}\right)\;,\label{kmunu1}
\end{eqnarray}
where $\epsilon'_0$ is a loop factor and $\delta_{\rm em}$ is an electromagnetic correction term.
 $R_{l23}$ is defined as~\cite{Antonelli:2008jg}
\begin{equation}
R_{l23} = \Big | \frac{V_{us}(K_{l2})~V_{ud}(0^+\rightarrow0^+)}{ V_{us}(K_{l3}) ~V_{ud}(\pi_{ l2})}\Big |=\left|1-\frac{m^2_{K^+}}{M^2_{H^+}}\left(1 - \frac{m_d}{m_s}\right)\frac{\tan^2\beta}{1+\epsilon'_0\tan\beta}\right|\;,
\end{equation}
where $K_{l2}$ and $K_{l3}$ denote helicity-suppressed ($K\rightarrow l\nu$) and helicity-allowed ($K\rightarrow \pi l\nu$) decays. $0^+\rightarrow0^+$ denotes nuclear beta-decay. Note that in the SM $R_{l23}=1$.
 \par
 There has been some discussion regarding $Br(B\rightarrow\tau\nu)$ and $Br(B\rightarrow D\tau\nu)$ in the context of a relatively light charged Higgs and a possible tension between the current measurements of the two and with $(g-2)_\mu$ (For a recent paper discussing this aspect see \cite{Bhattacherjee:2010ju}). The charged Higgs contribution in the MSSM may be written as~\cite{Isidori:2006pk,Hou:1992sy,Akeroyd:2003zr}
 \be
 Br(B\to \tau\nu)
=\frac{G_F^2|V_{ub}|^2}{8 \pi} m^2_\tau f^2_B m_B
\left(1-\frac{m^2_\tau}{m_B^2}\right)^2 \times
\left\vert1+\frac{m_B^2}{m_b
m_{\tau}}\,C^{\tau}_{NP}\right\vert^2 \; ,  
\eq
where
\bea
C^{\ell}_{NP}&=&-\frac{m_b
m_{\ell}}{m^2_{H^{+}}}\frac{\tan^2\beta}{1+\epsilon_0\tan\beta} \; .
\eea
$R_{\tau\nu_\tau}$ is the ratio between the measured $Br(B_u \to \tau \nu_\tau)$ and the SM prediction. The present $Br(B\rightarrow\tau\nu)$ measurements seem to suggest no/very small MSSM contribution. The SUSY contribution is generally substantial for large $\tan\beta$ or for light charged Higgs.  
%%%%%%%%%%%%%%%%%%%%%%%%%%%%%%%%%
\FIGURE
{
\includegraphics[scale=0.2]{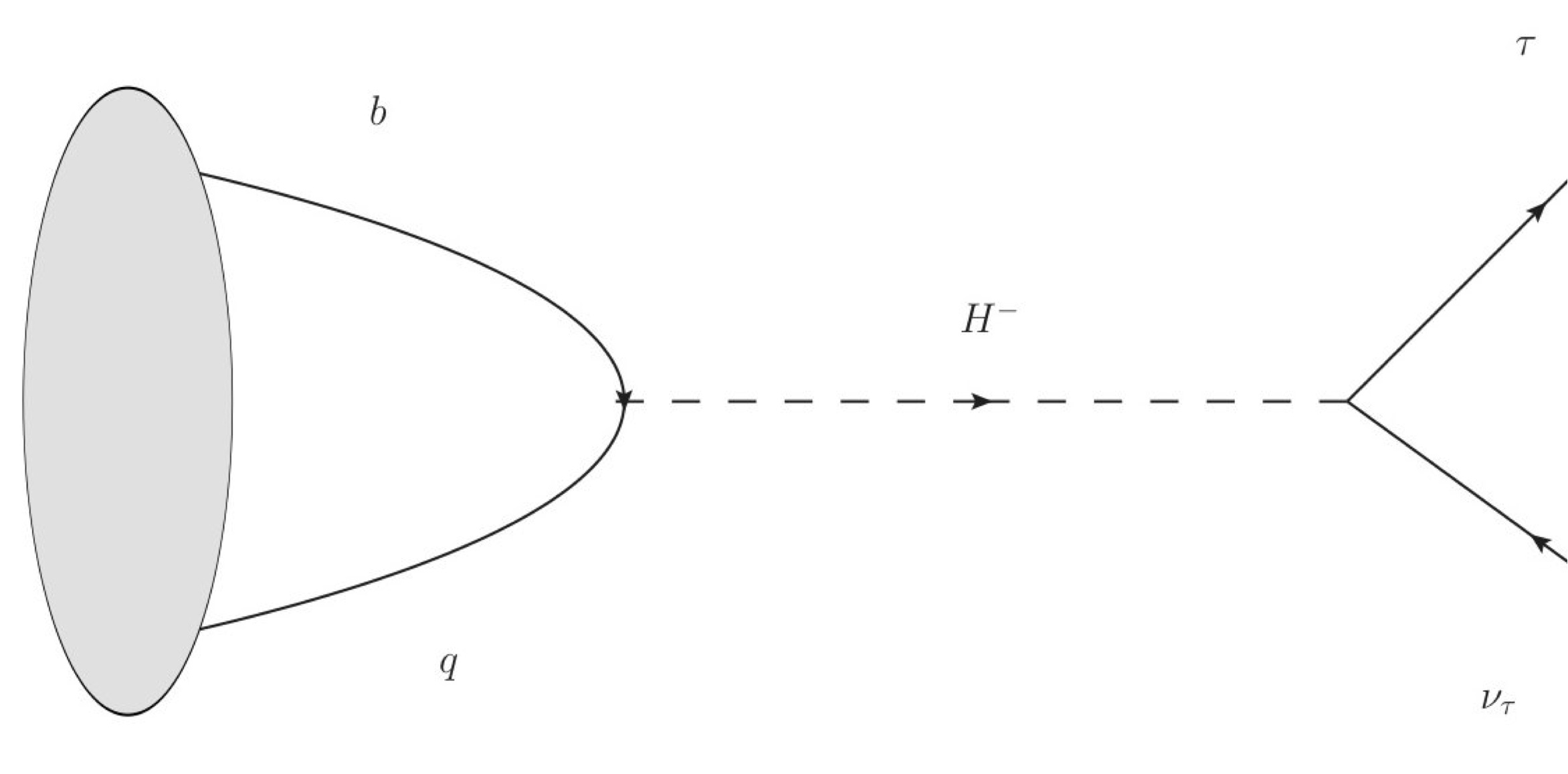}
\includegraphics[scale=0.6]{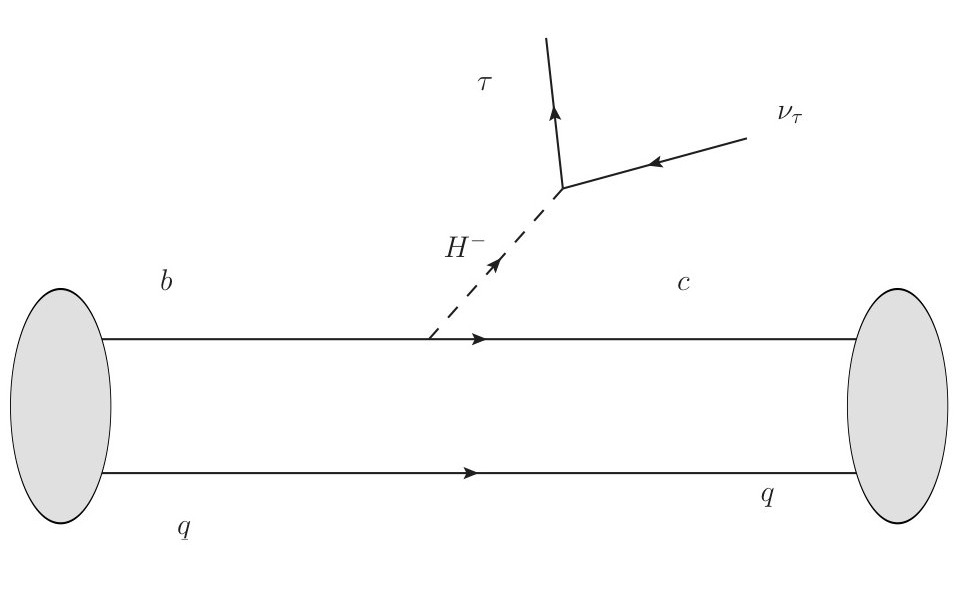}
\caption{In the MSSM, the charged-Higgs ($H^{\pm}$) contributes to both $B\rightarrow \tau \nu$ and $B\rightarrow D\tau \nu$ in a similar way. The relevant Cabibbo-Kobayashi-Maskawa (CKM) factor in the former case is $V_{ub}$ which has a larger uncertainty than $V_{cb}$, which is the relevant one for $B\rightarrow D\tau \nu$. Apart from this $B\rightarrow \tau \nu$ involves two powers of the QCD form factor $f_B$ which must be obtained from lattice calculations and presently has a large uncertainty. The latter process involves two QCD form factors which are better constrained~\cite{Nierste:2008qe}.}
}
%%%%%%%%%%%%%%%%%%%%%%%%%%%%%%%%%%
\par
The differential decay rate of $B\to D\ell \nu_\ell$ can be expressed as \cite{Kamenik:2008tj}
\begin{eqnarray}
\frac{d\Gamma(B\to D\ell \nu_\ell)}{dw} &=& \frac{G_F^2|V_{cb}|^2 m_B^5}{192\pi^3}\rho_V(w) \label{partialbdtaunu}\\
&& \times\left[1 - \frac{m_{\ell}^2}{m_B^2}\, \left\vert 1- t(w)\, \frac{m_b}{(m_b-m_c)m^2_{H^{+}}}\,\frac{\tan^2\beta}{1+\epsilon_0\tan\beta} \right\vert^2 \rho_S(w) \right]\;,\nonumber 
\end{eqnarray}
where  $p_D$, $p_B$ are the meson four-momenta, $w$ is defined as
\begin{equation}
w = \frac{1 + (m_D/m_B)^2-(p_B - p_D)^2/m_B^2}{2 m_D/m_B}\;,
\end{equation}
$t(w) = m_B^2 + m_D^2 - 2 w \,m_D \,m_B$ and $\rho_V(w)$, $\rho_S(w)$ are the vector and scalar Dalitz density distributions, respectively. 

The  $Br(B\rightarrow D\tau\nu)$ may be parametrized using the same Wilson coefficient as~\cite{Kamenik:2008tj}
\begin{eqnarray}
\xi_{D \tau \nu_\tau}=\frac{Br(B \to D \tau \nu)}{Br(B\to D e \nu)} &=& \left(0.28 \pm
0.02\right)\left[1 + 1.38(3) Re(C_{NP}^{\tau}) +
 0.88(2) |C_{NP}^{\tau}|^2\right] \; .~~~~~
\end{eqnarray}
\par
We would like to emphasize that the persistent indications from the muon anomalous magnetic moment discrepancy suggests some new additional contribution (maybe just from low-energy QCD, hadronic light-by-light or BSM, to speculate on a few sources). To achieve this in the purview of the MSSM/NMSSM one generally needs a sufficient $\tan\beta$ (not necessarily very large though; we use $\tan\beta=10$, for instance, and still get sufficiently large contributions to $(g-2)_{\mu}$ in the $3\sigma$ interval). Hence we expect some tension between $(g-2)_{\mu}$ and some of the flavor observables in  general. Thus imposing both of these will flush out the most interesting regions of the parameter space.

% ----------------------------------------------------------------------------------------------------------
\subsection{Collider bounds}
% ****************************************************
\begin{table}[!h]
\begin{center}
\hspace*{1.5cm}\begin{tabularx}{15.cm}{|X|X||X|X|X} 
\cline{1-4}
\multicolumn{4}{|c|}{}\\[-1mm]
\multicolumn{4}{|c|}{\textbf{Lower bounds on the Higgs and sparticle masses (GeV)}}\\[3mm]
\cline{1-4}
\multicolumn{4}{c}{\vspace*{-4mm}}\\
\cline{1-4}
$h^0$  & 111& $\tilde{e}_R$ & 73 &\\[3mm]
\cline{1-4}
$H^+$  & 79.3 & $\tilde{\mu}_R$ & 94 &\\[3mm]
\cline{1-4}
$A^0$  & 93.4 & $\tilde{\tau}_1$ & 81.9 &\\[3mm]
\cline{1-4}
$\chi^0_1$  & 46 & $\tilde{\nu}$  & 94 &\\[3mm]
\cline{1-4}
$\chi^0_2$  & 62.4 & $\tilde{t}_1$ & 95.7 &\\[3mm]
\cline{1-4}
$\chi^0_3$  & 99.9 & $\tilde{b}_1$  & 89 &\\[3mm]
\cline{1-4}
$\chi^0_4$ & 116 & $\tilde{q}$ & 379 &\\[3mm]
\cline{1-4}
$\chi^\pm_1$ & 94 & $\tilde{g}$ & 308 &\\[3mm]
\cline{1-4}
\end{tabularx}
\end{center}
\caption{Mass bounds from collider studies~\cite{Amsler:2008zzb}.}
\label{mass_bounds}
\end{table}
% ***********************************************************

\par
The direct collider bounds come from various channels. For instance, a promptly decaying NLSP places a limit on $\Lambda=F/M_m$ in RGM of about 100 TeV. For non-prompt NLSPs the lower bound comes from charged sparticle masses and indirect constraints such as the inclusive tri-lepton signal. In both cases, but especially for promptly decaying NLSPs, a heavy spectrum results.
\par 
The trilepton process can put indirect mass bounds. The process is
\be
p\bar{p}\rightarrow(\chi_2^0\rightarrow\chi_1^0 l^{+}l^{-})(\chi_1^{\pm}\rightarrow\chi^0 l^{\pm}\nu_{l}) \; ,  
\eq
and proceeds mainly via an s-channel virtual $W$ exchange (and a small t-channel squark exchange) and hence its cross section peaks at $M_{\chi}\approx M_W/2$. Otherwise the cross section falls with increasing $\chi^\pm$ mass with a weak $\tan\beta$ dependence. This puts indirect mass limits on the $\chi^\pm$ masses.
\par
The current bounds on the superpartner masses~\cite{Amsler:2008zzb} are shown in Table\,\ref{mass_bounds}. For the sleptons, constraints from the $Z^0$ decay width already put a lower bound of $\sim40~\rm{GeV}$ independent of decay modes. The squark case includes cascade decays which gives a slightly lower bound than if the squarks were assumed to decay directly to photinos. Also, in deriving the squark mass bound with cascade decays, fixed values for $\mu$ and $\tan\beta$ are taken. This is also true while deriving the $\tilde{g}$ mass bound. It is important to note that these experimental bounds have been derived after imposing certain assumptions some of which are not strictly true in many studies, and definitely not in ours. Hence some of these must be viewed as indicative. Some of these assumptions include ~\cite{Amsler:2008zzb} : 
\bi
\item  Conservation of R-parity. 
\item Gaugino mass unification at the GUT scale.
\item  $\tilde{\chi}_{1}^0$ is the LSP.
\item All scalar quarks are assumed to be degenerate in mass except for $\tilde{t}$ and $\tilde{b}$. 
\item $m_{\tilde{q}_R} = m_{\tilde{q}_L}$
\item The mass of $\tilde{G}$, in the decay final states, is neglected relative to the other masses.
\item Some of the mixing angles are small. For instance, bounds from $e^{+}-e^{-}$ collisions depend on the mixing angle of the lightest squark mass eigenstate. It is assumed that only $\tilde{b}$ and $\tilde{t}$ have non-trivial mixing angles in this case.
\ei
There are also model dependences, for instance, in NMSSM the mass of the lightest CP-odd Higgs can be quite small as opposed to a bound of $93.4~\rm{GeV}$. In Table\,\ref{mass_bounds} also note that we have taken the bound on the lightest CP-even Higgs, $m_h$, to be $111\,\text{GeV}$ rather than the usual LEP bound value of $114\,\text{GeV}$, to accommodate a possible $\sim3\,\text{GeV}$ uncertainty in the theoretical Higgs mass calculation.
\draftnote{}

% ----------------------------------------------------------------------------------------------------------

\section{Analysis}

\subsection{Numerical analysis and strategy }
We have taken the mediation/messenger scale ($M_{m}$) to be high, in the vicinity of the GUT scale
\be
M_m\simeq\Lambda_{\rm{GUT}} \; .
\eq
The study may be extended for other messenger scales in a straightforward manner. Imposing the messenger scale to be $\Lambda_{GUT}$ makes comparison with mSUGRA and gaugino mediation (with a compactification scale $\sim \Lambda_{GUT}$) convenient and to lowest order removes any effects which are solely due to RG running from GUT scale to low scale (electroweak). In this context the reader is also referred to \cite{{Jaeckel:2011ma},{Jaeckel:2011qj}}. The characteristic scale associated with the hidden sector is $M_s$. With this definition the correlation functions $(\tilde{A}_r,\tilde{B}_r)$ become dimensionless. The GGM masses, Eq. (\ref{GGMmass}), are imposed at the messenger scale as functions of the correlation functions ($\tilde{A}_r,\tilde{B}_r$) and then RG evolved to the low scale.  For completeness, we give the 1-loop RG equations for the MSSM and NMSSM in the Appendices. For the complete 2-loop RG equations the reader is directed, for example, to  \cite{{Martin:1993yx},{Yamada:1993ga}}.
\par
One point we would also like to emphasize is that we do not take $B\mu$ as the fundamental quantity, in contrast to an interesting study in~\cite{Abel:2009ve} for instance, and instead choose to keep $\tan\beta$ fixed at $10$ to more readily compare with various benchmark scenarios in the literature. If $B\mu$ is taken as the fundamental quantity and to be vanishing at the GUT scale (by adopting a strict interpretation of gauge mediation) then $\tan \beta$ is no longer an input and comes out to be large ($\sim\,15-65$) as in~\cite{Abel:2009ve}. A higher value of $\tan\beta$ can fit $\Delta \,a_{\mu}$ for a heavier sparticle spectrum but for small to intermediate values for $\tan\beta$ it is easier to accommodate flavor-physics constraints. We also adopt a pure gauge mediation setting where the assumption is that the generation of $\mu$ is independent of the SUSY-breaking and the GGM scenario. As we previously mentioned the MSSM Higgs sector has 4 free parameters and the NMSSM Higgs sector has 7 free parameters. We use the minimization conditions to trade $B\mu$ for $\tan\beta$. The $\mu$ parameter is calculated at the low scale, using the minimization conditions, to be consistent with the input $Z^0$ mass ($M_{Z^0}$). Thus the actual inputs we choose in the MSSM Higgs sector are
\be
M_{Z^0}, \tan\beta, m_{H_u}^2,m_{H_d}^2 \; .
\eq
Here $\tan\beta$ is chosen at the scale of EWSB. Notice that $H_u$ and $H_d$ carry the same quantum numbers as $\tilde{l}_L$. Thus we will assume that the predominant contribution to $m_{H_u}^2$ and $m_{H_d}^2$ at the high scale are same as the  $\tilde{l}_L$ soft mass contributions. As previously mentioned benchmark value of $\tan\beta=10$ is adopted throughout our study, since our main focus is on the effects of the hidden-sector correlation function ratios. 
\par
In the scale-invariant NMSSM case we take a slightly different set of inputs. The three Higgs potential minimization conditions in the NMSSM may be written as ~\cite{Ellwanger:2009dp}
\bea
v_{d} \left( m^2_{H_d}+\mu_{\text{\textit{eff.}}}^2+\lambda^2 v_u^2 +\frac{\bar{g}^2}{4}(v_d^2-v_u^2) \right)-v_u ~\mu_{\text{\textit{eff.}}} \left(A_\lambda+\frac{\kappa}{\lambda}\mu_{\text{\textit{eff.}}}\right)&=&0  \; ,  ~~\\ \nn
v_{u} \left( m^2_{H_u}+\mu_{\text{\textit{eff.}}}^2+\lambda^2 v_d^2 +\frac{\bar{g}^2}{4}(v_u^2-v_d^2) \right)-v_d ~\mu_{\text{\textit{eff.}}} \left(A_\lambda+\frac{\kappa}{\lambda}\mu_{\text{\textit{eff.}}}\right)&=&0 \; ,  \\ \nn
\frac{\mu_{\text{\textit{eff.}}}}{\lambda}\left(m_N^2+\kappa A_\kappa \frac{\mu_{\text{\textit{eff.}}}}{\lambda}+2 \kappa^2 \left(\frac{\mu_{\text{\textit{eff.}}}}{\lambda}\right)^2+\lambda^2(v_u^2+v_d^2)-2\lambda v_u v_d\right)-\lambda v_u v_d A_\lambda &=&0 \; .
\eea
Here
\be
\mu_{\text{\textit{eff.}}}=\lambda \langle N \rangle
\eq
The above minimization conditions may be used to eliminate three of the soft scalar mass terms. Using this freedom we choose the inputs 
\be
M_{Z^0}, \tan\beta,\lambda,\kappa,A_\lambda,A_\kappa,\mu_{\text{\textit{eff.}}} \; ,  
\eq
at the low scale. For our study we take $\tan\beta=10$ as before, $\lambda=0.1$, $\kappa=0.1$, $A_\lambda=150$, $A_\kappa=0$ and $\mu_{\text{\textit{eff.}}}$ is taken to be $210~\rm{GeV}$. Once again other suitable choice of parameters may be made, but our main focus in this study is on the dependence of the low-energy observables and spectra on the hidden-sector correlation function ratios. A very comprehensive and detailed study of the NMSSM parameter space with various choices of the parameters may be found in ~\cite{{Ellwanger:2009dp}, {Djouadi:2008uw}} and references therein.
\par 
Since the MSSM contribution to the muon anomalous magnetic moment depends on $sgn(\mu)$, currently $\mu > 0$ seems to be slightly favored and we focus on a positive value throughout.
\par
We have used the following codes in their original and modified forms : $\mathtt{SuSpect}$~\cite{Djouadi:2002ze}, $\mathtt{SuperIso}$~\cite{{Mahmoudi:2008tp},{Mahmoudi:2007vz}} and $\mathtt{micrOMEGAS}$~\cite{ {Belanger:2001fz}, {Belanger:2004yn},{Belanger:2006is}}. We have cross-checked the theoretical predictions of the low-energy observables between the programs and also independently estimated some of their values directly from the low-energy spectra. For the NMSSM we use our code to implement GGM soft masses at the GUT scale and evolve it down to a low scale (Electro-weak). The NMSSM RGEs are implemented only to 1-loop (see Appendix B.) and are found to introduce an error in the spectra of atmost $\sim 10-15\%$, mainly in the third generation, in comparison with mSUGRA 2-loop NMSSM implementations (for example in $\mathtt{NMSSMTools}$~\cite{{Ellwanger:2006rn},{Ellwanger:2005dv}}). At the low scale we have also used $\mathtt{NMSSMTools}$~\cite{{Ellwanger:2006rn},{Ellwanger:2005dv}} to calculate and cross-check some of the low-energy observables. 
\subsection {Features of the parameter space}
%  *****************************
\FIGURE[!h]{
\includegraphics[scale=0.575]{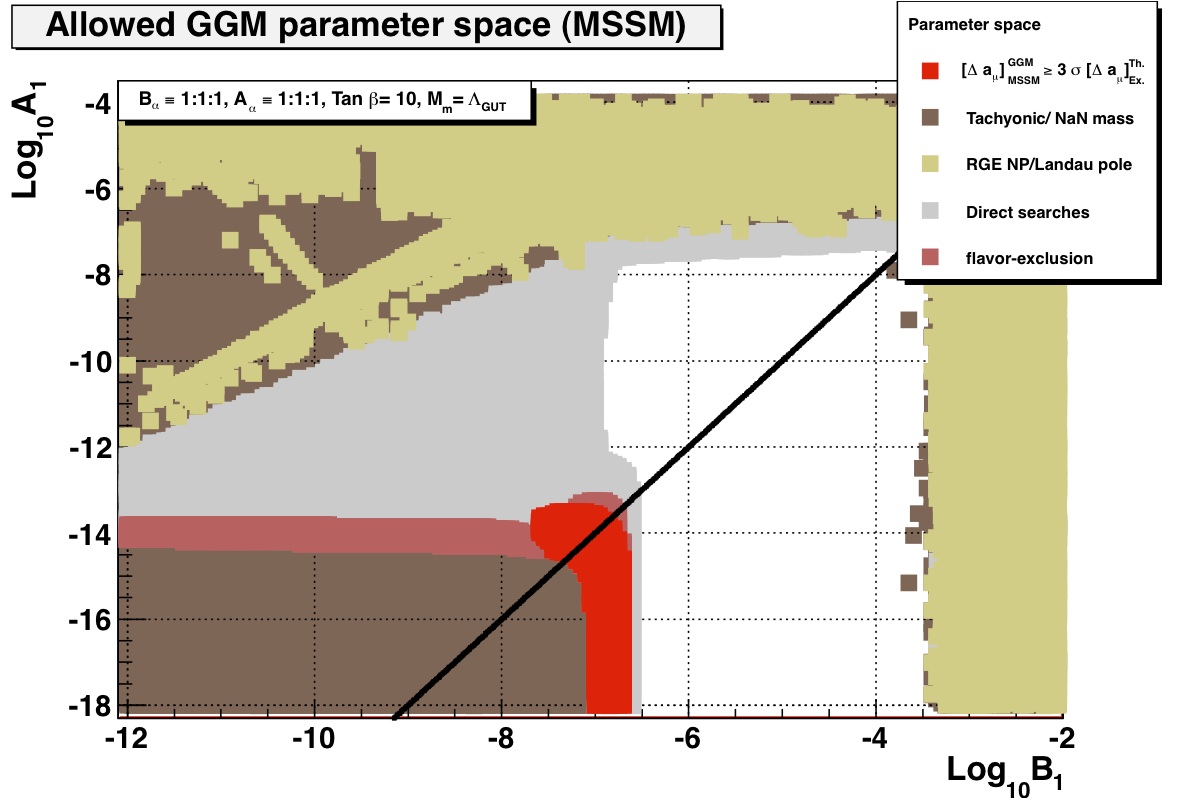}
\includegraphics[scale=0.575]{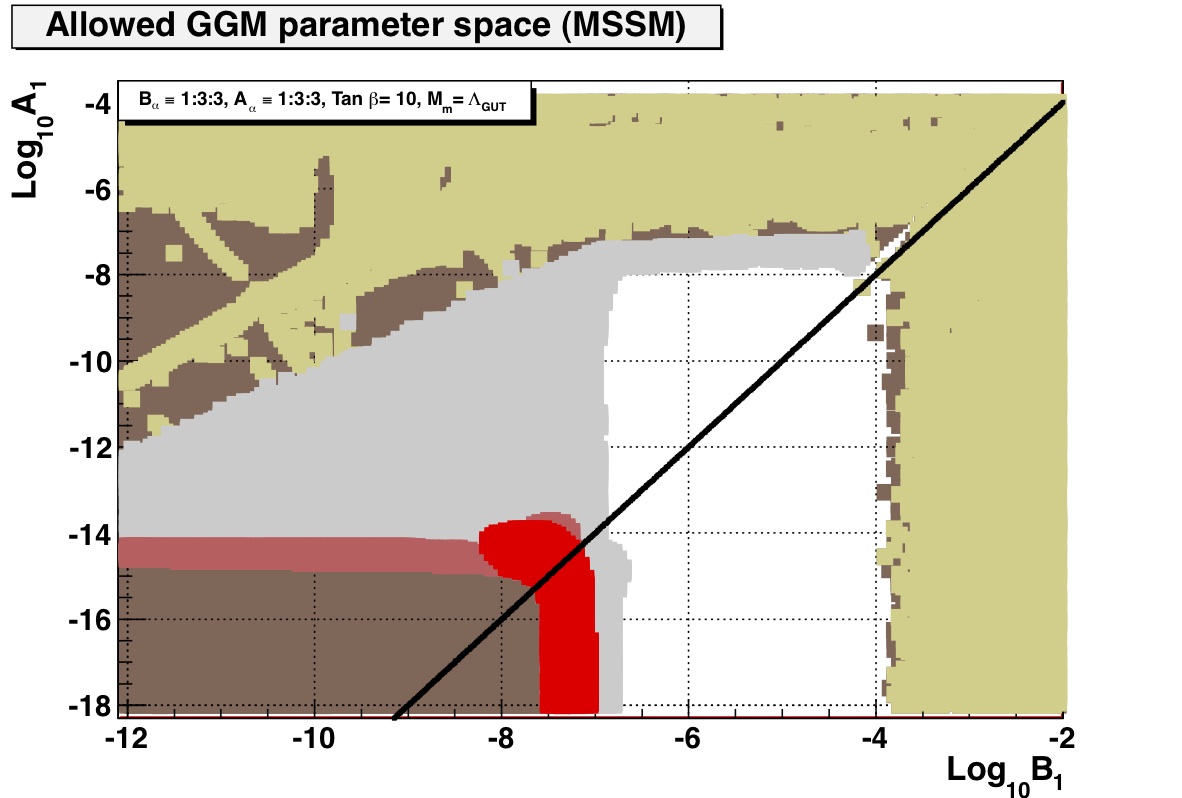} 
\caption{ Constraints on the $(1:1:1\vert1:1:1)$ (top) and $(1:3:3\vert1:3:3)$ (bottom) GGM parameter spaces (MSSM). Points in the vicinity of the linear band (black) for $(1:1:1\vert1:1:1)$, where $\log_{10}B_\alpha~\simeq~(1/2)\,\log_{10} A_\alpha $, correspond to RGM scenarios. For the $(1:3:3\vert1:3:3)$ case, regions near the linear band have no RGM interpretation and the band is merely given for comparison. In the figure the acronym RGE NP stands for non-perturbative renormalization group evolution. The exclusions due to $(g-2)_\mu$ are shown in red, those due to flavor observables in magenta/pink and those due to direct collider bounds in grey. The white portions represent the allowed regions.}
\label{111callowed}
 }
 % *********************************
% ----------------------------------------------------------------------------------------------------------
\par
We parametrize the various correlation function ratios with the notation
\be
(\tilde{B}_1:\tilde{B}_2:\tilde{B}_3~\vert~ \tilde{A}_1:\tilde{A}_2:\tilde{A}_3) \; ,  
\eq
where the $\tilde{B}_r$ and $\tilde{A}_r$ are defined as in Eq. \ref{GGMmass}. There are a couple of caveats. Note that there could be a hierarchy between the $\tilde{B}_{r}$ and $\tilde{A}_{r}$ correlation functions even when the case under consideration is  $(1:1:1\vert1:1:1)$. We elaborate on this below. We also remind ourselves that in order to make the $\tilde{B}_r$ and $\tilde{A}_r$ dimensionless, we have factored out a common scale $M_s=~2\times10^9~\rm{GeV}$, that may be thought of as a characteristic scale for the hidden sector. But this is just our convention.
\par
 To keep the investigation of low-energy observables manageable we will mainly focus on scenarios where the ratios are mirrored between the $\tilde{B}_r$ and $\tilde{A}_r$ ( i.e. the ratio associated with the $SU(2)_L$ and $SU(3)_C$ correlation functions with respect to the $U(1)_Y$ correlation functions are the same for both $\tilde{B}_r$ and $\tilde{A}_r$). Other scenarios may be considered in a straightforward manner. The viable GGM parameter space for three correlator ratios $(1:1:1\vert1:1:1)$, $(1:3:3\vert1:3:3)$ and $(1:1/3:1/3\vert1:1/3:1/3)$ are shown in Figs. \ref{111callowed} and \ref{311callowed}. As previously mentioned, the RGM limit can be obtained from the  $(1:1:1\vert1:1:1)$ GGM scenario by considering points where
\be
\tilde{B}_{r}=\sqrt{\frac{\tilde{A}_{r}}{2}} \; .
\eq
At these points, in the $(1:1:1\vert1:1:1)$ case, there is no hierarchy between the $\tilde{B}_{r}$ and $\tilde{A}_{r}$.
Large parts of the GGM parameter space are excluded purely from requirements of having no tachyonic masses, potential to be bounded from below, absence of charge breaking and perturbative RG evolution. Nevertheless there are viable regions different from RGM that have interesting observables and spectra. 
\par
There are also non-viable regions in Figs. \ref{111callowed} and \ref{311callowed} where the value of the MSSM contribution to $(g-2)_\mu$ is above $3\sigma$ relative to the current discrepancy between theory and observation. It is also found that large regions are excluded due to flavor-physics observables with a large overlap with exclusions due to direct collider bounds. Direct collider searches, flavor-physics and muon anomalous magnetic moment give interesting bounds on the gaugino and sfermion masses in the GGM scenario. 
\par 
Larger gaugino masses can cause the sfermion masses to run to larger values due to RGE and hence lead to more phenomenologically viable regions. Looking at the plots, it is interesting that $\log_{10}\tilde{A}_r$ does not seem to be bounded from below. This may be understood by noting that, for fixed gaugino soft mass terms, the $\tilde{A}_r\rightarrow 0$ limit implies that at $\Lambda_{GUT}$ we are basically setting all $m_{\tilde{f}}\rightarrow 0$ (since we have set the Fayet-Iliopoulos term $\zeta=0$, by say messenger parity arguments). The initial RG running of the sfermion masses from the GUT scale are therefore completely determined by the gaugino soft terms (see Appendix A) and hence the sfermion masses at the low scale approach a constant value as we decrease $\tilde{A}_r$.
%  *****************************
\FIGURE[!h]{
\includegraphics[scale=0.575]{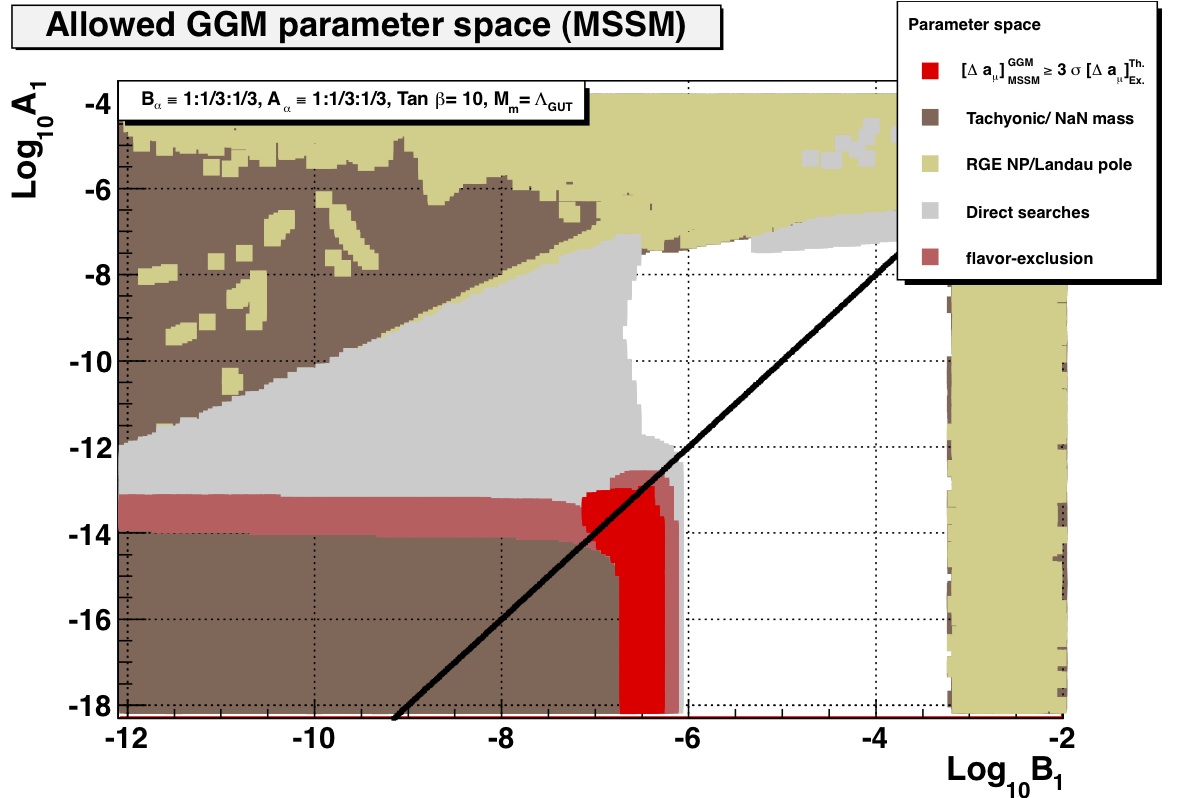}
\caption{ Constraints on the $(1:1/3:1/3\vert 1:1/3:1/3)$ GGM parameter space (MSSM). Note again that for this ratio, regions near the linear band (black) do not correspond to RGM. The white portion again represents the viable region. }
\label{311callowed}
 }
 % *********************************
 %  *****************************
\FIGURE{
\includegraphics[scale=0.360]{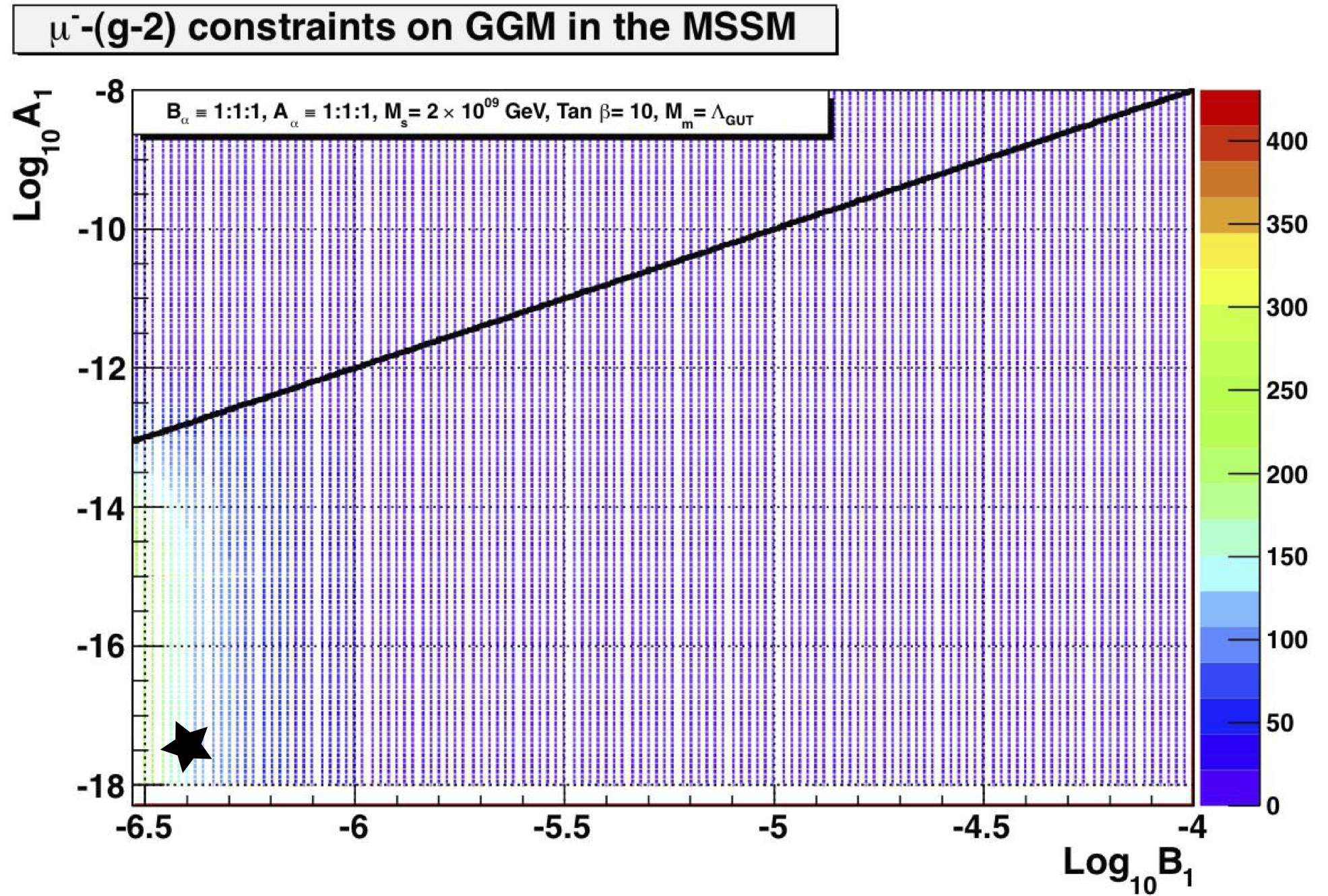}
 \includegraphics[scale=0.360]{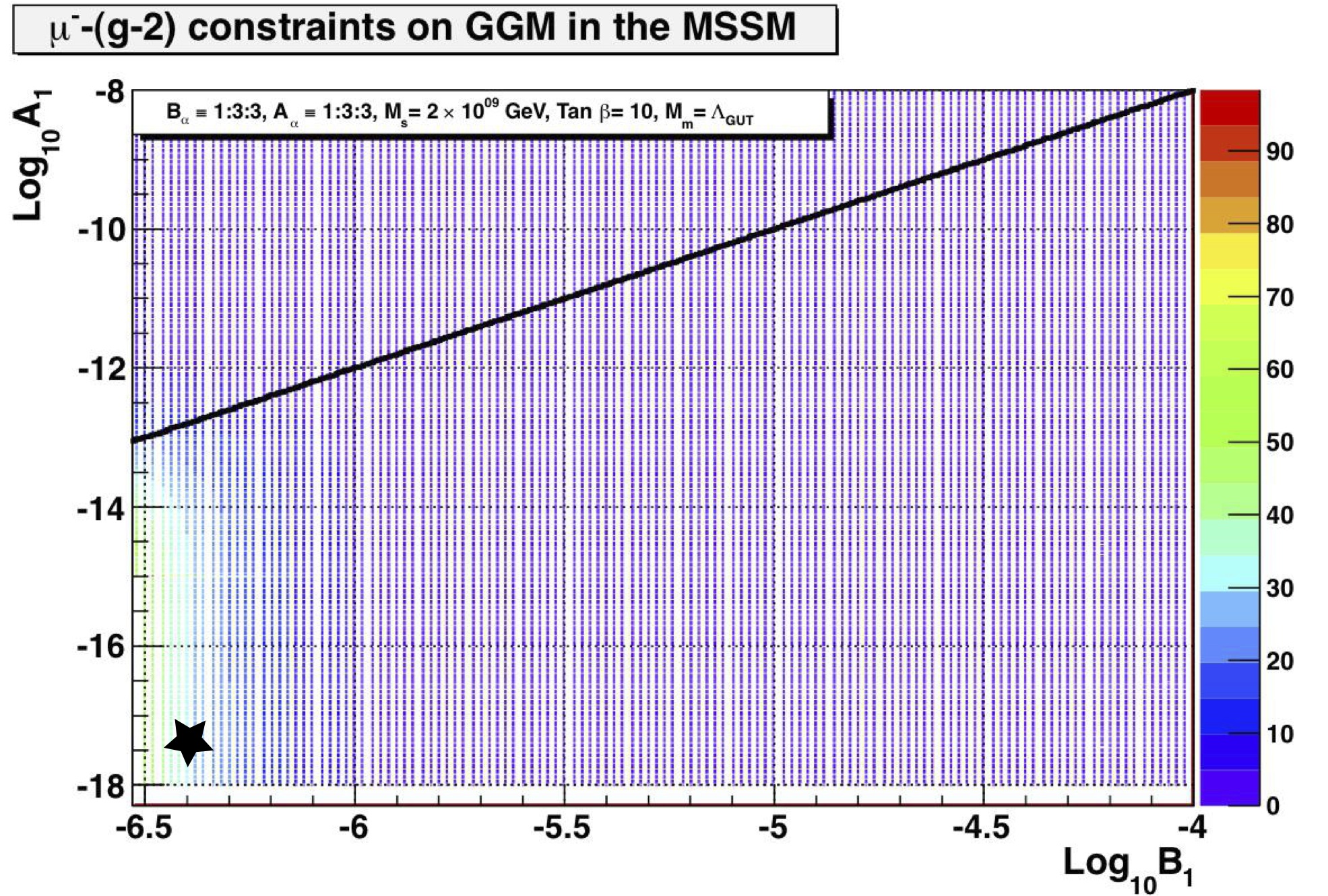}
\caption{ Points that give a positive contribution to $(g-2)_\mu$, for allowed regions in the $(1:1:1\vert1:1:1)$ (top) and $(1:3:3\vert1:3:3)$ (bottom) cases (MSSM). Points in the vicinity of the linear band (black) for $(1:1:1\vert1:1:1)$ correspond to RGM as before. The color scale to the right shows the MSSM contribution to $(g-2)_\mu$ in units of $10^{-11}$. Note that the color scales are different in each case. The point marked with the star symbol is an example point that we will explore later, as a benchmark, in the context of low-energy observables and mass spectra.}
\label{111mallowed}
 }
 % *********************************

\par 
In RGM there is very little freedom to change the sfermion to gaugino mass ratio except by $\mathcal{O}(1)$ factors. The ratio may be tuned by varying the number of messenger fields (non-minimal RGM), but larger ratios may be disfavored since they may spoil gauge coupling unification. In GGM we can explore larger ratios between them generically and even among the three gauge groups, not restricted just by ratios of $g_a^2$, due to the possibility of hierarchy among the three $\tilde{B}_r$. 
\par
Let us take up the question of precision gauge coupling unification in the MSSM. There is a mismatch in $\alpha_3$ of about $3\%$ at the GUT scale derived from the unification of $\alpha_1$ and $\alpha_2$. The RG evolution of the coupling constants, taking into account threshold effects, obey
\begin{equation}
\frac{\text{d}g_i}{\text{d}t} ~=~ \frac{b_i}{16\pi^2} \Theta_i\,  g_i^3\;, \quad (b_1,b_2,b_3)=(\tfrac{33}{5},1,-3) \;,
\end{equation}
with $t\equiv \text{ln}(q/\mu)$. The $\Theta_i$ take into account the thresholds due to the superpartner masses and are given by
\begin{subequations}\label{eq:theta}
\begin{eqnarray}
 \Theta_1
 & = & \frac{1}{33}\left[20 + \theta_{\widetilde{h}_1} + \theta_{\widetilde{h}_2}+\frac{1}{2}\left( \theta_{h_1} + \theta_{h_2}\right)+\sum_{i=1}^3\left( \tfrac{1}{2}  \theta_{\widetilde{l}_i}
 + \theta_{\widetilde{e}_i}  + \tfrac{1}{6}  \theta_{\widetilde{q}_i}  + \tfrac{4}{3}  \theta_{\widetilde{u}_i}
 + \tfrac{1}{3}  \theta_{\widetilde{d}_i}   \right)    \right] \; ,  ~~~~~~~ \\
 \Theta_2
 & = &
 -\frac{10}{3}+\frac{4}{3}\theta_{\widetilde{w}}+\frac{1}{3}\left(\theta_{\widetilde{h}_1}
 +\theta_{\widetilde{h}_2}\right)  +\frac{1}{6}\left( \theta_{h_1} + \theta_{h_2}\right)
 + \frac{1}{6}\sum_{i=1}^3\left(3 \theta_{\widetilde{q}_i} +
 \theta_{\widetilde{l}_i}\right) \; ,  \\
 \Theta_3
 & = & \frac{7}{3}-\frac{2}{3} \theta_{\widetilde{g}} -\frac{1}{18}\sum_{i=1}^3\left(2 \theta_{\widetilde{q}_i}
 + \theta_{\widetilde{d}_i}+ \theta_{\widetilde{u}_i}\right) \; ,  
\end{eqnarray}
\end{subequations}
where the Heaviside functions above are defined as $\theta_{\tilde{f}}=\theta(q^2-m_{\tilde{f}}^2)$. At a scale much above all the thresholds, the $\Theta_i\rightarrow 1$ (see Appendix A ). 
\par
A straightforward examination of the above expressions tells us that a heavy wino ($\tilde{w}$) and/or a light gluino ($\tilde{g}$) can lead to precision gauge coupling unification by changing the running of $\alpha_2$ and/or $\alpha_3$.  The other colored particles can also slow down the running of $\alpha_3$ but they come with a smaller coefficient and moreover will also lead to changes in the $\alpha_1$ and $\alpha_2$ running which makes it more tricky to implement. It is more efficient to work with a light $\tilde{g}$ as opposed to a heavier $\tilde{w}$ to achieve precision gauge coupling unification since this way one can utilize the larger value of $\alpha_3$ at lower energies.
%  *****************************
\FIGURE{
 \includegraphics[scale=0.360]{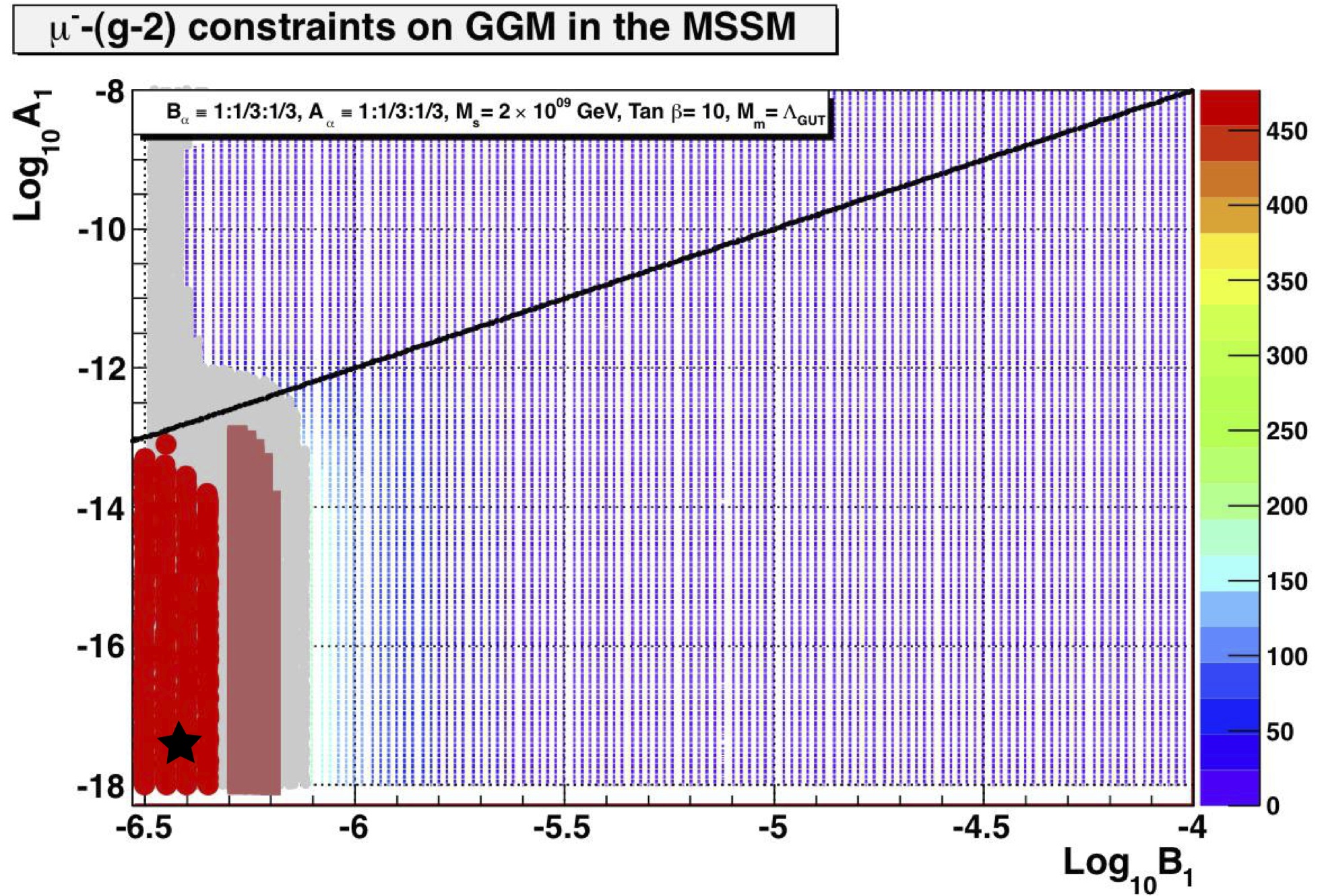}
\caption{ MSSM contribution to $(g-2)_\mu$ for the allowed region  in the $(1:1/3:1/3\vert 1:1/3:1/3)$ case. The color scale to the right show the value of $(g-2)_\mu$ again in units of $10^{-11}$. The regions ruled out by collider bounds are shown in grey, those ruled out by flavor physics by pink/magenta and those with a $(g-2)_\mu~\geq\,3\sigma$ by red. As in the $(1:3:3\vert 1:3:3)$ case the linear black band is merely given for comparison with the $(1:1:1\vert 1:1:1)$ case and does not have an RGM interpretation. The star symbol again denotes the benchmark point. It is ruled out in the present case from collider and low-scale observable bounds. In fact, note that for the $(1:1/3:1/3\vert 1:1/3:1/3)$  case regions below $\log_{10}B_1\lesssim-6.1$ are ruled out in contrast to the $(1:1:1\vert 1:1:1)$ and $(1:3:3\vert 1:3:3)$ cases. }
\label{311mallowed}
 }
 % *********************************
 \par
To lowest order the achievement of precision gauge coupling unification therefore crucially depends on the mass ratios of the gauginos (hence on $\tilde{B}_1:\tilde{B}_2:\tilde{B}_3$), specifically that of the $\tilde{w}$ and $\tilde{g}$ (note that the bino ($\tilde{B}$) does not contribute to the running). Thus in the GGM scenario there is more room for variations in these ratios to possibly improve precision gauge coupling unification at the GUT scale.
%  *****************************
\FIGURE{
 \includegraphics[scale=0.374]{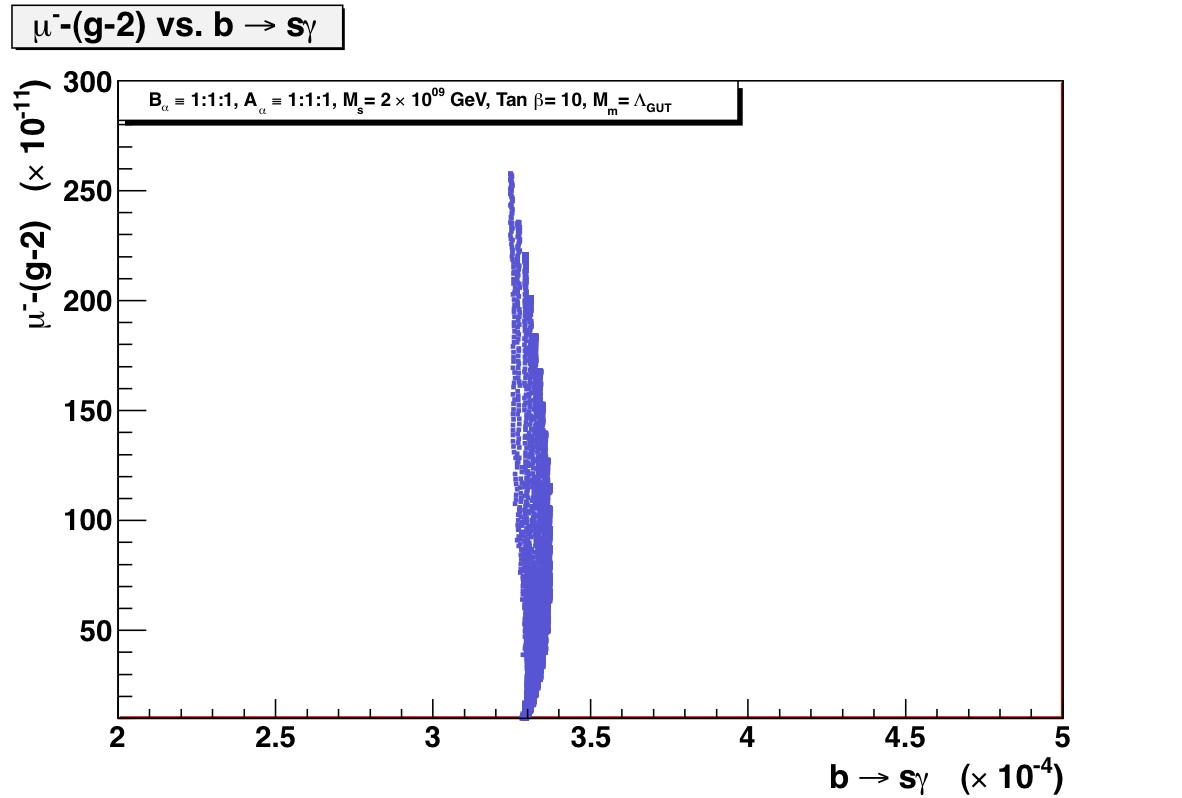}
 \includegraphics[scale=0.374]{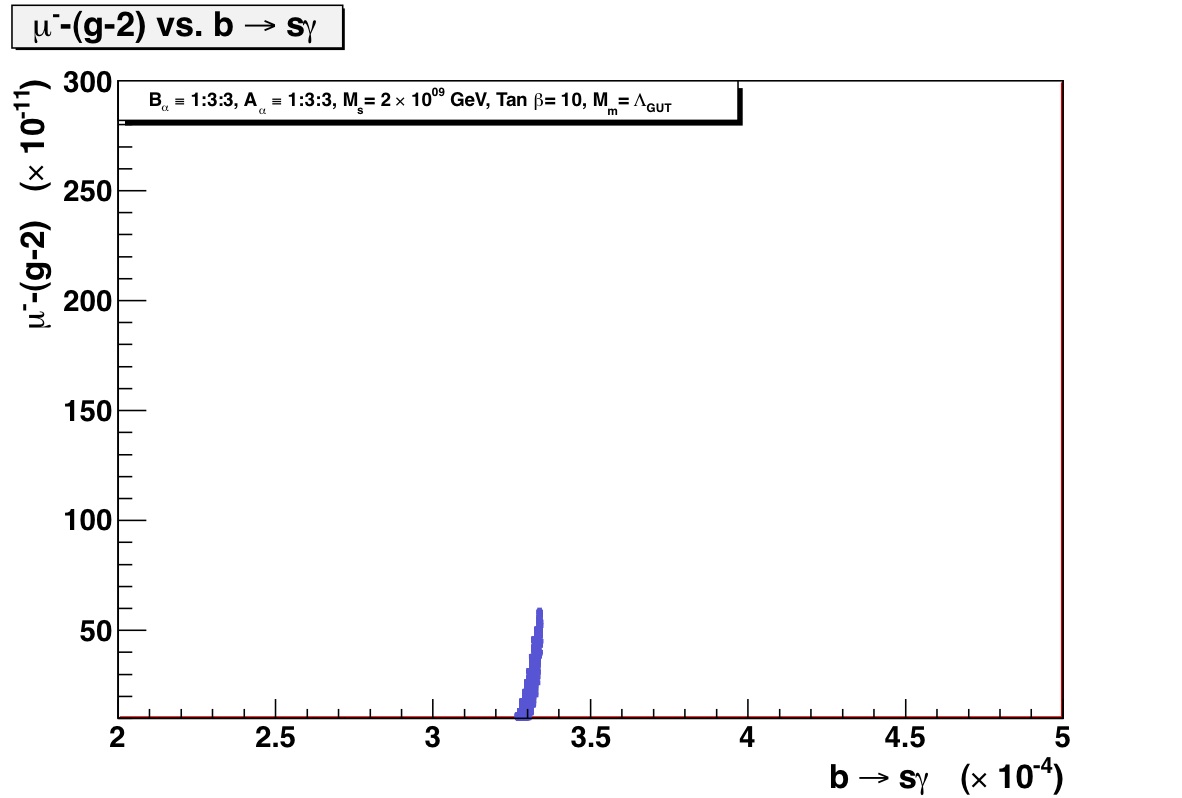}
  \includegraphics[scale=0.374]{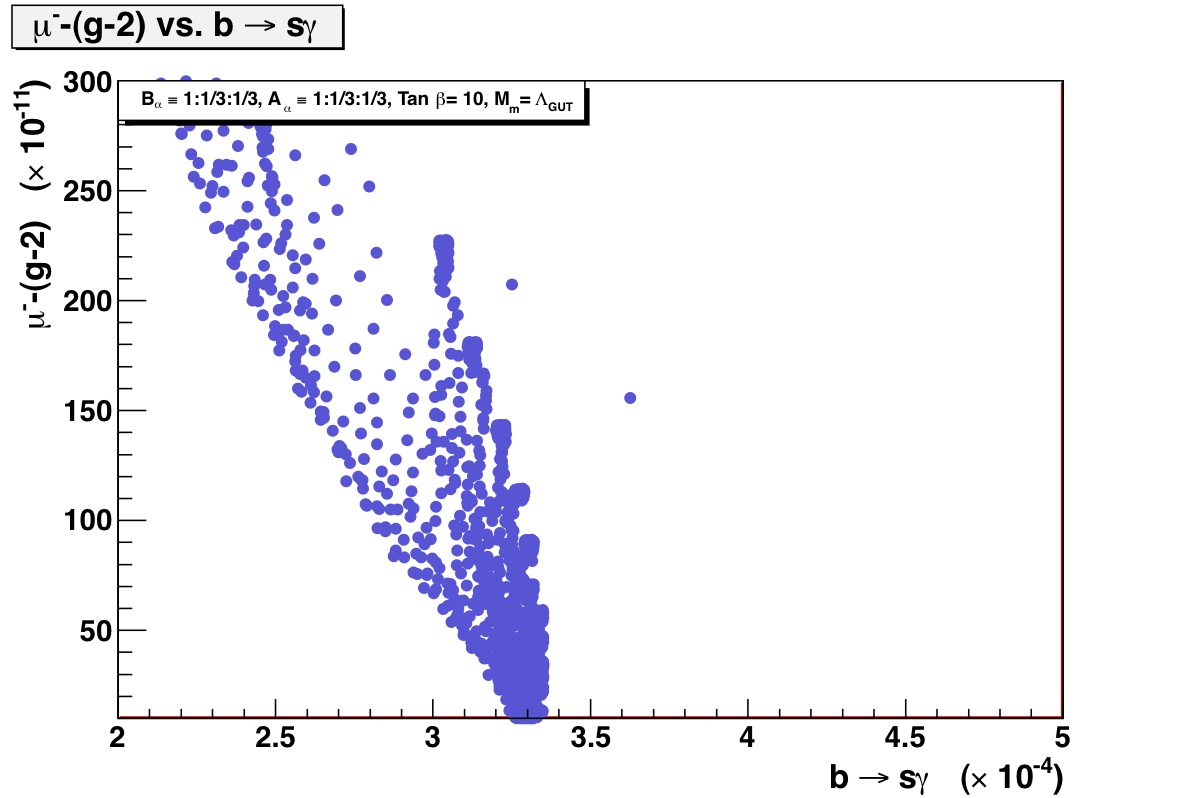}
 \caption{$\Delta a_{\mu}$ vs. $\text{BR}(b\rightarrow s \gamma )$ for the $(1:1:1\vert1:1:1)$ (top), $(1:1/3:1/3\vert1:1/3:1/3)$ (right) and $(1:3:3\vert1:3:3)$ (left) cases (MSSM). Note again that the direct collider bounds have not been imposed. We note that there is a slight anti-correlation between the $(g-2)_\mu$ contribution and the $\text{BR}(b\rightarrow s \gamma )$, which is most obvious in the $(1:1/3:1/3\vert1:1/3:1/3)$ case.}
  \label{bsgnlsp}
  }
 % *********************************
\par
Figures \ref{111mallowed} and \ref{311mallowed} show the contribution to the muon anomalous magnetic moment for the same correlation function ratios $(1:1:1\vert1:1:1)$, $(1:3:3\vert1:3:3)$ and $(1:1/3:1/3\vert1:1/3:1/3)$. Note that $(g-2)_{\mu}$ places strong constraints on the space if it is assumed that the MSSM contributions are viable explanations for the current discrepancy (i.e. the contribution is within at least $3\sigma$ of the current discrepancy between SM theory and experiment). 
\par
In the $(1:1:1\vert 1:1:1)$ case  
the region in the interval $\log_{10}B_1~\epsilon~[-6.5,-6.3]$ and $\log_{10}A_1~\epsilon~[-18,-13.5]$ have $(g-2)_\mu$ contributions that are in the interesting range for explaining the current discrepancy. Regions with $\log_{10}B_1\lesssim-6.5$ have even larger $(g-2)_\mu$ values but are ruled out by collider and flavor-physics bounds.  In regions outside $\log_{10}B_1~\epsilon~[-6.5,-6.3]$ and $\log_{10}A_1~\epsilon~[-18,-13.5]$ the $(g-2)_\mu$ contribution drops rapidly and have low values that are not sufficient to explain the discrepancy. 
\par
Changing the correlation function ratios from
\ba
(1:1:1\vert1:1:1)_{\text{\tiny{MSSM}}}\rightarrow (1:3:3\vert 1:3:3)_{\text{\tiny{MSSM}}} \; ,  
\ea
drastically reduces the $(g-2)_\mu$ values in the $\log_{10}B_1~\epsilon~[-6.5,-6.3]$ and $\log_{10}A_1~\epsilon~[-18,-13.5]$ intervals. The values are now lower by a factor of 3-6 and unlikely to explain the discrepancy by themselves.
%  *****************************
\FIGURE{
\includegraphics[scale=0.345]{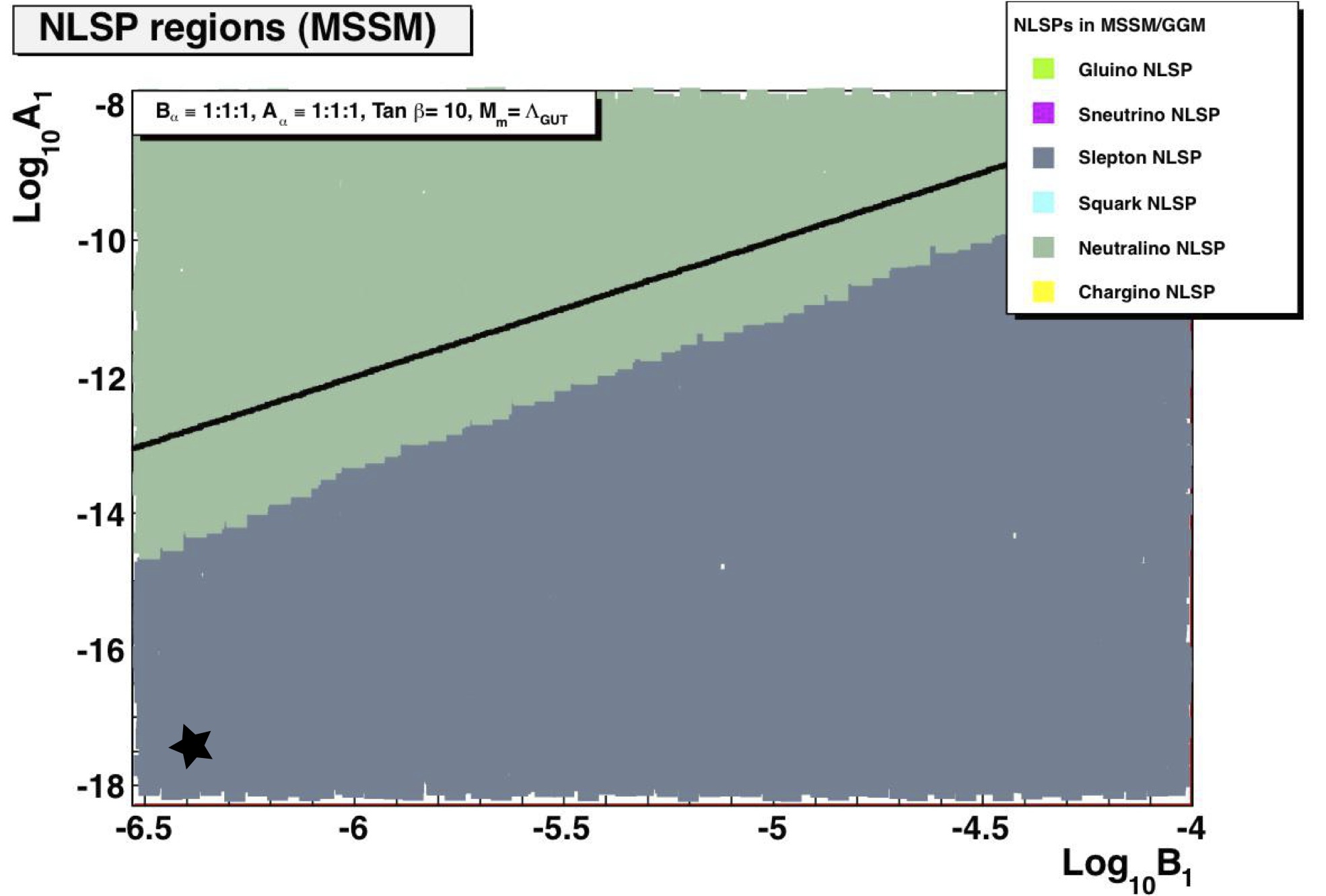}
\includegraphics[scale=0.345]{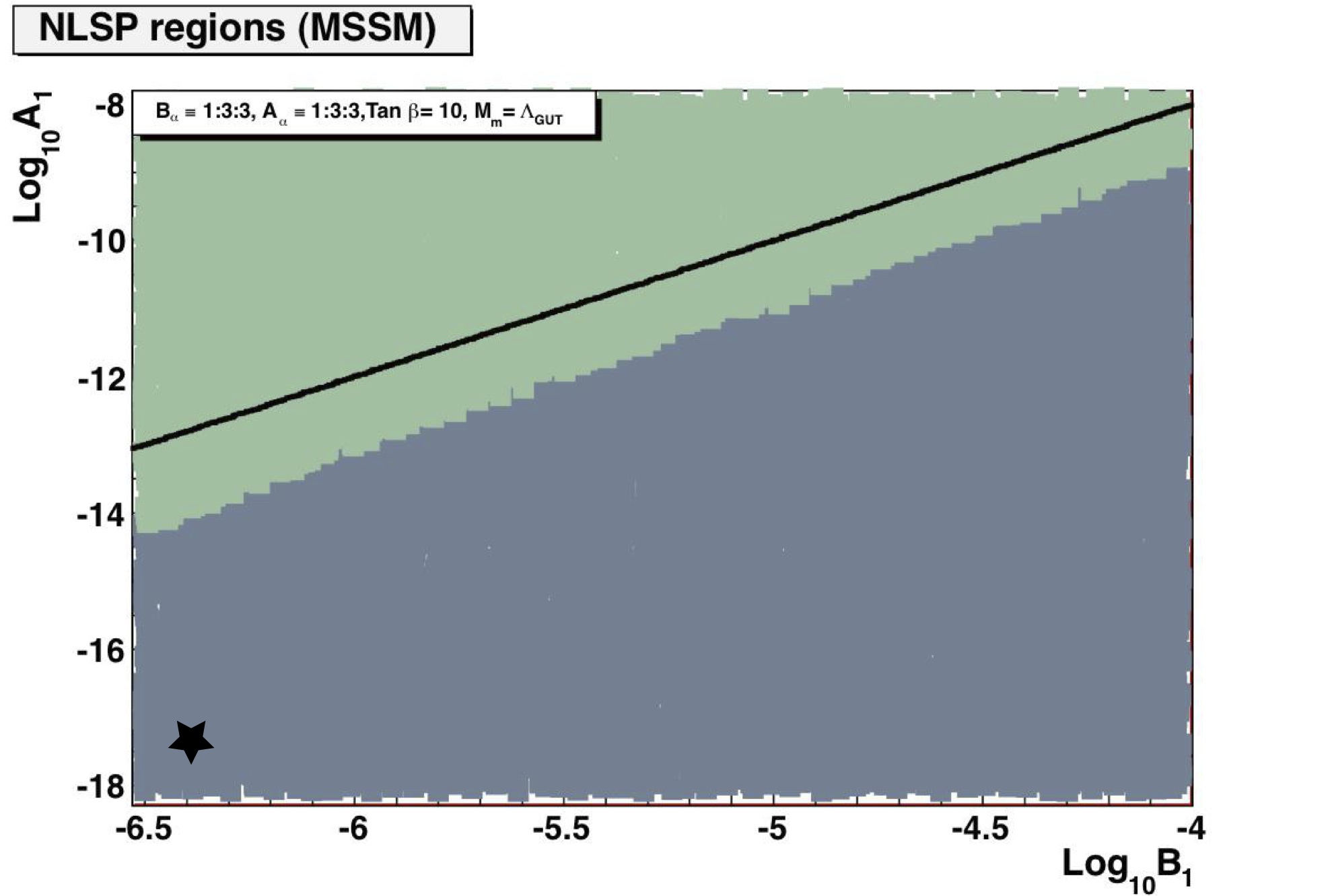}
\caption{ NLSP species in the $(1:1:1\vert1:1:1)$ (top) and  $(1:3:3\vert1:3:3)$ (bottom) GGM cases (MSSM). It is interesting that the NLSP topography looks very similar in the  $(1:1:1\vert1:1:1)$ and $(1:3:3\vert1:3:3)$ cases in spite of differences in the over-all mass spectra and other low-energy observables. The benchmark point we take as an example is again shown by the star symbol.   }
\label{111nallowed}
 }
 % *********************************
\par
For the $(1:1/3:1/3\vert 1:1/3:1/3)$ case regions below $\log_{10}B_1\lesssim-6.1$ become non-viable due to $(g-2)_\mu\geq~3\sigma$, flavor-physics exclusions and collider bounds. The interesting region is now shifted slightly to $\log_{10}B_1~\epsilon~[-6.1,-5.9]$ and $\log_{10}A_1~\epsilon~[-18,-13.0]$.
\par
Thus in the MSSM a combination of low-energy observables and consistency conditions place severe restrictions on the GGM parameter space. We will later look at the NMSSM case and come to a similar conclusion. In conclusion, for the $(1:1:1\vert 1:1:1)$ and $(1:1/3:1/3\vert 1:1/3:1/3)$ cases, there is a very narrow band where the $(g-2)_{\mu}$ contribution is within the $2\sigma$ interval of the current discrepancy. 
\par
The $\Delta a _\mu$ vs. $Br(B\to X_s \gamma)$ plot is shown in Fig. \ref{bsgnlsp}, for the three correlation function ratios. It is noticed that in the $(1:1:1\vert 1:1:1)$ and $(1:3:3\vert 1:3:3)$ cases the $Br(B\to X_s \gamma)$ is confined to a thin sliver between the values $[3.2,3.4]\times10^{-4}$. For intermediate values of $(g-2)_\mu$ the allowed $Br(B\to X_s \gamma)$ interval is the broadest. In the $(1:1/3:1/3\vert 1:1/3:1/3)$ case the $Br(B\to X_s \gamma)$ interval enlarges appreciably to $[2.0,3.4]\times10^{-4}$ for allowed values of $(g-2)_\mu$. We note a slight anti-correlation between $\Delta a _\mu$ and $Br(B\to X_s \gamma)$ in this case. There is also an aggregation of points for $Br(B\to X_s \gamma)$ in the interval $[3.2,3.4]\times10^{-4}$.
%  *****************************
\FIGURE{
\includegraphics[scale=0.345]{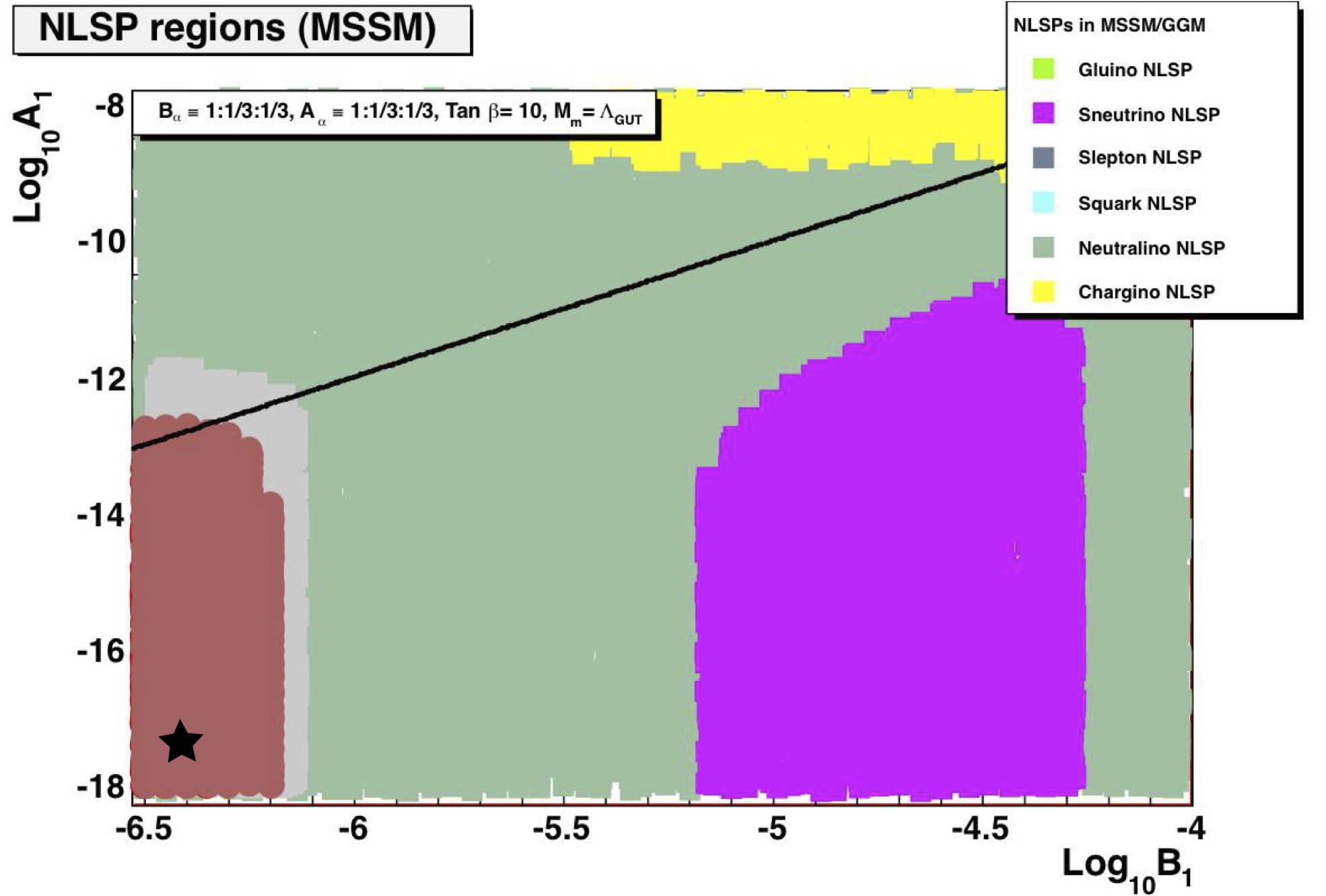}
\caption{ NLSP species in the $(1:1/3:1/3\vert 1:1/3:1/3)$ case (MSSM). The NLSP topography is quite distinct from the  $(1:1:1\vert1:1:1)$ and  $(1:3:3\vert1:3:3)$ cases. The collider exclusions are shown in grey and the flavor exclusions in magenta/pink. The flavor excluded regions have an overlap with regions excluded by $(g-2)_\mu\geq\,3\sigma$ and are not visible in the figure. }
\label{311nallowed}
 }
 % *********************************
 \par
Let us now consider the NLSP species in each of the cases. Figs. \ref{111nallowed} and \ref{311nallowed} survey the topography of the NLSP species. In the $(1:1:1\vert1:1:1)_{\text{\tiny{MSSM}}}$ and $(1:3:3\vert1:3:3)_{\text{\tiny{MSSM}}}$   cases the most favored NLSP candidate in the region with interesting values of $(g-2)_\mu$ (i.e. within $2\sigma$ of the current discrepancy) is the stau ($\tilde{\tau}$) with a small region (around $\log_{10}\tilde{B}_1\sim\,[-6.5,-6.3]$, $\log_{10}\tilde{A}_1\sim\,[-15,-13]$)  that also accommodates neutralinos ($\tilde{\chi}^0$). There are also regions with small contributions to $(g-2)_\mu$ that have both the slepton and neutralino as the NLSP. Observe that in the RGM case the NLSP would have been the $\tilde{\chi}^0$ solely and would give a very low value for $(g-2)_{\mu}$ in that region. In this context also note that there are viable regions with $M_{\lambda} \ll m_{\widetilde{f}}$ with a $\tilde{\chi}^0$ NLSP which nevertheless give a very low contribution to  $(g-2)_{\mu}$. Do note that for the above two cases there are boundaries where the $\tilde{\chi}^0$ and $\tilde{l}$ are almost degenerate in mass. In these regions $\tilde{\chi}^0$-$\tilde{l}$ coannihilations might be important \cite{Rajaraman:2009ga} and could lead to interesting phenomenology. Also note that there are no viable regions with a chargino ($\chi^\pm$) or sneutrino ($\tilde{\nu}$) NLSP in these two cases.
 %  *****************************
\FIGURE{
 \includegraphics[scale=0.25]{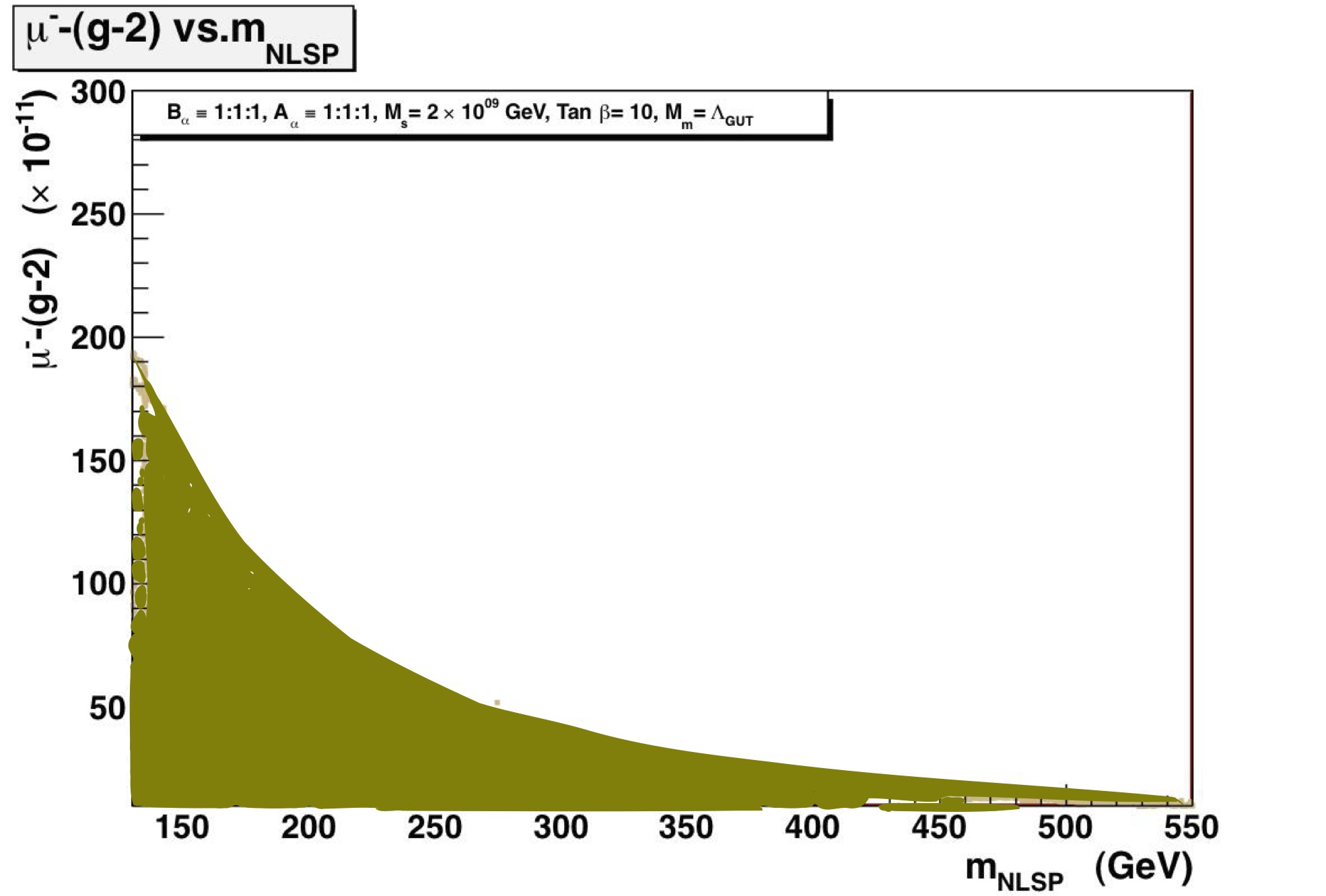}
 \includegraphics[scale=0.25]{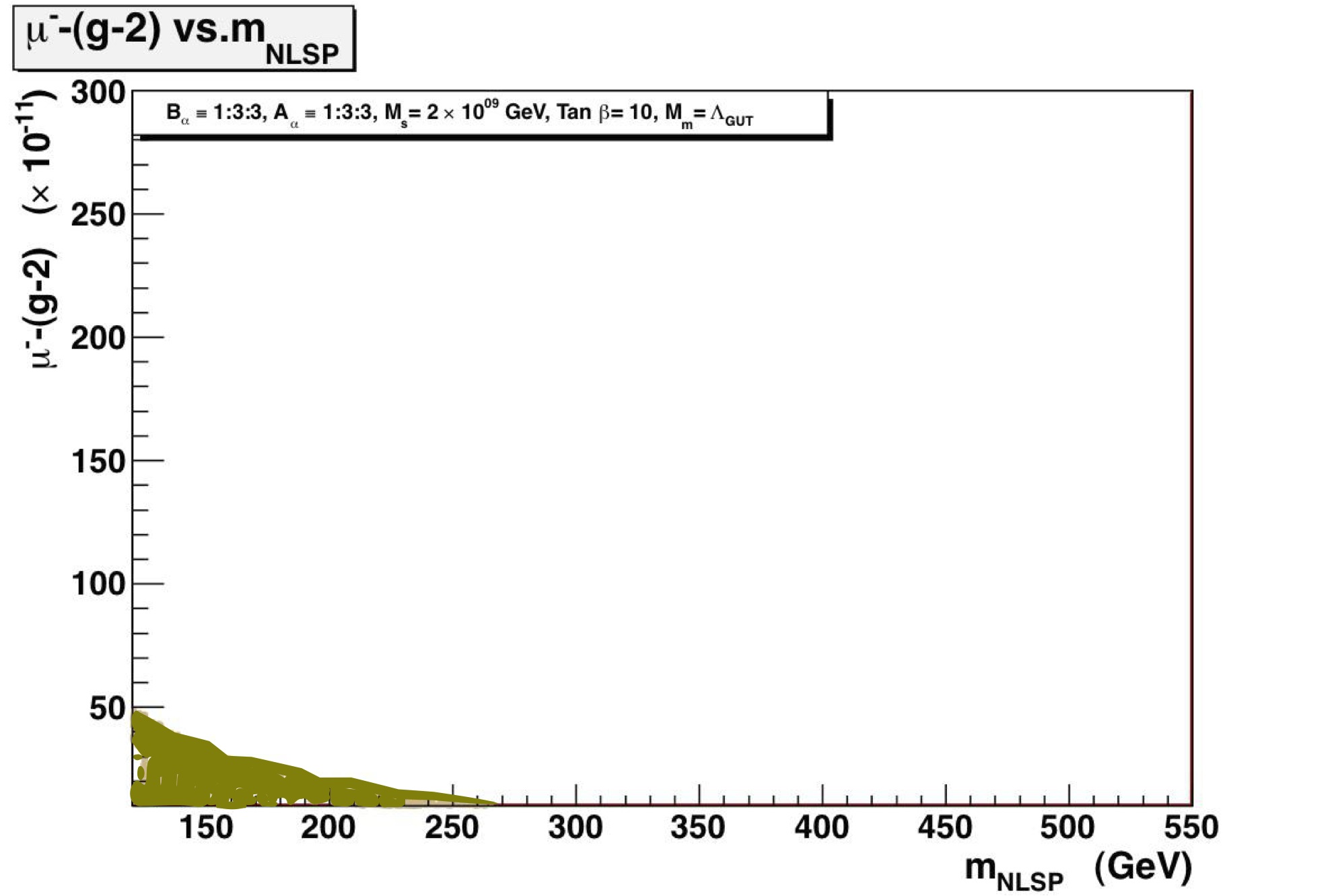}
 \includegraphics[scale=0.25]{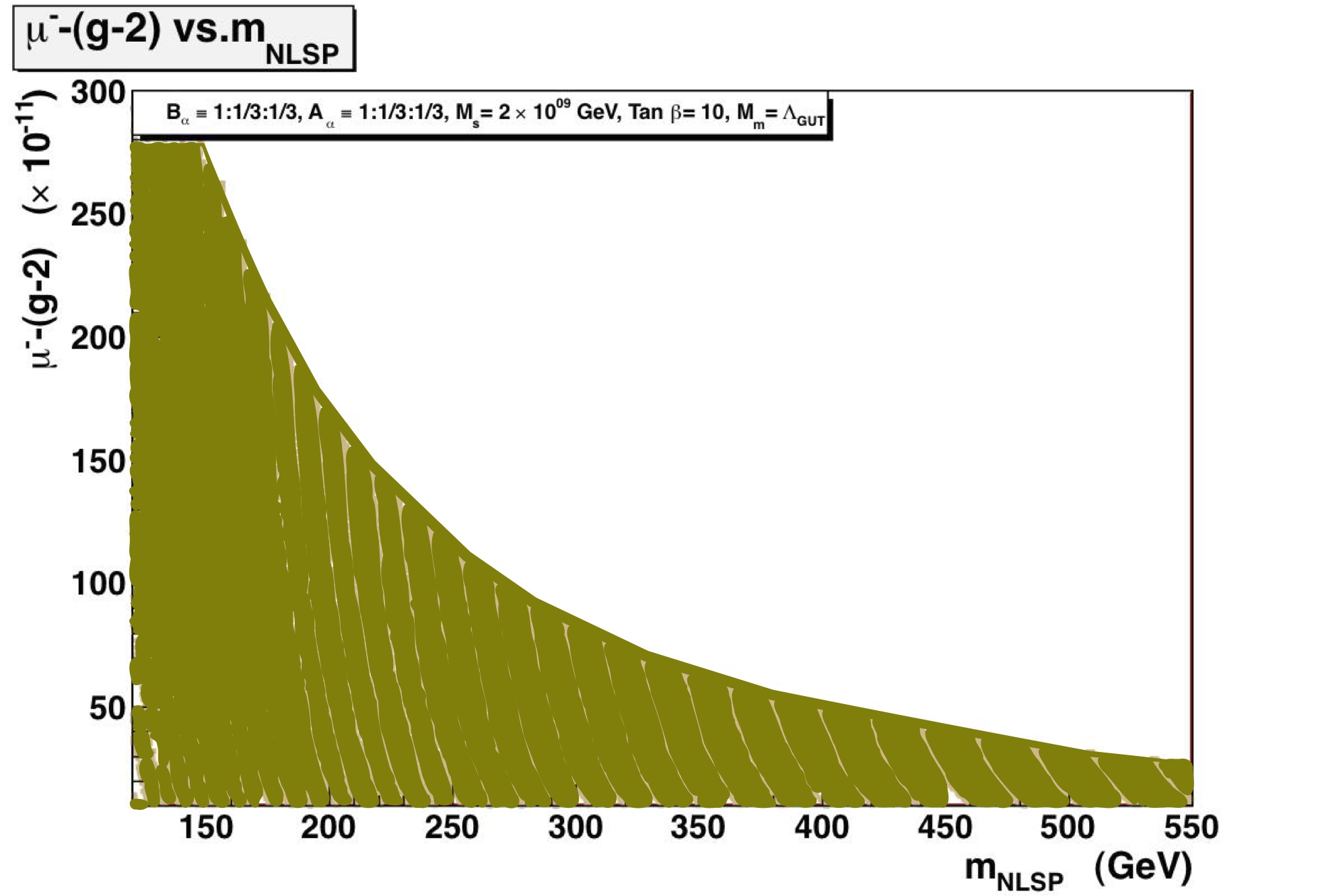}
  \caption{$\Delta a_{\mu}$ vs. $m_{\text{\tiny{NLSP}}}$ for the $(1:1:1\vert1:1:1)$ (top), $(1:3:3\vert1:3:3)$ (middle) and$(1:1/3:1\vert1:1/3:1/3)$ (bottom) cases (MSSM). As previously mentioned we only focus on positive $(g-2)_\mu$ contributions. Note that the direct collider bounds have not been imposed in the plot. The constraints from the collider bounds, in each of these cases, can nevertheless be easily read off by comparison to the $(g-2)_\mu$ plots presented earlier. The filled regions denote allowed values.}
 \label{amunlsp}
  }
 % *********************************
\par 
In the $(1:1/3:1/3\vert1:1/3:1/3)_{\text{\tiny{MSSM}}}$ case the NLSP topography is completely transformed. In contrast to the $(1:1:1\vert1:1:1)_{\text{\tiny{MSSM}}}$ and $(1:3:3\vert1:3:3)_{\text{\tiny{MSSM}}}$ cases there are viable regions with a $\chi^\pm$ or $\tilde{\nu}$ NLSP. In the regions with a significant $(g-2)_\mu$ contribution, the favored NLSP is now solely the $\tilde{\chi}^0$. In the approximate interval $\log_{10}B_1~\epsilon~[-5.2,-4.3]$ and $\log_{10}A_1~\epsilon~[-18,-11.0]$ the $\tilde{\nu}$ is the NLSP. For large values of $\log_{10}A_1\sim -8$, between the intervals $\log_{10}B_1~\sim[-5.5,-4.1]$ the $\tilde{\chi}^\pm$ \textit{along with} the $\tilde{\chi}^0$ are degenerate in mass. Thus in contrast to $(1:1:1\vert1:1:1)_{\text{\tiny{MSSM}}}$ and $(1:3:3\vert1:3:3)_{\text{\tiny{MSSM}}}$ now there are regions and boundary regions with $\tilde{\chi}^0-\tilde{\chi}^\pm$ and $\tilde{\chi}^0-\tilde{\nu}$ degenerate in mass. In the vicinity of these regions again co-annihilations among the almost degenerate NLSP candidates become important and will lead to distinct phenomenology.
\par
A plot of $\Delta a _\mu$ vs. $m_{\text{\tiny{NLSP}}}$ is shown in Fig. \ref{amunlsp} for the three correlation function ratios. The filled regions denote the allowed values. It is clear that for each of the cases the allowed values of the NLSP mass are correlated, albeit weakly, with the $(g-2)_\mu$ value. We observe that for large values of the $(g-2)_\mu$ contribution the allowed range of the NLSP masses is smaller than for lower values of $(g-2)_\mu$. In the  $(1:3:3\vert1:3:3)$ case the NLSP mass is within a small interval of $[100,275]~\rm{GeV}$.  For the $(1:1:1\vert1:1:1)$ and  $(1:1/3:1/3\vert1:1/3:1/3)$ cases the NLSP mass is within a broader interval stretching all the way from $100~\rm{GeV}$ to $\sim 600~\rm{GeV}$. For a given value of $(g-2)_\mu$ the 
$(1:1/3:1/3\vert1:1/3:1/3)$ case accommodates a wider range of allowed NLSP masses than the $(1:1:1\vert1:1:1)$ case.
\par
 The mass of the lightest CP-even Higgs is plotted in Figs. \ref{111hallowed} and \ref{311hallowed}. For the $(1:1:1\vert1:1:1)$, $(1:3:3\vert1:3:3)$ and $(1:1/3:1/3\vert1:1/3:1/3)$ cases in the region with interesting values of $(g-2)_\mu$, the mass of the lightest CP-even Higgs (including the leading 2-loop corrections) is found to be typically in the interval $[114,~125]~\rm{GeV}$. The actual value at a given point in parameter space is seen to depend sensitively on the correlator ratios. It is found in general that increasing the ratio  of the other correlation functions with respect to $\tilde{B}_1$ and $\tilde{A}_1$ increases the Higgs boson mass and conversely decreasing the ratio decreases the mass.
 %  *****************************
\FIGURE{
 \includegraphics[scale=0.335]{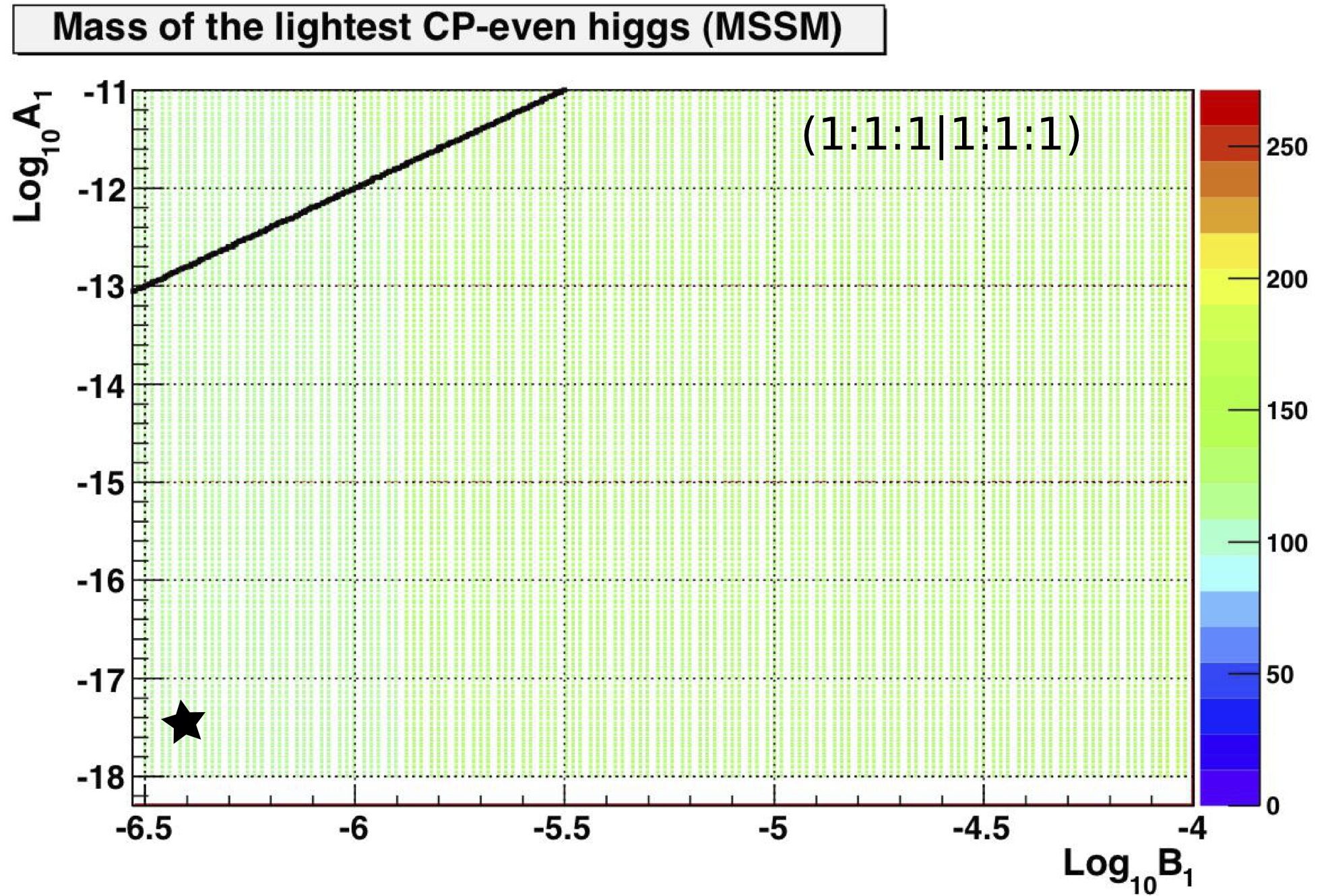}
  \includegraphics[scale=0.335]{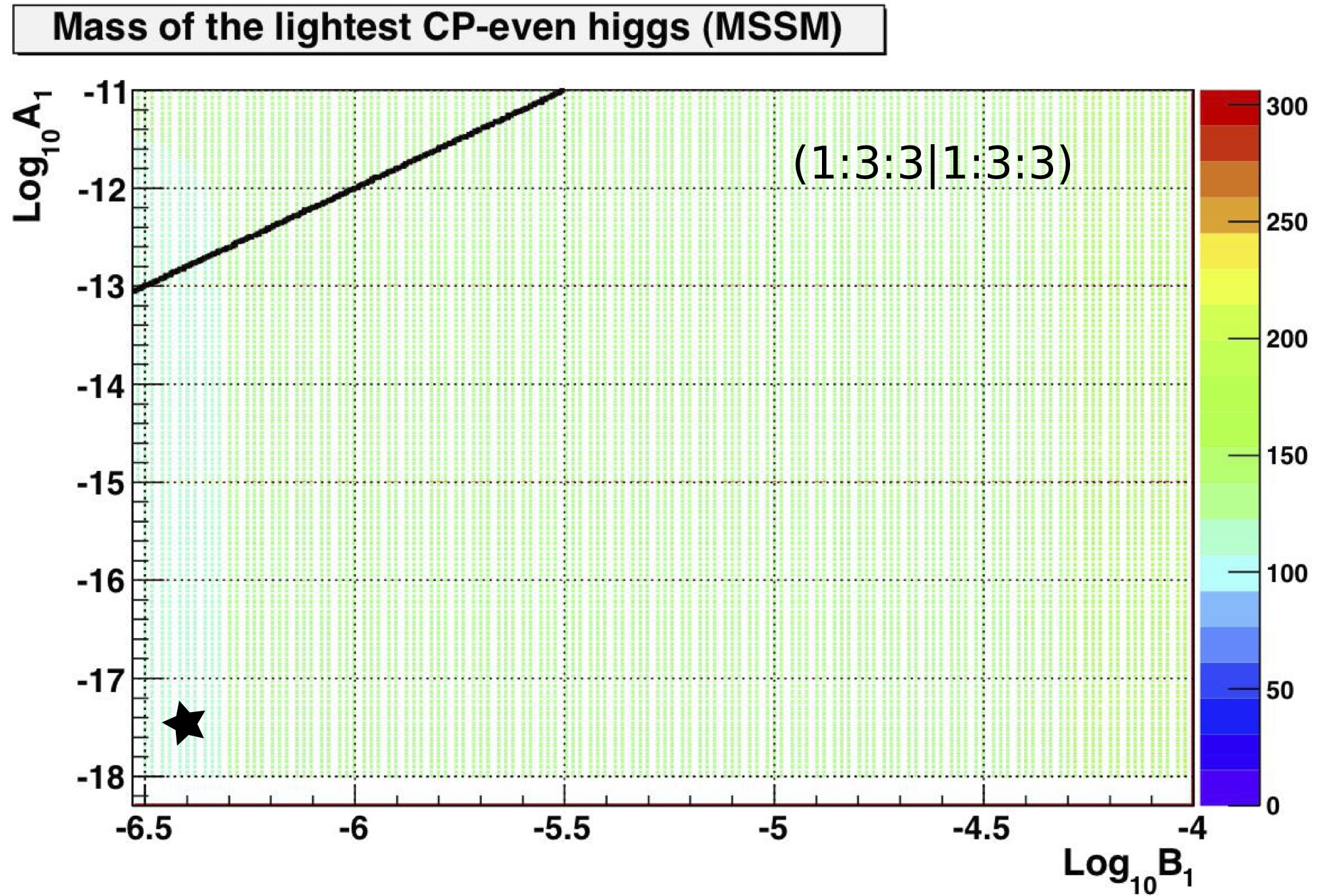}
 \caption{Mass of the lightest CP-even Higgs in the MSSM for the $(1:1:1\vert1:1:1)$ (top) and $(1:3:3\vert1:3:3)$ (bottom) GGM cases. The color scales are in units of GeV and different in each case.}
 \label{111hallowed}
  }
 % *********************************

  %  *****************************
\FIGURE{
 \includegraphics[scale=0.335]{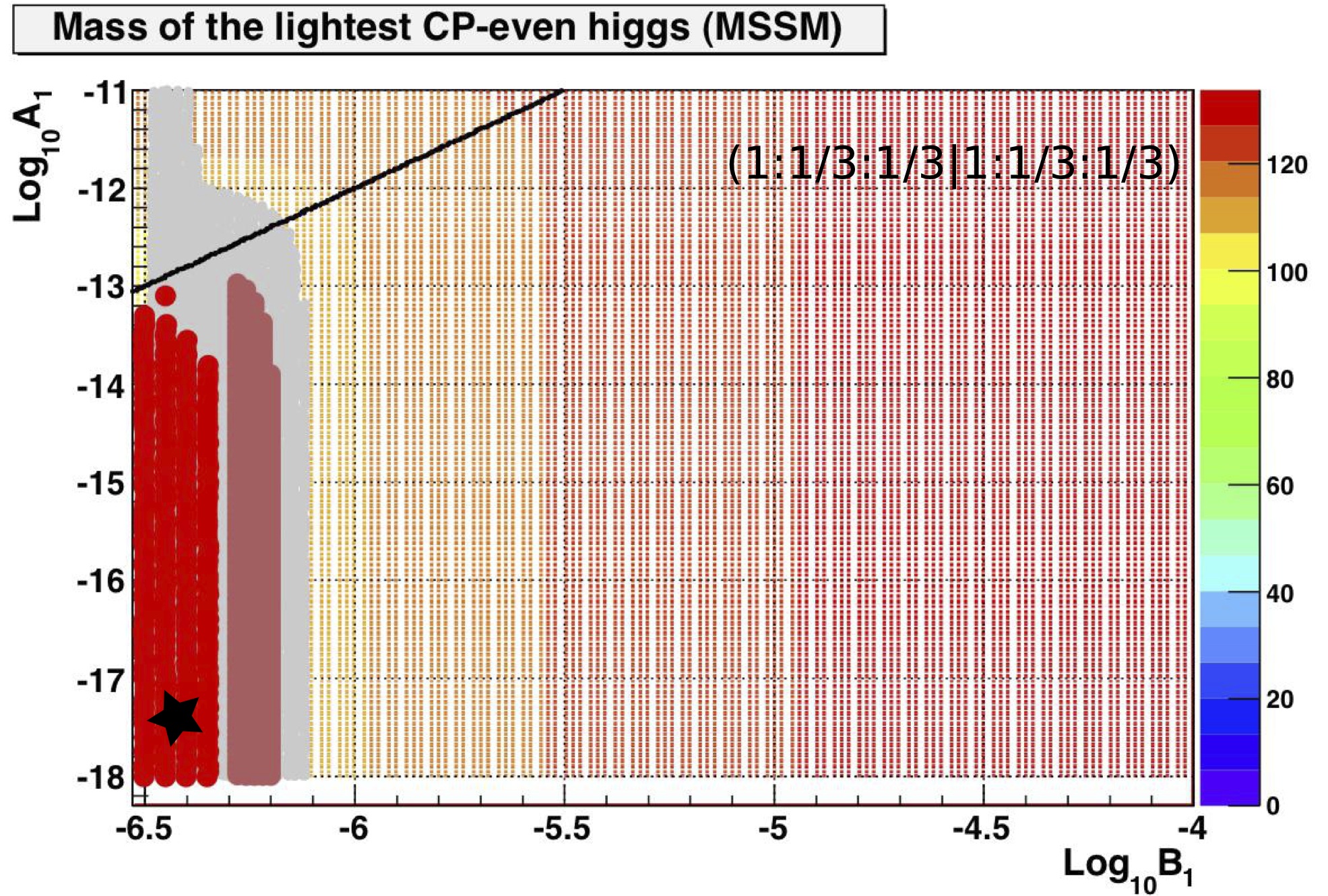}
 \caption{ Mass of the lightest CP-even Higgs in the $(1:1/3:1/3\vert1:1/3:1/3)$ case. Note that the color scales are again in units of GeV and different from the $(1:1:1\vert1:1:1)$ and $(1:3:3\vert1:3:3)$ cases. Once again the collider bounds are in grey, the flavor bounds in magenta/pink and $(g-2)_\mu$ exclusions in deep red (unhatched).}
 \label{311hallowed}
  }
 % *********************************
\par 
RGM models with very large $N_{m}$ have a superpartner spectrum similar to gaugino mediation (where the scale is typically taken close to the GUT scale). GGM can therefore be thought of as an interpolation between RGM and gaugino mediation~\cite{Rajaraman:2009ga}. This is also another motivation for taking our messenger scale to be near $\Lambda_{GUT}$, so as to make it more convenient to compare GGM scenarios with mSUGRA or gaugino mediation scenarios.
\par 
Owing to the freedom of current-current correlator ratios there may be scenarios where GGM gives a squeezed superpartner spectrum alleviating the little hierarchy problem. If $\tilde{B}_{3}$ is lower with respect to $\tilde{B}_2$, for example, this will compress the low-energy spectrum (making for instance the $\tilde{g}$-$\tilde{\chi}$ mass ratios smaller) while if it is higher it will further increase the hierarchy. For example, at the point $(-6.5,-17.5)$ (MSSM case) in the $(\log_{10} \tilde{B}_1,\log_{10} \tilde{A}_1)$ space (with $\tan\beta=10$ and $M_{m}=\Lambda_{GUT}$) going from 
\ba
(1:3:3\vert1:3:3)_{\text{\tiny{MSSM}}}\rightarrow (1:3:1\vert1:3:3)_{\text{\tiny{MSSM}}} \; ,  
\ea
changes the $\tilde{g}$-$\tilde{\chi}$ mass ratio by
\ba
\left(\frac{m_{\tilde{g}}}{m_{\tilde{\chi^0_4}}}\right)_{\text{\tiny{EW}}}~&\approx&~~~~2 \rightarrow ~1  \; ,  \\
\left(\frac{m_{\tilde{g}}}{m_{\tilde{\chi^0_1}}}\right)_{\text{\tiny{EW}}}~&\approx&~~~~16 \rightarrow ~6  \; .\\
\ea
Lowering $\tilde{B}_{2}$ with respect to $\tilde{B}_{3}$ can similarly make the mass ratio selectively larger.
 \par
The RGM intuition of large squark ($\tilde{q}$) to slepton ($\tilde{l}$) mass ratios need not again necessarily hold true in GGM. This is because generically the $\tilde{l}$ have larger hypercharges and for large enough $\tilde{A}_1$ the hypercharge term may selectively give a larger contribution to $\tilde{l}$ masses as compared to the squarks.  Going from 
\ba
(1:1:1\vert1:1:1)_{\text{\tiny{MSSM}}}\rightarrow (1:1:1\vert \mathcal{K}:\mathcal{K}:\mathcal{K})_{\text{\tiny{MSSM}}} \; ,  
\ea
at the high scale with $\mathcal{K}\gg1$ is seen to decrease the $\tilde{q}$ to $\tilde{l}$ low-scale mass ratio appreciably in large parts of the viable GGM space. Note that this is equivalent to raising the over-all scale of the $\tilde{A}_r$ correlation functions relative to the scale of the $\tilde{B}_r$ correlation functions. For example starting at $(-6.0,-12.0)$ (which for  $(1:1:1\vert1:1:1)$ is an RGM like point) in the $(\log_{10} \tilde{B}_1,\log_{10} \tilde{A}_1)$ space, with $\tan\beta=10$ and $M_{m}=\Lambda_{GUT}$ as before, going from $(1:1:1\vert1:1:1)_{\text{\tiny{MSSM}}}$ to $(1:1:1\vert \mathcal{K}:\mathcal{K}:\mathcal{K})_{\text{\tiny{MSSM}}}$ with $\mathcal{K}=20$
changes the $\tilde{q}$ to $\tilde{l}$ ratios at the low-energy scale from
\bea
\left(\frac{m_{\tilde{q}}}{m_{\tilde{l}}}\right)~&\approx&~ 2\rightarrow 1  \; .\\ \nn
\eea
 \par 
Existing models of gaugino-mediation have large hierarchy between sfermion and gaugino masses at the compactification scale (typically $\sim \Lambda_{GUT}$) which gets washed out at EW scales due to RG running. 
For instance a typical spectrum in gaugino mediation at the compactification scale ($\sim \Lambda_{GUT}$) may be (see for example \cite{{McGarrie:2010kh},{Chacko:1999mi}})
\ba
M_a &=& m_{1/2} ~~~\forall \, a\, \epsilon\, 1,2,3 \; ,  \\[1mm]
m_{\tilde{f}}^2 &\simeq& \frac{m_{1/2}^2}{16\pi^2} \; ,  \\[1mm]
A&\sim&  \frac{m_{1/2}}{16\pi^2} \; ,  \\[1mm]
\mu &\sim& m_{1/2} \; ,  \\
M^2_{H_i}&\sim&~m_{1/2}^2~~~\forall \, i\, \epsilon\, u,d  \; ,  \\[1mm]
B\mu &\sim& m_{1/2} \; ,  
\ea
which exhibits a huge gaugino to sfermion mass ratio of $\mathcal{O}(4\pi)$. This ratio gets reduced to $\mathcal{O}(1)$ when we run the masses from $\Lambda_{GUT}$ to $\Lambda_{EW}$~\cite{Chacko:1999mi} . If the compactification scale is taken to be low, then such a ``low-scale" gaugino-mediation usually prefers a $\tilde{l}$ NLSP over a $\tilde{B}$ NLSP. This is due to the short RG running scale.
%  *****************************
\FIGURE{
 \includegraphics[scale=0.3]{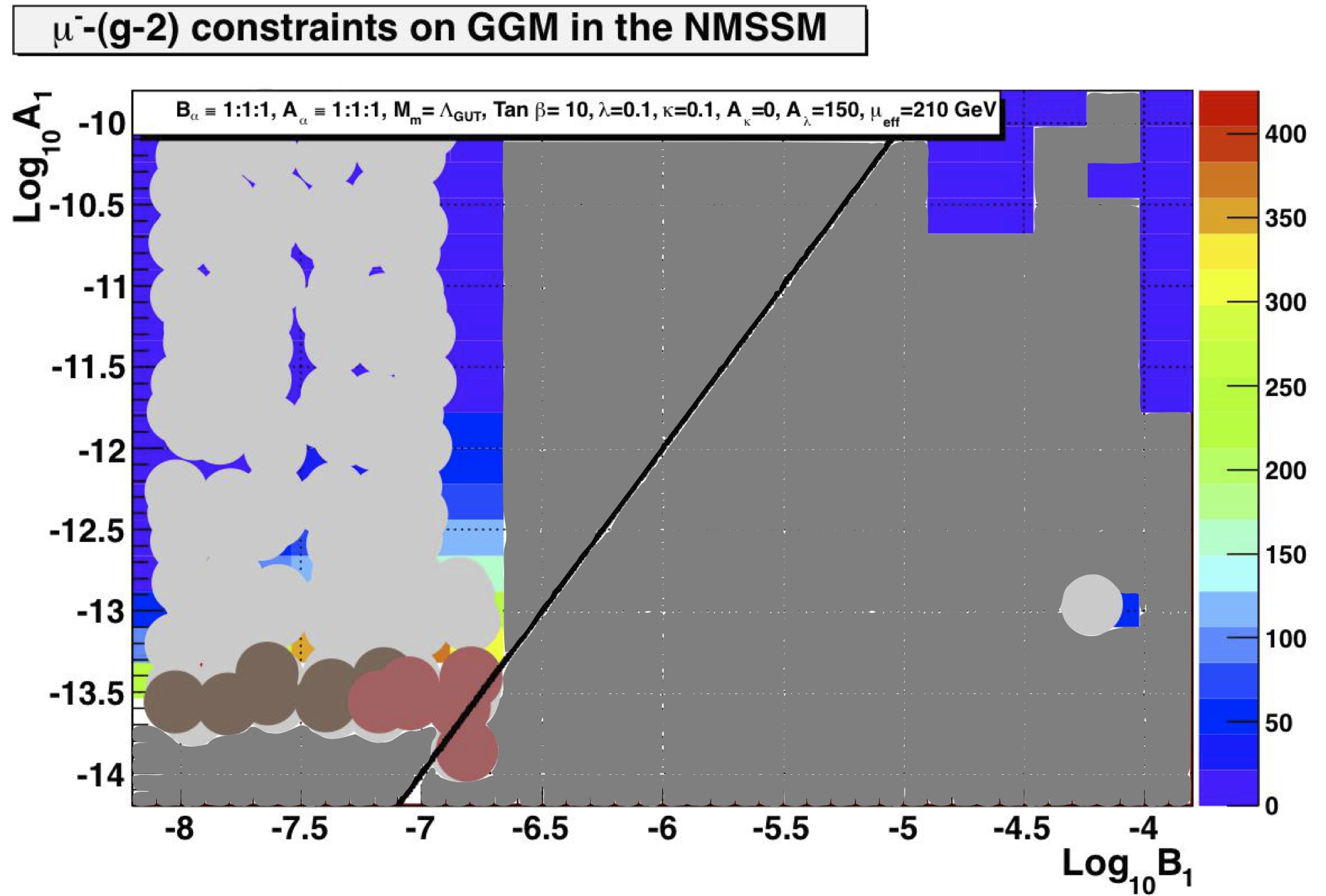}
  \includegraphics[scale=0.3]{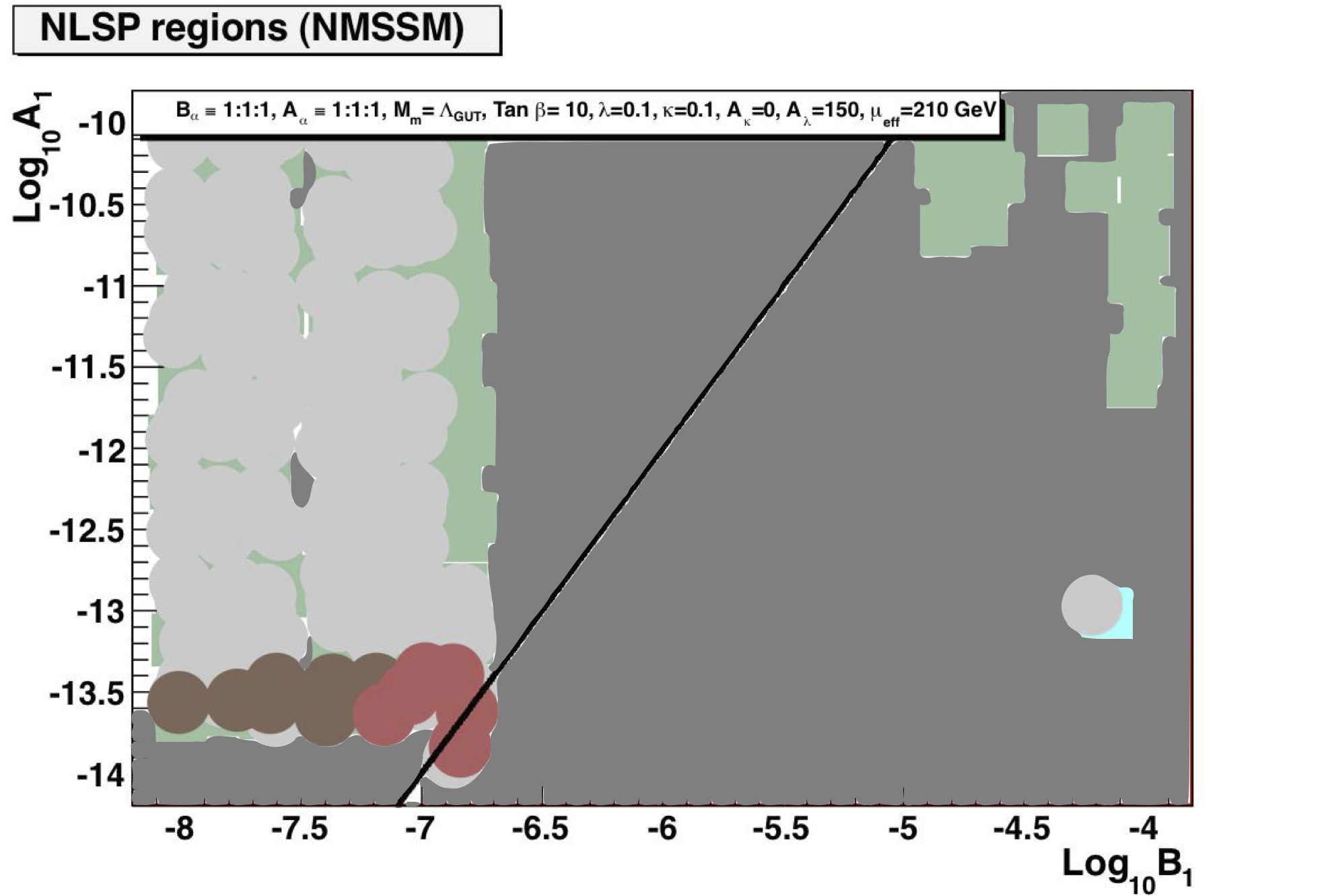}
 \caption{A section of the viable parameter space in the NMSSM for the $(1:1:1\vert1:1:1)$ case showing the $\Delta a_{\mu}$ contribution (top) and the NLSP candidates (bottom). Regions with tachyonic masses or where the solutions to the 1-loop RG equations failed to converge are shown in dark grey. Light grey represents regions excluded by collider bounds, brown denotes regions where $(g-2)_\mu$ is above $3\sigma$ and magenta represents regions ruled out by flavor-physics constraints. The $\Delta a_{\mu}$ contribution for this choice of parameters is relatively small. Note that the lightest CP-odd scalar in this case generally has a low mass (see for example the benchmark point in Table\,\ref{bmp2}) and its 1-loop contribution is negative. The regions near the linear black band correspond to RGM and it is clear from the plots that, with the current choice of model parameters, most regions with an RGM-like spectra are already excluded. The NLSP is the neutralino (represented in green) in the allowed regions shown.}
 \label{NMSSM1}
  }
 % *********************************
 
 %  *****************************
\FIGURE{
 \includegraphics[scale=0.3]{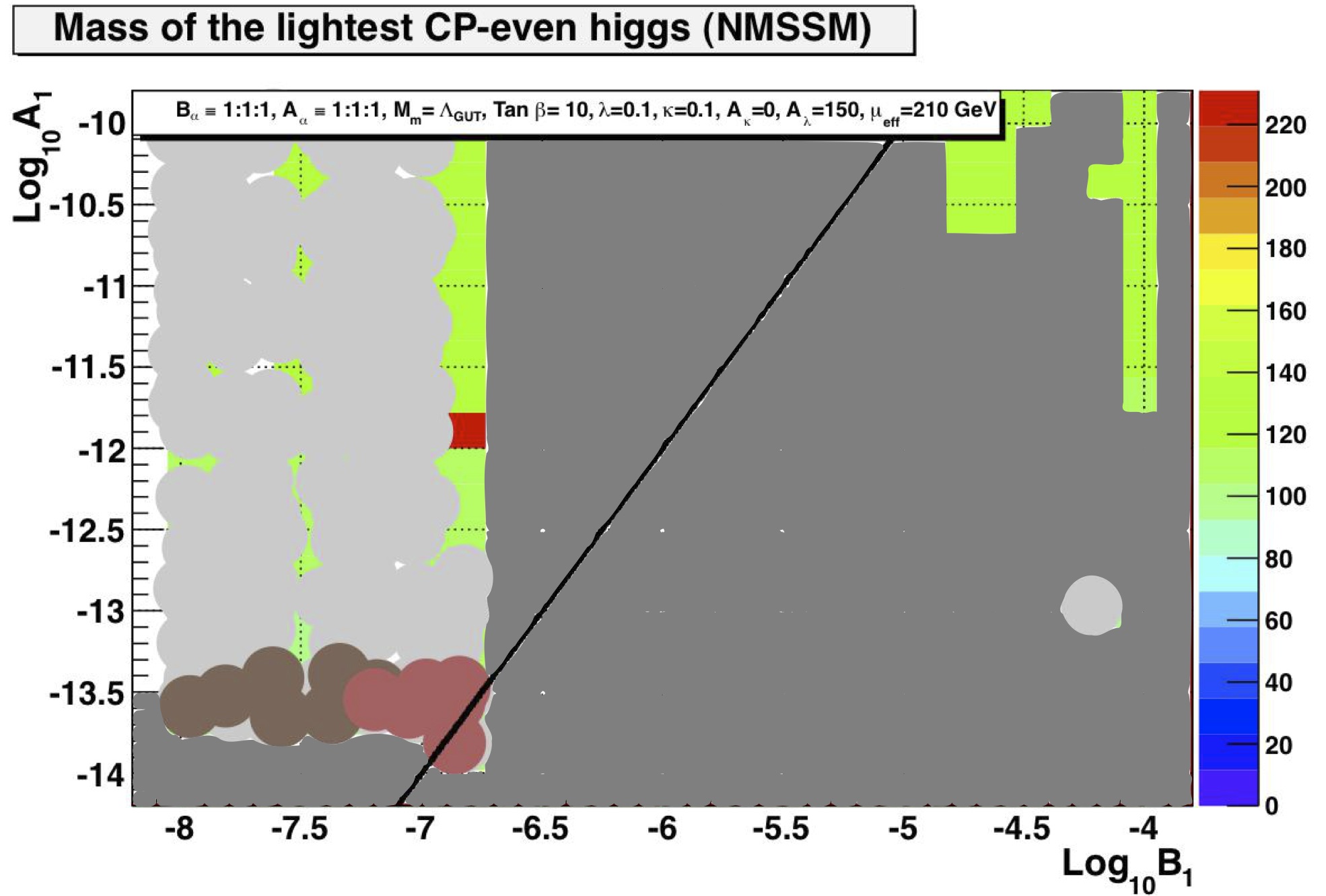}
 \caption{The lightest CP-even Higgs mass in the $(1:1:1\vert1:1:1)$ case. The mass is relatively small and in the $[111,125] ~\rm{GeV}$ window. As in the previous figure, regions with tachyonic masses or where the solutions to the 1-loop RG equations failed to converge are shown in dark-grey. Light-grey represents regions excluded by collider bounds, brown denotes regions where $(g-2)_\mu$ is above $3\sigma$ and magenta represents regions ruled out by flavor-physics constraints.}
 \label{NMSSM1higgs}
  }
 % *********************************

%  *****************************
\FIGURE{
 \includegraphics[scale=0.3]{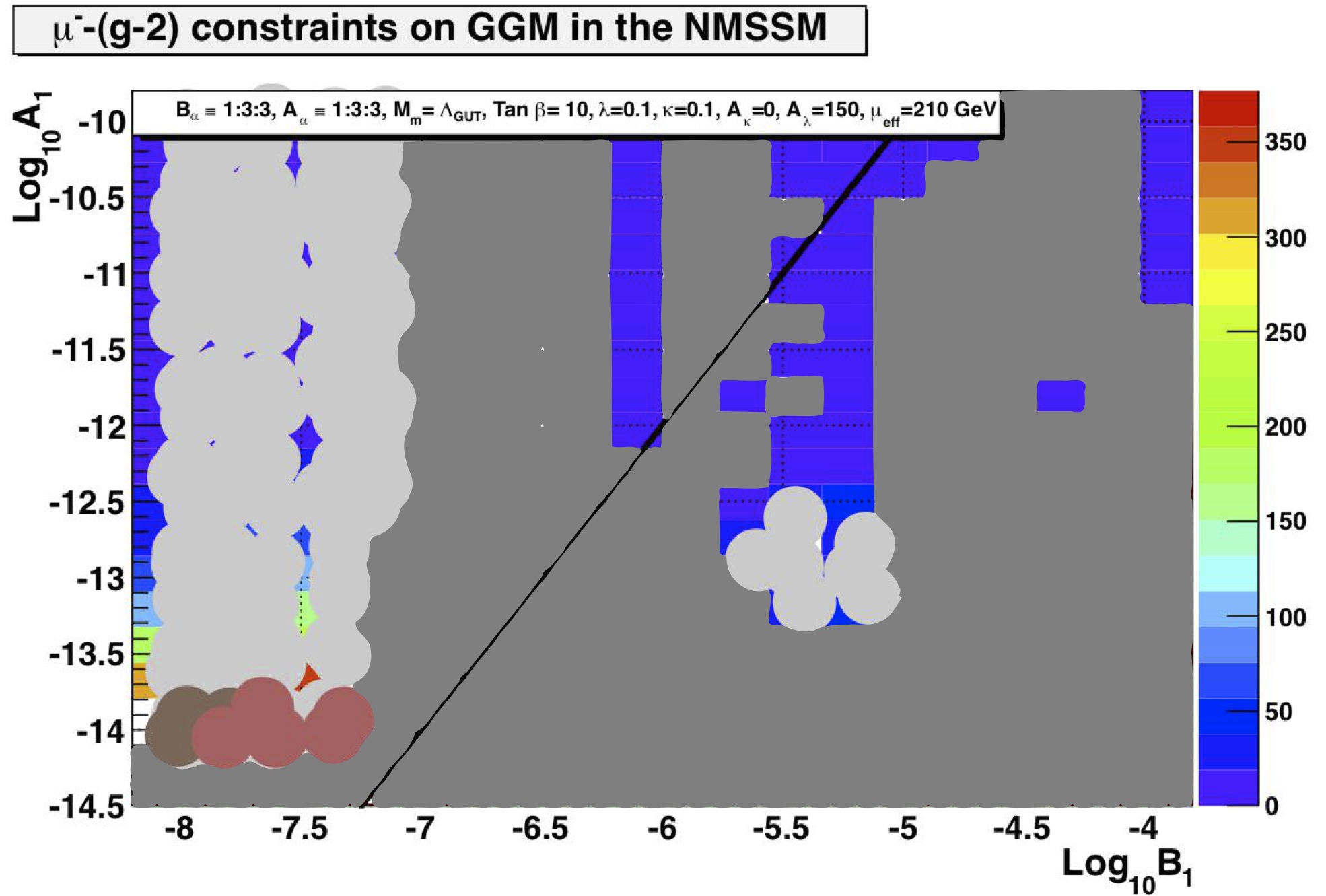}
  \includegraphics[scale=0.3]{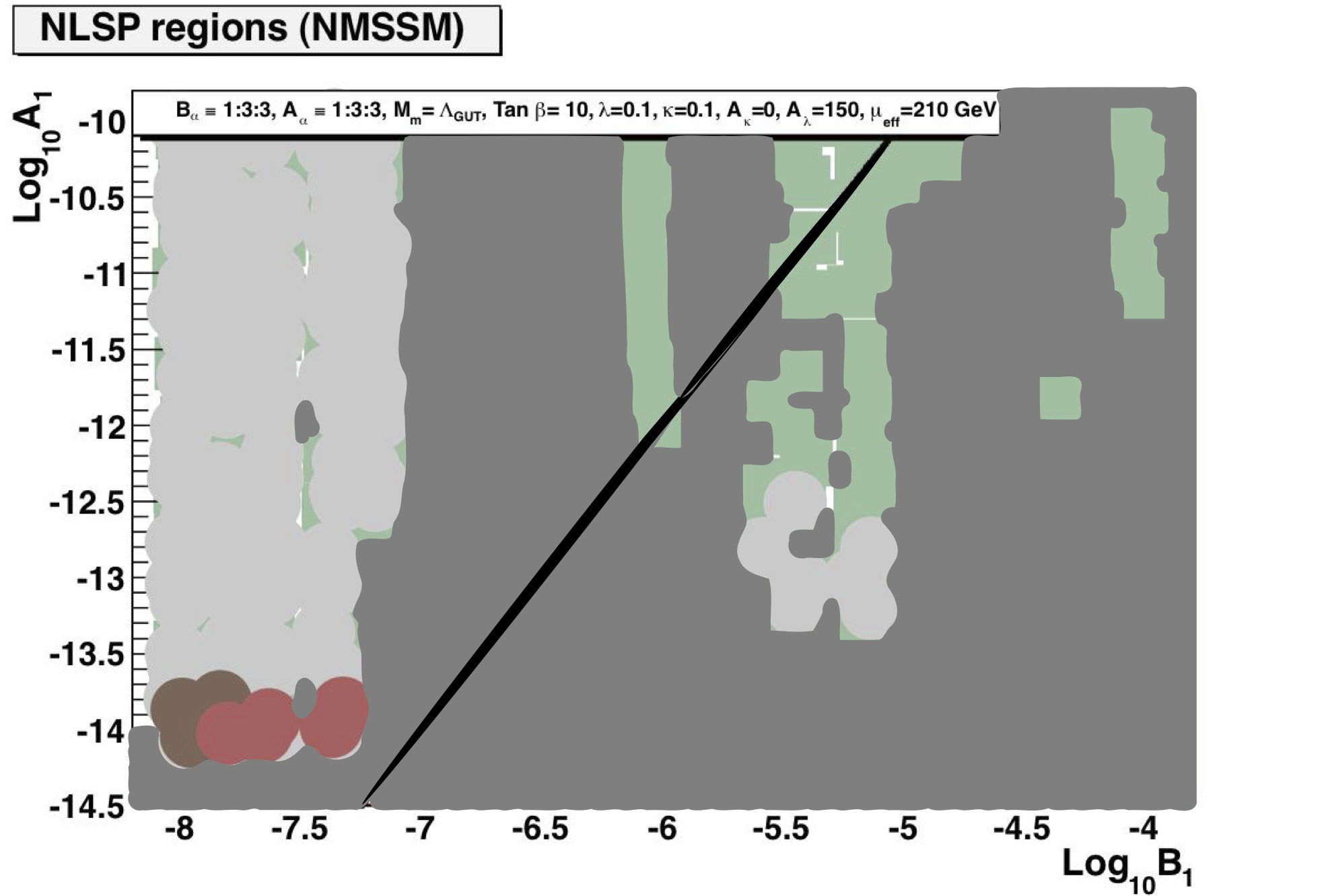}
 \caption{A section of the viable parameter space in the NMSSM for the $(1:3:3\vert1:3:3)$ case, showing the $\Delta a_{\mu}$ contribution (top) and the NLSP candidate (bottom). The color schemes are identical to the $(1:1:1\vert1:1:1)_{\rm{NMSSM}}$ case. Changing the correlation function ratios from $(1:1:1\vert1:1:1)$ to  $(1:3:3\vert1:3:3)$ has further decreased the $(g-2)_\mu$ contribution but not appreciably. Also note that certain regions previously allowed in the $(1:1:1\vert1:1:1)$ case are now non-viable and there are new regions that are viable. The NLSP species in this section of the viable parameter space is still the neutralino.}
 \label{NMSSM2}
  }
 % ******************************************************************

 %  *****************************
\FIGURE{
 \includegraphics[scale=0.3]{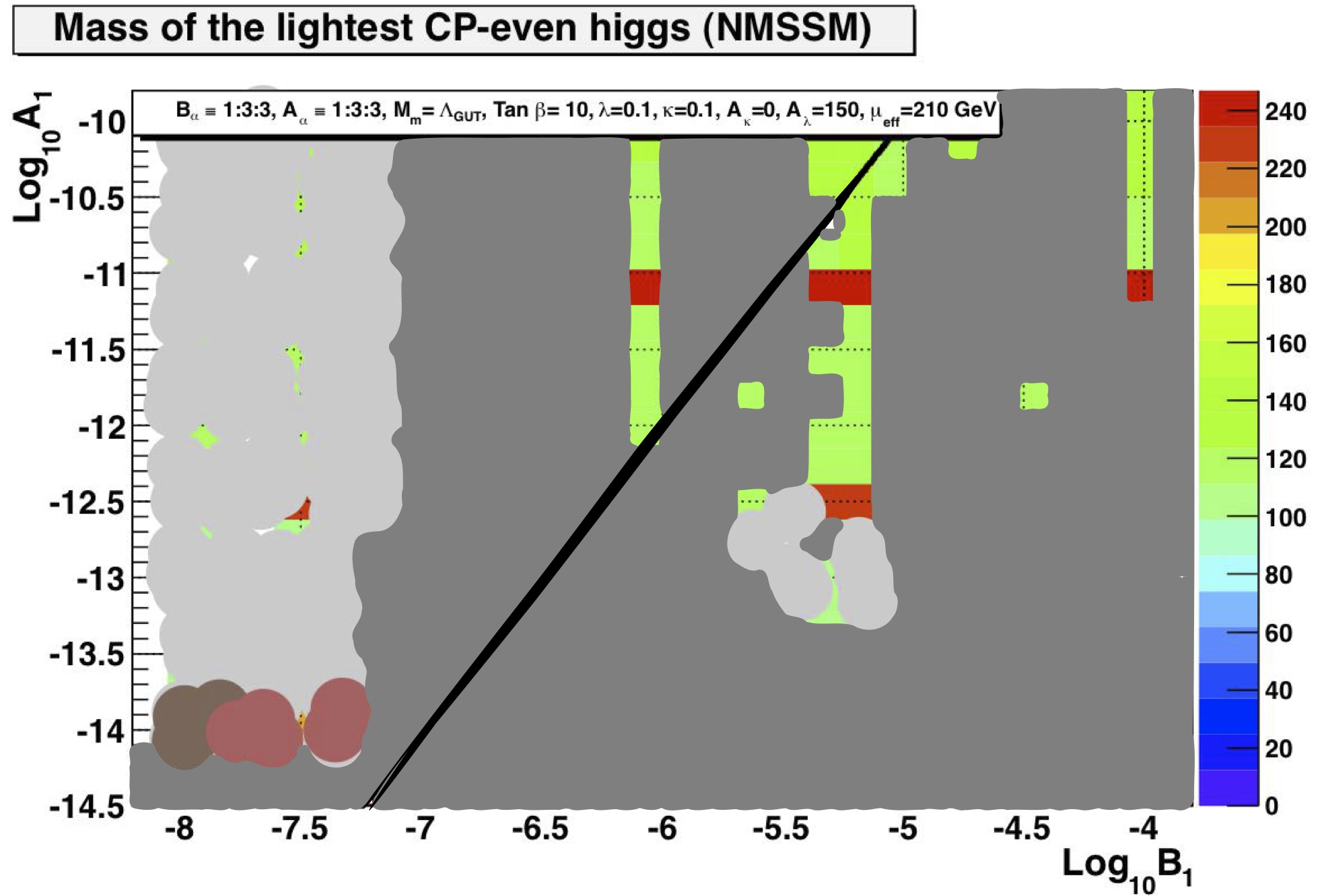}
 \caption{The lightest CP-even Higgs mass in the $(1:3:3\vert1:3:3)$ case.}
 \label{NMSSM2higs}
  }
 % ******************************************************************

\par
Generically, GGM may favor non-split ($M_{\lambda}\simeq m_{\tilde{f}}$) or mildly split SUSY (in split SUSY we have $M_{\lambda}\lesssim m_{\tilde{f}}$), characteristic of RGM and direct/hybrid gauge mediation models respectively. For instance as pointed out already in \cite{Rajaraman:2009ga} and is clear from our  figures there are points in the GGM space that have a spectrum with gauginos lighter than sfermions. These regions in contrast to RGM could give rise to very different phenomenologies. This is characteristic of `direct gauge mediation' models (where messengers participate in SUSY breaking) which usually have gaugino masses suppressed with respect to scalar masses. 
\par
In Figs.\,\ref{NMSSM1} and \ref{NMSSM2} we show for the NMSSM the distribution of $(g-2)_\mu$,  NLSP and lightest CP-even Higgs mass in a section of the viable GGM parameter space. For our choice of parameters it is observed that the $(g-2)_\mu$ is not very large and in the viable regions the $\tilde{\chi}^0$ is the NLSP.
 \par
 In Table\,\ref{bmp1} we show the dependence on the correlation function ratios for a benchmark point, in the MSSM, chosen so that the lightest CP-even Higgs mass is near the LEP bound. It is noticed that for this benchmark point in going from $(1:1:1\vert1:1:1)_{\text{\tiny{MSSM}}}\,\rightarrow\,(1:3:3\vert1:3:3)_{\text{\tiny{MSSM}}}$ the flavor-physics contributions are unchanged but the value of the $a_\mu$ contribution drops roughly by a factor of $4$. This may be understood partially by noting that the $\tilde{\chi}^0,\,\tilde{\chi}^\pm$ masses have increased. Thus as alluded to, while discussing the similar NLSP topography in the two cases, the low-energy observables are quite different at each point in the parameter space. Table\,\ref{bmp2} shows similarly a benchmark point for the NMSSM case. The choice of the benchmark points in both the MSSM and NMSSM were guided mainly by requirements of a viable mass spectra and low-energy observables, but we have specifically picked a point which gives a relatively light CP-even Higgs (close to the current LEP bound).

 % **************************************************************************
\begin{center}
{
\begin{table}
  \begin{tabular}{ |c|c||c|c|}
  \hline
 Description/Units&Quantity & $(1:1:1\vert1:1:1)$ & $(1:3:3\vert1:3:3)$\\
\hline
Inputs (High-scale): 
       & $\log_{10} B_1$ & -6.4   & -6.4   \\
        & $\log_{10} A_1$ & -17.5   & -17.5    \\
       & $M_m$    & $\Lambda_{GUT}$ & $\Lambda_{GUT}$  \\ 
       \hline
       Inputs (Low-scale)
       & $\tan\beta$ & 10     & 10      \\[1.5mm]
        \hline
Low energy & $\Delta a_{\mu}$  & $139\times 10^{-11}$& $34.9\times 10^{-11}$ \\
contributions: &${\rm BR}(B \to X_s \gamma) $  & $3.02\times 10^{-4}$ & $3.10\times 10^{-4}$  \\
          & ${\rm BR}(B_s \to \mu^+ \mu^-)$   & $3.4\times 10^{-9}$& $3.16\times 10^{-9}$ \\
          & ${\rm BR}(B_u \to \tau \nu_\tau)$  & $9.3\times 10^{-5}$& $9.41\times 10^{-5}$  \\          
          &${\rm BR}(B \to D^0 \tau \nu_\tau)$ & $6.9\times 10^{-3}$&$6.91\times 10^{-3}$ \\[1.5mm]
 \hline
gluino (GeV) : & $M_3$ & 945.3 & 2608  \\
\hline
charginos: & $m_{\chi_1^{\pm}}$   & 308.2& 980.7 \\
(GeV) & $m_{\chi_2^{\pm}}$ & 548.7& 1399.4 \\[1.5mm]
\hline
neutralinos: & $m_{\chi_1^0}$ & 164.2 & 167.1 \\
          (GeV)   & $m_{\chi_2^0}$ & 308.4 & 980.7 \\
             & $m_{\chi_3^0}$ & 533.2 & 1391 \\
             & $m_{\chi_4^0}$ & 549& 1399.3 \\[1.5mm]
\hline
Higgs: & $m_{h^0}$     & 114.3& 121\\
    (GeV)   & $m_{H^0}$     & 588.6 & 1578.8 \\
       & $m_{A^0}$     & 588.3& 1578.7\\
       & $m_{H^{\pm}}$ & 594 & 1581 \\[1.5mm]
\hline
squarks: & $m_{\tilde{t}_1}$ & 654.3& 1841\\
   (GeV)      & $m_{\tilde{t}_2}$ & 848.9& 2204.7  \\
         & $m_{\tilde{u}_L}$ & 862.5 & 2355 \\
         & $m_{\tilde{u}_R}$ & 833 & 2243 \\
         & $m_{\tilde{d}_L}$ & 866 & 2356.2  \\
         & $m_{\tilde{d}_R}$ & 830.6& 2244 \\[1.5mm]
\hline
sleptons: & $m_{\tilde{\tau}_1}$ & 147.4& 128.5\\
   (GeV)       & $m_{\tilde{\tau}_2}$ & 276.8& 763.7\\
          & $m_{\tilde{\mu}_L}$    & 274.3 & 764.6 \\
          & $m_{\tilde{\mu}_R}$    & 155.7& 149.8 \\
          & $m_{\tilde{\nu}_{\mu_L}}$  & 263.1 & 760.7  \\
          & $m_{\tilde{\nu}_{\tau_L}}$  & 262.3 & 759.1 \\[1.5mm]
\hline
\end{tabular}
\vspace{0.125in}
\caption{A benchmark point in the GGM parameter space for two correlation function ratios (\textbf{MSSM}). The point was chosen so as to have a light Higgs very near the LEP limit in the $(1:1:1\vert1:1:1)$ case. This point is completely ruled out in the $(1:1/3:1/3\vert1:1/3:1/3)$ case from collider, flavor and $(g-2)_\mu$ bounds. In both the cases above, $\tilde{\tau}$ is the NLSP. Among other things, also note that changing the ratio from $(1:1:1\vert1:1:1)\rightarrow (1:3:3\vert1:3:3)$ causes the $(g-2)_\mu$ contribution to decrease by a factor of $4$, with a corresponding increase in the $\tilde{\chi}^\pm$ masses. }
\label{bmp1}
\end{table}
}
\end{center}

 %  ************************************************************

 % **************************************************************
\begin{center} 
{
\begin{table}
  \begin{tabular}{|c|c||c|c|}
  \hline
  Description/Units& Quantity & $(1:1:1\vert1:1:1)$ & $(1:2:2\vert1:2:2)$ \\
\hline
Inputs (High-scale): 
       & $\log_{10} B_1$ & -6.8  & -6.8  \\
        & $\log_{10} A_1$ & -12.1   & -12.1 \\
         & $M_m$    &  $\Lambda_{GUT}$ &  $\Lambda_{GUT}$  \\   
       \hline
       Inputs (Low-scale)
       & $\lambda$ & 0.1     & 0.1    \\
       & $\kappa$ & 0.1     & 0.1     \\   
       & $A_{\lambda}$   & 150     & 150  \\
       & $A_{\kappa}$   & 0     & 0  \\
        & $\mu_{\text{\textit{eff.}}}$ (GeV)  & 210     & 210  \\
         & $\tan\beta$ & 10     & 10     \\
 \hline
Low energy & $\Delta a_{\mu}$& $41\times 10^{-11}$& $27\times 10^{-11}$ \\
contributions: &${\rm BR}(B \to X_s \gamma) $& $3.34\times 10^{-4}$ & $3.45\times 10^{-4}$  \\
          & ${\rm BR}(B_s \to \mu^+ \mu^-)$ & $3.51\times 10^{-9}$& $3.53\times 10^{-9}$ \\
          & ${\rm BR}(B_u \to \tau \nu_\tau)$& $13.1\times 10^{-5}$& $13.1\times 10^{-5}$ \\         
 \hline
gluino (GeV): & $M_3$ & 542.8 & 1134  \\
\hline
charginos: & $m_{\chi_1^{\pm}}$ & 113.8& 191.1\\
(GeV) & $m_{\chi_2^{\pm}}$ & 257& 353.3 \\
\hline
neutralinos: & $m_{\chi_1^0}$ & 62 & 38.7 \\
          (GeV)   & $m_{\chi_2^0}$ & 116 & 192.5 \\
             & $m_{\chi_3^0}$ & 223.7 & 223.4\\
             & $m_{\chi_4^0}$ & 256 & 353 \\
             & $m_{\chi_5^0}$ & 428.0 & 432 \\
\hline
Higgs: & $m_{h_1^0}$     & 114.1& 117 \\
 (GeV) & $m_{h_2^0}$     & 420.2& 420.3 \\
             & $m_{h_3^0}$     & 853.7 & 846.4  \\
       	   & $m_{a_1^0}$     & 10.7& 9.54 \\
	    & $m_{a_2^0}$     & 853.3& 846 \\
       	   & $m_{H^{\pm}}$ & 857.3 & 850 \\
\hline
squarks: & $m_{\tilde{t}_1}$ & 1084 & 2060.5\\
   (GeV)      & $m_{\tilde{t}_2}$ & 1309.7 & 2112 \\q.t.
         & $m_{\tilde{u}_L}$ & 1391.5& 1893\\
         & $m_{\tilde{u}_R}$ & 1281.7  & 1524 \\
         & $m_{\tilde{d}_L}$ & 1393.6 & 1894 \\
         & $m_{\tilde{d}_R}$ & 1170.2 & 1562 \\
\hline
sleptons: & $m_{\tilde{\tau}_1}$ & 910.6 &655.3 \\
   (GeV)       & $m_{\tilde{\tau}_2}$ & 915.5 & 935.7\\
          & $m_{\tilde{\mu}_L}$    & 915.2 & 1106.2 \\
          & $m_{\tilde{\mu}_R}$    & 916 & 1128\\
          & $m_{\tilde{\nu}_{\mu_L}}$  &912.7 & 1125.5  \\
          & $m_{\tilde{\nu}_{\tau_L}}$  & 911& 932.5\\
\hline
\end{tabular}
\caption{Benchmark points for two correlation function ratios (\textbf{NMSSM}). The point was again chosen so as to have the mass of the lightest CP-even Higgs boson near the LEP limit. For larger correlation function ratios, for instance  $(1:3:3\vert1:3:3)$ or $(1:1/3:1/3\vert1:1/3:1/3)$, this point is again found to be non-viable. Note that we have only implemented 1-loop RGEs (Appendix B ) for this estimation as mentioned before.}
\label{bmp2}
\end{table}
}
\end{center}

% ****************************************************************

\par
Regions where $m_{\tilde{f}} \ll M_{\lambda} $ may imply that the Poppitz-Trivedi type mass terms (these are R-symmetry preserving terms accounted for by a non-vanishing messenger supertrace) in these regions are small~\cite{Carpenter:2008he}. In our specific context the $M_{\lambda}\lesssim m_{\tilde{f}}$ may be understood as a consequence of a hierarchy between the $\tilde{B}_r$ and $\tilde{A}_r$ correlation functions, for all other parameters remaining the same. Thus by raising the overall scale of the $\tilde{A}_r$ relative to the $\tilde{B}_r$ the spectra may be inverted from an RGM like spectra. For example, consider the starting point $(-6.5,-13.0)$ in the $(\log_{10} \tilde{B}_1,\log_{10} \tilde{A}_1)$ space (again with $\tan\beta=10$ and $M_{m}=\Lambda_{GUT}$) which would roughly correspond to RGM in the $(1:1:1\vert1:1:1)_{\text{\tiny{MSSM}}}$ case. Then if we consider the re-scaling of the $\tilde{A}_r$ correlation functions as
\ba
(1:1:1\vert1:1:1)_{\text{\tiny{MSSM}}}\rightarrow (1:1:1\vert\, \mathcal{K}: \mathcal{K}: \mathcal{K})_{\text{\tiny{MSSM}}} \; ,  
\ea
for $ \mathcal{K} \gg 1$ as before, this changes the gluino-sfermion mass ratios to give an inverted spectrum
\ba
\left(\frac{m_{\lambda}}{m_{\tilde{f}}}\right)_{\text{\tiny{EW}}}~&\approx&~1~~\rightarrow ~~\left(\frac{m_{\lambda}}{m_{\tilde{f}}}\right)_{\text{\tiny{EW}}}~\ll~1  \; .\\
\ea
\par
There also exist even at the low scale other very approximate scaling relations, between the superpartner masses and the hidden sector correlation function ratios, in large parts of the viable parameter space that are not completely washed out by the RG running. For instance it is found that in regions with small values of $\tilde{A}_r$ for
\be
(1:1:1\vert \,1:1:1)\rightarrow (\mathcal{K}:\mathcal{K}:\mathcal{K}\vert\, \,1:1:1)  \; ,  
\eq
at the high scale, we have the approximate scaling
\be
(m_{\tilde{g}},m_{\tilde{\chi_0}},m_{\tilde{\chi_{\pm}}},m_{\tilde{q}},m_{\tilde{l}},m_{\tilde{\nu}})_{\text{\tiny{EW}}} \rightarrow ~\sim (\mathcal{K}\,m_{\tilde{g}},\mathcal{K}\,m_{\tilde{\chi_0}},\mathcal{K}\,m_{\tilde{\chi_{\pm}}} ,\mathcal{K}\,m_{\tilde{q}},\mathcal{K}\,m_{\tilde{l}},\mathcal{K}\,m_{\tilde{\nu}}) _{\text{\tiny{EW}}} \; .
\eq
This may be understood by noting that when the $\tilde{A}_r$ are relatively small the initial RG running of the sfermion masses is dominated by the gaugino mass terms (see Appendix A and B). The above scaling is just tantamount to changing the over-all scale of the $\tilde{B}_r$ correlation functions, and hence the gaugino masses at the high-scale. To give a concrete example consider in the $(1:1:1\vert \,1:1:1)$ case the point $(-6.0,-16.0)$ in the $(\log_{10} \tilde{B}_1,\log_{10} \tilde{A}_1)$ space. Then we have the following at the low scale :
\ba
m_{\tilde{g}}~(\rm{GeV})&:&~ 2215\xrightarrow{\mathcal{K}=1/2}1166\xrightarrow{\mathcal{K}=2} 4223 \xrightarrow{\mathcal{K}=5} 9951 \; ,  \\
m_{\tilde{\chi}^0_1}~(\rm{GeV})&:&~ 434\xrightarrow{\mathcal{K}=1/2}209\xrightarrow{\mathcal{K}=2} 898 \xrightarrow{\mathcal{K}=5} 2339 \; ,  \\
m_{\tilde{\chi}^0_4}~(\rm{GeV})&:&~ 1208\xrightarrow{\mathcal{K}=1/2}665\xrightarrow{\mathcal{K}=2} 2206 \xrightarrow{\mathcal{K}=5} 4903 \; ,  \\
m^{\text{\tiny{lightest}}}_{\tilde{\chi}^\pm}~(\rm{GeV})&:&~ 816\xrightarrow{\mathcal{K}=1/2}394\xrightarrow{\mathcal{K}=2} 1663 \xrightarrow{\mathcal{K}=5} 4226 \; ,  \\
m^{\text{\tiny{heaviest}}}_{\tilde{\chi}^\pm}~(\rm{GeV})&:&~ 1207\xrightarrow{\mathcal{K}=1/2}665\xrightarrow{\mathcal{K}=2} 2206 \xrightarrow{\mathcal{K}=5} 4903 \; ,  
\ea

\ba
m^{\text{\tiny{lightest}}}_{\tilde{q}}~(\rm{GeV})&:&~ 1570\xrightarrow{\mathcal{K}=1/2}815\xrightarrow{\mathcal{K}=2} 2991 \xrightarrow{\mathcal{K}=5} 6983 \; ,  \\
m^{\text{\tiny{heaviest}}}_{\tilde{q}}~(\rm{GeV})&:&~ 2007\xrightarrow{\mathcal{K}=1/2}1065\xrightarrow{\mathcal{K}=2} 3794 \xrightarrow{\mathcal{K}=5} 8835 \; ,  \\
m^{\text{\tiny{lightest}}}_{\tilde{l}}~(\rm{GeV})&:&~ 367\xrightarrow{\mathcal{K}=1/2}184\xrightarrow{\mathcal{K}=2} 729 \xrightarrow{\mathcal{K}=5} 1809 \; ,  \\
m^{\text{\tiny{heaviest}}}_{\tilde{l}}~(\rm{GeV})&:&~ 667\xrightarrow{\mathcal{K}=1/2}342\xrightarrow{\mathcal{K}=2} 1312 \xrightarrow{\mathcal{K}=5} 3207 \; ,  \\
m^{\text{\tiny{lightest}}}_{\tilde{\nu}}~(\rm{GeV})&:&~ 662\xrightarrow{\mathcal{K}=1/2}331\xrightarrow{\mathcal{K}=2} 1306 \xrightarrow{\mathcal{K}=5} 3198 \; ,  \\
m^{\text{\tiny{heaviest}}}_{\tilde{\nu}}~(\rm{GeV})&:&~ 663\xrightarrow{\mathcal{K}=1/2}332\xrightarrow{\mathcal{K}=2} 1310 \xrightarrow{\mathcal{K}=5} 3205 \; .
\ea

\par
To conclude, in this section we explored the effects of different hidden-sector correlation function ratios on low-energy observables and the SUSY mass spectra. Our main guiding points were observables from  flavor physics and the muon anomalous magnetic moment. It was found that large regions of the GGM parameter space were disfavored (though not necessarily ruled out) from the viewpoint of these and there were interesting regions that gave sizeable contributions to $(g-2)_\mu$ on par with the current discrepancy.

\section{Summary}
It is indicative from the present study and some of the earlier ones~\cite{Abel:2009ve} that low-energy observables can play a very complementary role to collider studies in deducing viable and interesting regions in the GGM parameter space. Specifically, we saw that there are strong relationships between the correlation function ratios in the hidden sector and the values of the low-energy observables and mass spectra at the low scale. Let us recapitulate some of the salient features noticed in the GGM case: 
\bi
\item Muon anomalous moment and flavor physics place strong constraints. The detailed features of the parameter space are found to depend sensitively on the correlation function  ratios.
\item The GGM scenario can more easily accommodate precision gauge coupling unification. This is due to the larger freedom of non-universal gaugino masses (due to different $\tilde{B}_r$) that could lead to a light $\tilde{g}$ or heavy $\tilde{w}$ relative to RGM scenarios.
\item The topography of NLSP is found to depend on the correlation function ratios. Even in cases where the NLSP topography is naively similar there are significant differences in the mass spectra and low-energy observable values. A case in point is the $(1:1:1\vert1:1:1)$ and $(1:3:3\vert1:3:3)$ cases that we considered. Also, it is found that there are interesting regions and boundaries in the parameter space with multiple NLSPs. In these regions co-annihilations may be important and could lead to interesting phenomenology.
\item We find that in most regions with interesting values of $(g-2)_\mu$ (i.e. within $2\sigma$ of the current discrepancy) the mass of the lightest CP-even Higgs boson is relatively small.
\item There are very approximate, but interesting, scaling relations observed in large parts of the GGM parameter space (between the correlation function ratios at the high scale and sparticle masses at the low scale) that are not completely washed out by RG running.
\item There are viable regions where the gauginos are lighter than sfermions. These regions may have interesting phenomenological implications. This may be understood as a hierarchy between the $\tilde{B}_r$ and $\tilde{A}_r$ correlation functions.
\item There are also allowed regions where the RGM intuition of large $\tilde{q}$ to $\tilde{l}$ ratios is no longer true. These are regions where the correlation functions $\tilde{A}_r$ are large. These regions favor larger $\tilde{l}$ masses relative to $\tilde{q}$ due to the fact that $\tilde{l}$ hypercharges are typically larger than the corresponding $\tilde{q}$ hypercharges. Thus there is a slight enhancement to the $\tilde{l}$ masses in these regions.
\item The $\tilde{g}$ to $\tilde{\chi}$ mass ratio may again be small in contrast to RGM expectations. This may be partially understood as a consequence of the hierarchy between $\tilde{B}_2$ and $\tilde{B}_3$ correlation functions.
\item The $\tilde{A}_r$ are unbounded from below. This is the limit where the sfermion soft masses at the GUT scale tend to zero, since we have put the Fayet-Iliopoulos term $\zeta=0$, and the initial sfermion mass generation is essentially due to the gaugino masses during RGE.

\ei

\acknowledgments
 I thank Jonathan L. Rosner for useful comments and a careful reading of the manuscript. Discussions with Carlos E. M. Wagner and David Shih are gratefully acknowledged.  I also thank Matthew Dolan, Patrick Draper, David Krohn and Arjun Menon for discussions. This work was supported in part by the United States Department of Energy under Grant No. DE-FG02-90ER40560.

\appendix
\section*{Appendix A. 1-Loop MSSM renormalisation group equations}
\addcontentsline{toc}{section}
{Appendix A. MSSM Renormalisation group equations}

For completeness we list the 1-loop RG equations for the MSSM and NMSSM. The RGEs are written in the third family dominant approximation with the definitions $t=\log(Q^2/Q_0^2)$ and $U(1)_Y$ gauge coupling constant $g_1^2=\frac{3}{5}(g_1^{\text{\tiny{GUT}}})^2$. The terms in the MSSM superpotential and the soft terms are as defined in Eqs. (\ref{WMSSM}) and (\ref{SOFTMSSM}). For the complete 2-loop RG equations, see for instance \cite{{Martin:1993yx},{Yamada:1993ga}}.

\subsection*{A.1 Gauge and Yukawa couplings}

\bea
16\pi^2 \frac{dg_1^2}{dt} &=& 11g_1^4 \; , \nn
\\
16\pi^2 \frac{dg_2^2}{dt} &=& g_2^4 \; , \nn \\
16\pi^2 \frac{dg_3^2}{dt} &=& - 3 g_3^4  \; , \nn
\\
\eea
\bea
16\pi^2 \frac{dy_t^2}{dt} &=& y_t^2\bigg( 6y_t^2 + y_b^2 - \frac{13}{9}g_1^2 - 3g_2^2 - \frac{16}{3}g_3^2 \bigg) \nn \\
16\pi^2 \frac{dy_b^2}{dt} &=& y_b^2\bigg( 6y_b^2 + y_t^2 + y_\tau^2 - \frac{7}{9}g_1^2 - 3g_2^2 - \frac{16}{3}g_3^2 \bigg) \nn \\
16\pi^2 \frac{dy_\tau^2}{dt} &=& y_\tau^2\Big( 4y_\tau^2 + 3y_b^2 - 3g_1^2 - 3g_2^2 \Big) 
\eea

\subsection*{A.2 Gaugino masses}
\bea
16\pi^2 \frac{dM_1}{dt} &=& 11g_1^2M_1\; , \nn
\\
16\pi^2 \frac{dM_2}{dt} &=& g_2^2M_2\; , \nn
\\
16\pi^2 \frac{dM_3}{dt} &=& -3g_3^2M_3 
\eea

\subsection*{A.3 Squark and slepton masses}

Let 
\bea
\xi &=& {\rm Tr}\big[{\bf m}_Q^2 - 2{\bf m}_U^2 + {\bf m}_D^2
- {\bf m}_L^2 + {\bf m}_E^2\big]
+ m_{H_u}^2 - m_{H_d}^2\; , \nn \\
M_t^2 &= & m_{Q_3}^2+m_{U_3}^2+m_{H_u}^2+A_t^2\; , \nn \\
M_b^2 &= & m_{Q_3}^2+m_{D_3}^2+m_{H_d}^2+A_b^2\; , \nn \\
M_\tau^2 &= & m_{L_3}^2+m_{E_3}^2+m_{H_d}^2+A_\tau^2\; ,
\eea
where the terms in bold are matrices with respect to generations.
Then the RG equations are then

\bea
16\pi^2 \frac{dm_{Q_a}^2}{dt} &=&
\delta_{a3}y_t^2M_t^2 + \delta_{a3}y_b^2M_b^2 - \frac{1}{9}g_1^2M_1^2
- 3g_2^2M_2^2 - \frac{16}{3}g_3^2M_3^2 + \frac{1}{6}g_1^2\xi \nn \\
16\pi^2 \frac{dm_{U_a}^2}{dt} &=&
2\delta_{a3}y_t^2M_t^2 - \frac{16}{9}g_1^2M_1^2
- \frac{16}{3}g_3^2M_3^2 - \frac{2}{3}g_1^2\xi \nn \\
16\pi^2 \frac{dm_{D_a}^2}{dt} &=&
2\delta_{a3}y_b^2M_b^2 - \frac{4}{9}g_1^2M_1^2
- \frac{16}{3}g_3^2M_3^2 + \frac{1}{3}g_1^2\xi \nn \\
16\pi^2 \frac{dm_{L_a}^2}{dt} &=&
\delta_{a3}y_\tau^2M_\tau^2 - g_1^2M_1^2 - 3g_2^2M_2^2 - \frac{1}{2}g_1^2\xi \nn \\
16\pi^2 \frac{dm_{E_a}^2}{dt} &=&
2\delta_{a3}y_\tau^2M_\tau^2 - 4g_1^2M_1^2 + g_1^2\xi
\eea

\subsection*{A.4 MSSM Higgs masses}

\bea
16\pi^2 \frac{dm_{H_u}^2}{dt} &=&
3y_t^2M_t^2  - g_1^2M_1^2 - 3g_2^2M_2^2 + \frac{1}{2}g_1^2\xi \nn \\
16\pi^2 \frac{dm_{H_d}^2}{dt} &=&
3y_b^2M_b^2 + y_\tau^2M_\tau^2 - g_1^2M_1^2
- 3g_2^2M_2^2 - \frac{1}{2}g_1^2\xi 
\eea

\subsection*{A.5 MSSM Trilinear couplings (Rescaled)}

\bea
16\pi^2 \frac{dA_t}{dt} &=& 6y_t^2A_t + y_b^2A_b + \frac{13}{9}g_1^2M_1 + 3g_2^2M_2 + \frac{16}{3}g_3^2M_3\nn \\
16\pi^2 \frac{dA_b}{dt} &=& 6y_b^2A_b + y_t^2A_t + y_\tau^2A_\tau+\frac{7}{9}g_1^2M_1 + 3g_2^2M_2 + \frac{16}{3}g_3^2M_3\nn \\
16\pi^2 \frac{dA_\tau}{dt} &=& 4y_\tau^2A_\tau + 3y_b^2A_b+ 3g_1^2M_1 + 3g_2^2M_2  \nn \\
16\pi^2 \frac{dA_\mu}{dt} &=& 3y_b^2A_b + y_\tau^2A_\tau+ 3g_1^2M_1 + 3g_2^2M_2 
\eea

\subsection*{A.6 The $\mu$ and $B\mu$ terms}
\bea
16\pi^2 \frac{d\mu}{dt} &=&
\mu\Big( 3y_t^2 + 3y_b^2
+ y_\tau^2 - g_1^2 - 3g_2^2 \Big)\; ,\nn \\
16\pi^2 \frac{dB}{dt} &=&
\Big( 3y_t^2 A_t+ 3y_b^2 A_b
+ y_\tau^2 A_\tau- g_1^2M_1 - 3g_2^2 M_2 \Big)\; .
\eea
%
%%%%%%%%%%%%%%%%%%%%%%%%%%%%%%%%%%%%%%%%%%%%%%%

\section*{Appendix B. 1-Loop NMSSM renormalisation group equations}
\addcontentsline{toc}{section}
{Appendix B. NMSSM Renormalisation group equations}
Again the RGEs  \cite{{Martin:1993yx},{Yamada:1993ga}} are written assuming the running is dominated by the third family. We as before define $t=\log(Q^2/Q_0^2)$ and $g_1^2=\frac{3}{5}(g_1^{\text{\tiny{GUT}}})^2$. The NMSSM superpotential and soft terms are as defined in Eqs. (\ref{WNMSSM}) and (\ref{SOFTNMSSM}).

\subsection*{B.1 Gauge and Yukawa couplings}

\bea
16\pi^2 \frac{dg_1^2}{dt} &=& 11g_1^4 \; , \nn
\\
16\pi^2 \frac{dg_2^2}{dt} &=& g_2^4 \; , \nn \\
16\pi^2 \frac{dg_3^2}{dt} &=& - 3 g_3^4  \; , \nn
\\
\eea
\bea
16\pi^2 \frac{dy_t^2}{dt} &=& y_t^2\bigg( 6y_t^2 + y_b^2 + \l^2
- \frac{13}{9}g_1^2 - 3g_2^2 - \frac{16}{3}g_3^2 \bigg) \nn \\
16\pi^2 \frac{dy_b^2}{dt} &=& y_b^2\bigg( 6y_b^2 + y_t^2 + y_\tau^2 +
\l^2 - \frac{7}{9}g_1^2 - 3g_2^2 - \frac{16}{3}g_3^2 \bigg) \nn \\
16\pi^2 \frac{dy_\tau^2}{dt} &=& y_\tau^2\Big( 4y_\tau^2 + 3y_b^2 + \l^2
- 3g_1^2 - 3g_2^2 \Big) 
\eea

\bea
16\pi^2 \frac{d\l^2}{dt} &=& \l^2\Big(3y_t^2 + 3y_b^2 + y_\tau^2 +4\l^2
+ 2\k^2 - g_1^2 - 3g_2^2 \Big) \nn \\
16\pi^2 \frac{d\k^2}{dt} &=& \k^2\Big(6\l^2 +6\k^2\Big) 
\eea

\subsection*{B.2 Gaugino masses}
\bea
16\pi^2 \frac{dM_1}{dt} &=& 11g_1^2M_1\; , \nn
\\
16\pi^2 \frac{dM_2}{dt} &=& g_2^2M_2\; , \nn
\\
16\pi^2 \frac{dM_3}{dt} &=& -3g_3^2M_3 
\eea

\subsection*{B.3 Squark and slepton masses}

Let 
\bea
\xi &=& {\rm Tr}\big[{\bf m}_Q^2 - 2{\bf m}_U^2 + {\bf m}_D^2
- {\bf m}_L^2 + {\bf m}_E^2\big]
+ m_{H_u}^2 - m_{H_d}^2\; , \\
M_t^2 &= & m_{Q_3}^2+m_{U_3}^2+m_{H_u}^2+A_t^2\; , \nn \\
M_b^2 &= & m_{Q_3}^2+m_{D_3}^2+m_{H_d}^2+A_b^2\; , \nn \\
M_\tau^2 &= & m_{L_3}^2+m_{E_3}^2+m_{H_d}^2+A_\tau^2\; , \nn
\eea
Then the RG equations are

\bea
16\pi^2 \frac{dm_{Q_a}^2}{dt} &=&
\delta_{a3}y_t^2M_t^2 + \delta_{a3}y_b^2M_b^2 - \frac{1}{9}g_1^2M_1^2
- 3g_2^2M_2^2 - \frac{16}{3}g_3^2M_3^2 + \frac{1}{6}g_1^2\xi \nn \\
16\pi^2 \frac{dm_{U_a}^2}{dt} &=&
2\delta_{a3}y_t^2M_t^2 - \frac{16}{9}g_1^2M_1^2
- \frac{16}{3}g_3^2M_3^2 - \frac{2}{3}g_1^2\xi \nn \\
16\pi^2 \frac{dm_{D_a}^2}{dt} &=&
2\delta_{a3}y_b^2M_b^2 - \frac{4}{9}g_1^2M_1^2
- \frac{16}{3}g_3^2M_3^2 + \frac{1}{3}g_1^2\xi \nn \\
16\pi^2 \frac{dm_{L_a}^2}{dt} &=&
\delta_{a3}y_\tau^2M_\tau^2 - g_1^2M_1^2 - 3g_2^2M_2^2 - \frac{1}{2}g_1^2\xi \nn \\
16\pi^2 \frac{dm_{E_a}^2}{dt} &=&
2\delta_{a3}y_\tau^2M_\tau^2 - 4g_1^2M_1^2 + g_1^2\xi
\eea

\subsection*{B.4 NMSSM Higgs masses}
Let
\bea
M_\l^2 &= & m_{H_u}^2+m_{H_d}^2+m_N^2+A_\l^2\; , \nn \\
M_\k^2 &= & 3m_N^2+A_\k^2\; , \nn 
\eea

\bea
16\pi^2 \frac{dm_{H_u}^2}{dt} &=&
3y_t^2M_t^2 + \l^2M_\l^2 - g_1^2M_1^2 - 3g_2^2M_2^2 + \frac{1}{2}g_1^2\xi \nn \\
16\pi^2 \frac{dm_{H_d}^2}{dt} &=&
3y_b^2M_b^2 + y_\tau^2M_\tau^2 + \l^2M_\l^2 - g_1^2M_1^2
- 3g_2^2M_2^2 - \frac{1}{2}g_1^2\xi \nn \\
16\pi^2 \frac{dm_S^2}{dt} &=&
2\l^2M_\l^2 + 2\k^2M_\k^2
\eea

\subsection*{B.5 NMSSM Trilinear couplings (Rescaled)}

\bea
16\pi^2 \frac{dA_t}{dt} &=& 6y_t^2A_t + y_b^2A_b  + \l^2A_\l
+ \frac{13}{9}g_1^2M_1 + 3g_2^2M_2 + \frac{16}{3}g_3^2M_3\nn \\
16\pi^2 \frac{dA_b}{dt} &=& 6y_b^2A_b + y_t^2A_t + y_\tau^2A_\tau
+ \l^2A_\l +\frac{7}{9}g_1^2M_1 + 3g_2^2M_2 + \frac{16}{3}g_3^2M_3\nn \\
16\pi^2 \frac{dA_\tau}{dt} &=& 4y_\tau^2A_\tau + 3y_b^2A_b
+ \l^2A_\l + 3g_1^2M_1 + 3g_2^2M_2  \nn \\
16\pi^2 \frac{dA_\mu}{dt} &=& 3y_b^2A_b + y_\tau^2A_\tau
+ \l^2A_\l + 3g_1^2M_1 + 3g_2^2M_2 
\eea

\bea
16\pi^2 \frac{dA_\l}{dt} &=& 4\l^2A_\l + 3y_t^2A_t + 3y_b^2A_b
+ y_\tau^2A_\tau + 2\k^2A_\k + g_1^2M_1 + 3g_2^2M_2 \nn \\
16\pi^2 \frac{dA_\k}{dt} &=& 6\k^2A_\k  + 6\l^2A_\l
\eea

\subsection*{B.6 Other parameters of the NMSSM}
\bea
32\pi^2 \frac{d\mu}{dt} &=&
\mu\Big( 3y_t^2 + 3y_b^2
+ y_\tau^2 + 2\l^2 - g_1^2 - 3g_2^2 \Big)\; , \nn \\
16\pi^2 \frac{d\mu'}{dt} &=&
\mu'\Big( 2\l^2 + 2\k^2 \Big)\; .
\eea
\bea
32\pi^2 \frac{dm_3^2}{dt} &=&
3y_t^2\big(m_3^2 + 2\mu A_t\big) + 3y_b^2\big(m_3^2 + 2\mu A_b\big)
+ y_\tau^2\big(m_3^2 + 2\mu A_\tau\big) \nn \\
&+& 2\l^2\big(3m_3^2 + 2\mu A_\l\big) +  2\l\k m_N'^2
- g_1^2\big(m_3^2-2\mu M_1\big)
- 3g_2^2\big(m_3^2-2\mu M_2\big) \nn \\
16\pi^2 \frac{dm_N'^2}{dt} &=&
2\l^2\big(m_N'^2 + 2\mu'A_\l\big) + 4\k^2\big(m_N'^2 + \mu'A_\k\big)
+ 4\l\k m_3^2 \nn \\
\eea

\bea
16\pi^2 \frac{d\xi_F}{dt} &=&
\xi_F\Big( \l^2 + \k^2 \Big)\nn \\
16\pi^2 \frac{d\xi_N}{dt} &=&
\l^2\big(\xi_N+2A_\l\xi_F\big) + \k^2\big(\xi_N+2A_\k\xi_F\big) \nn \\
&+& 2\l\big(m_3^2(A_\l+\mu')+\mu(m_{H_u}^2+m_{H_d}^2)\big)
+ \k\big(m_N'^2(A_\k+\mu')+2\mu m_N^2\big) \nn \\
\eea

 % ----------------------------------------------------------------------------------------------------------

\bibliography{draft}
\bibliographystyle{jhep}

\end{document}